\documentclass[twocolumn,showpacs,amssymb,floatfix,deluxe,prd]{revtex4}

\usepackage{graphicx}
\usepackage{dcolumn}
\usepackage{bm}
\usepackage{epsfig}

\begin{document}


\title{A Theory of Cosmic Rays}

\author{Arnon Dar}%
\affiliation{Department of Physics and Space Research 
Institute, \\
Technion, Haifa 32000, Israel}

\author{A. De R\'ujula}
\affiliation{Theory Unit, CERN,
1211 Geneva 23, Switzerland;
Physics Department, Boston University, USA}%

\date{\today}

\begin{abstract}
We present a theory of non-solar cosmic rays (CRs) in which the bulk
of their observed flux is due to a single type of CR source at all energies. The 
total luminosity of the Galaxy, the broken power-law spectra with their
observed slopes, the position of the `knee(s)'  and `ankle', and the CR
composition and its variation with energy are all predicted in terms of
very simple and completely `standard' physics.  The source of CRs is 
extremely `economical': it has only one parameter to be fitted to 
the ensemble of all of the
mentioned data. All other inputs are `priors', that is, theoretical or
observational items of information independent of the properties of
the source of CRs, and chosen to lie in their pre-established ranges. 
The theory is part of a `unified view of high-energy astrophysics' 
---based on the `Cannonball' model of the relativistic ejecta of 
accreting black holes and neutron stars. 
The model has been extremely successful in predicting
all the novel properties of Gamma Ray Bursts recently
observed with help of the Swift satellite.
If correct, this model is only
lacking a satisfactory theoretical understanding of the `cannon' that
emits the cannonballs in catastrophic processes of accretion.
\end{abstract}

\maketitle

\section{Introduction and outlook}
\label{s1}

The field of cosmic-ray (CR) physics was born as a lucky failure. The 1912
attempt by Victor Hess to measure the {\it decrease} of the Earth's
radioactivity in an ascending balloon gave an opposite result: there was
an extra-terrestrial source of what are now known to be high-energy nuclei
and electrons. Almost a century later, the origin of non-solar CRs
is still a subject of intense research and little consensus 
\cite{Reviews}. We shall refer throughout to non-solar cosmic rays simply
as CRs.

\begin{figure}[]
\centering
 \epsfig{file=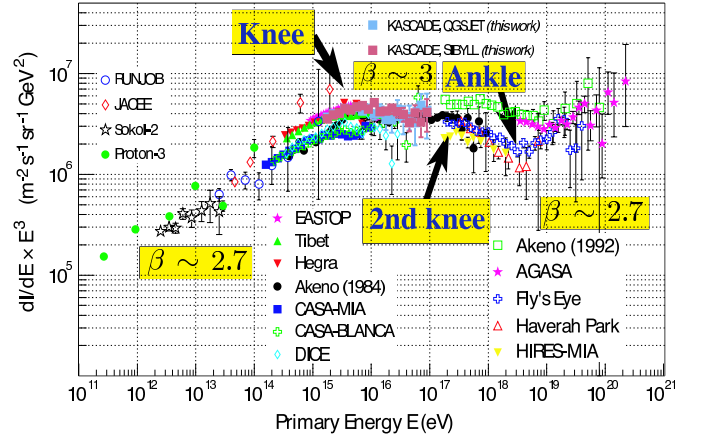,height=5.3cm,width=8.8cm}
\vspace*{5pt}
\caption{The all-particle  CR spectrum wheighted by $E^3$, with 
its three changes of spectral index: the `knee', the `second knee', and the
`ankle' \cite{Kampert}.}
 \label{features}
\end{figure}

Over almost a century, an impressive set of CR data have been gathered, e.g.~the
{\it all-particle} spectrum (of nuclei, without distinction of charge and mass)
has been measured over some 13 orders of magnitude in energy and more than 30
orders of magnitude in flux
(perhaps only Coulomb's law is measured over an
even wider range). 
It has become standard practice to present the spectral data
as the flux $dF/dE$ times a power of energy, which emphasizes the spectral
`features' and the discrepancies between experiments, while de-emphasizing the
pervasive
systematic errors in energy. The all-particle spectrum $E^3\,dF/dE$ is shown in
Fig.~\ref{features} for energies $E>10^{11}$ eV. The
figure shows that the spectrum is roughly describable as a broken power law
\cite{Reviews,Kampert,HIRES1}:
\begin{eqnarray}
&&\;\;\;\;\;\;\;\;\;\;\;\;\;\;{dF / dE} \propto E^{-\beta},\nonumber\\
\beta&\simeq& 2.7;\;\;\; E<E[{\rm knee}]\sim 3\times10^{15}\rm\;eV, \nonumber\\
\beta&\simeq& 3.0;\;\;\; E[{\rm knee}]\,<\,E<\,E[{\rm knee2}]
\sim 2\times10^{17}\rm\;eV, \nonumber\\
\beta&\sim& 3.1;\;\;\; E[{\rm knee2}]\,<\,E<\,E[{\rm ankle}], \nonumber\\
 \beta& \sim& 2.7;\;\;\; E>E[{\rm ankle}]\sim 3\times10^{18}\rm\;eV. 
\label{flux}
\end{eqnarray}

Below ${E[{\rm knee]}}$, protons constitute  $\sim 96\%$
of the CRs at fixed energy {\it per nucleon}. Their flux 
above ${E_p\sim 10}$ GeV is~\cite{Haino}: 
 \begin{equation}
{dF_p\over dE}\simeq {1.37\pm 0.13\over\rm cm^{2}~s~sr~GeV}\;  \left
[E \over \rm GeV \right]^{-2.73\pm 0.03}\, .
\label{protons}
\end{equation}

The conventional theory of CRs \cite{GP} posits that supernova 
remnants are the site of acceleration of (non-solar) CRs
for energies up to $E[{\rm knee}]$. No
consensus on a preferred accelerator site or mechanism exists for energies
between $E[{\rm knee}]$ and $E[{\rm ankle}]$. It has long being argued that CRs
of energy above $E[{\rm ankle}]$ are {\it extragalactic} in origin 
\cite{Cocconi,Morrison}: they cannot be isotropized by the Galactic magnetic fields,
but their observed arrival directions are  isotropic \cite{HIRES2}.
We refer to CRs with  $E>E[{\rm ankle}]$ as ultra-high energy cosmic rays
(UHECRs). They are the subject of great interest, considerable controversy and 
imaginative model-building; see, for instance, the review by J. Cronin in \cite{Reviews}.
In this paper we use many abbreviations. They are
listed in Table \ref{table0}.

\begin{table}
\caption{Frequently used abbreviations
\label{table0}}
\begin{tabular}{lclc}
\colrule
Afterglow(s) & AG(s) \\
Active Galactic Nucleus(i) & AGN(s) \\
Cosmic Background Radiation & CBR \\
Cosmic Microwave Background & CMB \\
$[$Non-solar$]$ Cosmic Ray(s) & CR(s)  \\
Cosmic-Ray Electron(s) & CRE(s) \\
Gamma Background Radiation & GBR \\
$[$Long-duration$]$ Gamma-Ray Burst(s)& GRB(s) \\
Inter-Galactic Medium & IGM \\
Inter-Stellar Medium & ISM \\
Inverse Compton Scattering & ICS \\
Greisen, Zatsepin \& Kuzmin & GZK \\
Lorentz Factor(s) & LF(s) \\
Magnetic Field(s) & MF(s) \\
Short Hard [$\gamma$-ray] Burst(s) & SHBs \\
Starlight & SL \\
Superbubble(s) & SB(s) \\
Supernova(e) & SN(e) \\
Supernova Remnant(s) & SNR(s) \\
Synchrotron Radiation & SR \\
Ultra-High Energy Cosmic Ray(s) & UHECR(s) \\
X-Ray Flash(es) & XRF(s) \\
\end{tabular}
\end{table}

Radio, X-ray and $\gamma$-ray observations of supernova remnants (SNRs) provide
clear evidence that electrons are accelerated to high energies in these sites. So
far, they have not provided unambiguous evidence that SNRs accelerate 
CR nuclei and are their main source in any energy range \cite{Plaga}. 
Moreover, SNRs cannot accelerate CRs to energies as large as 
$E[{\rm knee}]$ \cite{Hillas}, though this point is still debated.
 A direct proof ---such as a localized
source--- of an extragalactic origin of the UHECRs was lacking
until very recently \cite{Reviews,HIRES2}. The precise interpretation
of the recent results of Auger \cite{PAC} on the correlation
of directions between UHECRs and Active Galactic Nuclei
(AGNs), which we discuss in detail in 
Sections \ref{Auger} and \ref{AugerAGNcorr}, may still be
debatable.

There is mounting observational
evidence that, in addition to the ejection of a non-relativistic spherical
shell, the explosion of a core-collapse supernova (SN) results in the emission of highly
relativistic bipolar jets of plasmoids of ordinary matter, {\it Cannonballs}.
Evidence for the ejection of such jets in SN 
explosions is not limited to GRBs but comes also from
optical observations of SN 1987A \cite{NP}, from X-ray \cite{UH} and
infrared \cite{OK} observations of Cassiopeia A and, perhaps, from the
morphology of radio SNRs \cite{RNM}. These jets may be the main 
source of CR nuclei at all energies \cite{DP,DAD,Florence}.
They also explain long-duration GRBs
\cite{DETAL}, as advocated in the CB model \cite{GRB1,DD}.

In this paper we elaborate on a previous theory of CRs \cite{DP}, which
is very different from the conventionally accepted theories \cite{Reviews}.
For much of the required input, we exploit the subsequently acquired
information provided by the CB-model analysis of long-duration $\gamma$-ray
bursts (GRBs) and X-ray flashes (XRFs). 
The jets of CBs responsible for GRBs are akin to the 
jets of CBs emitted by quasars and microquasars. The former jets, we shall
argue, are also responsible for the generation of CRs.

\begin{figure}[]
\centering
\vbox{
 \epsfig{file=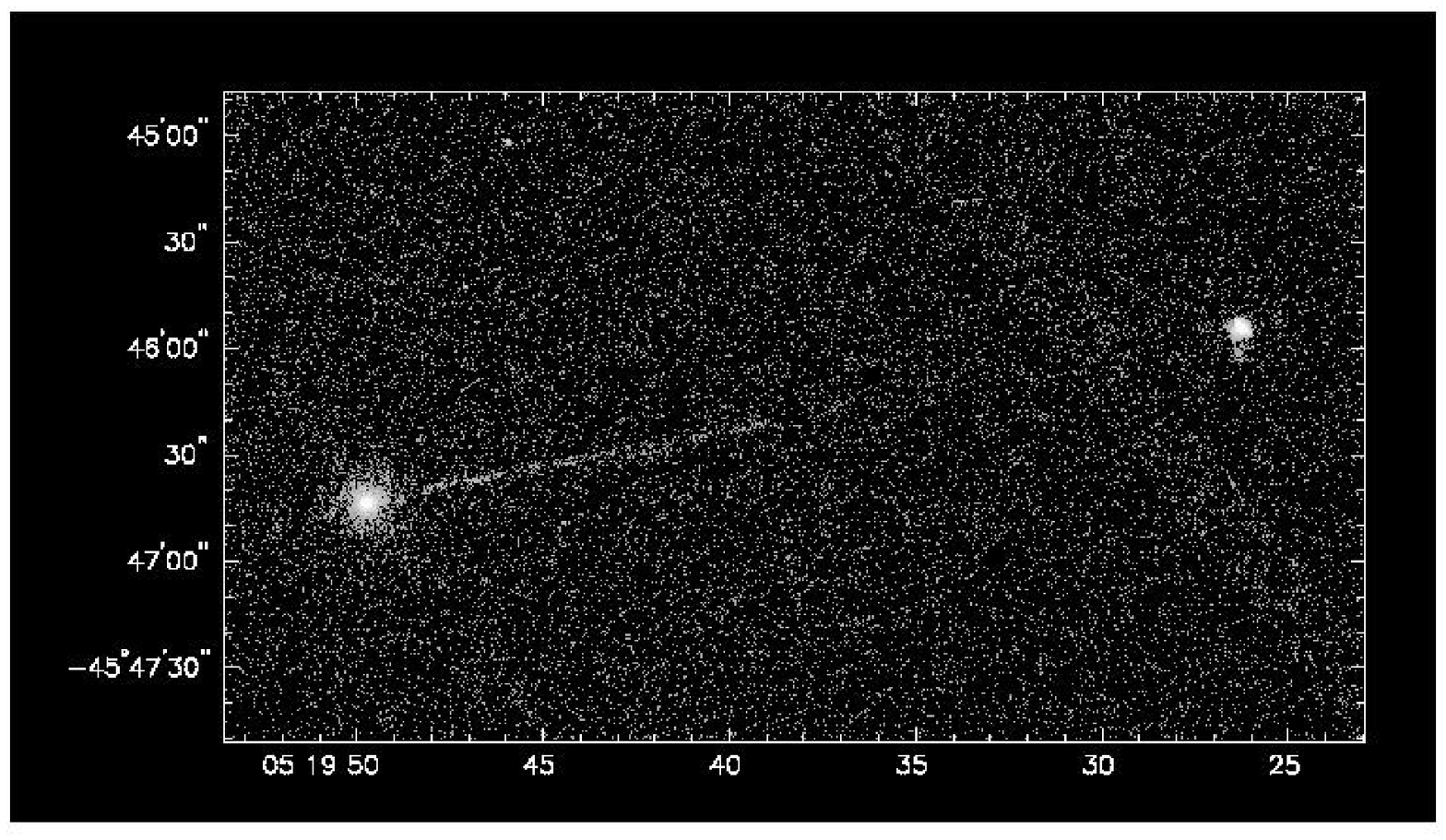,height=4.4cm,width=7.7cm}
}
\vbox{
 \epsfig{file=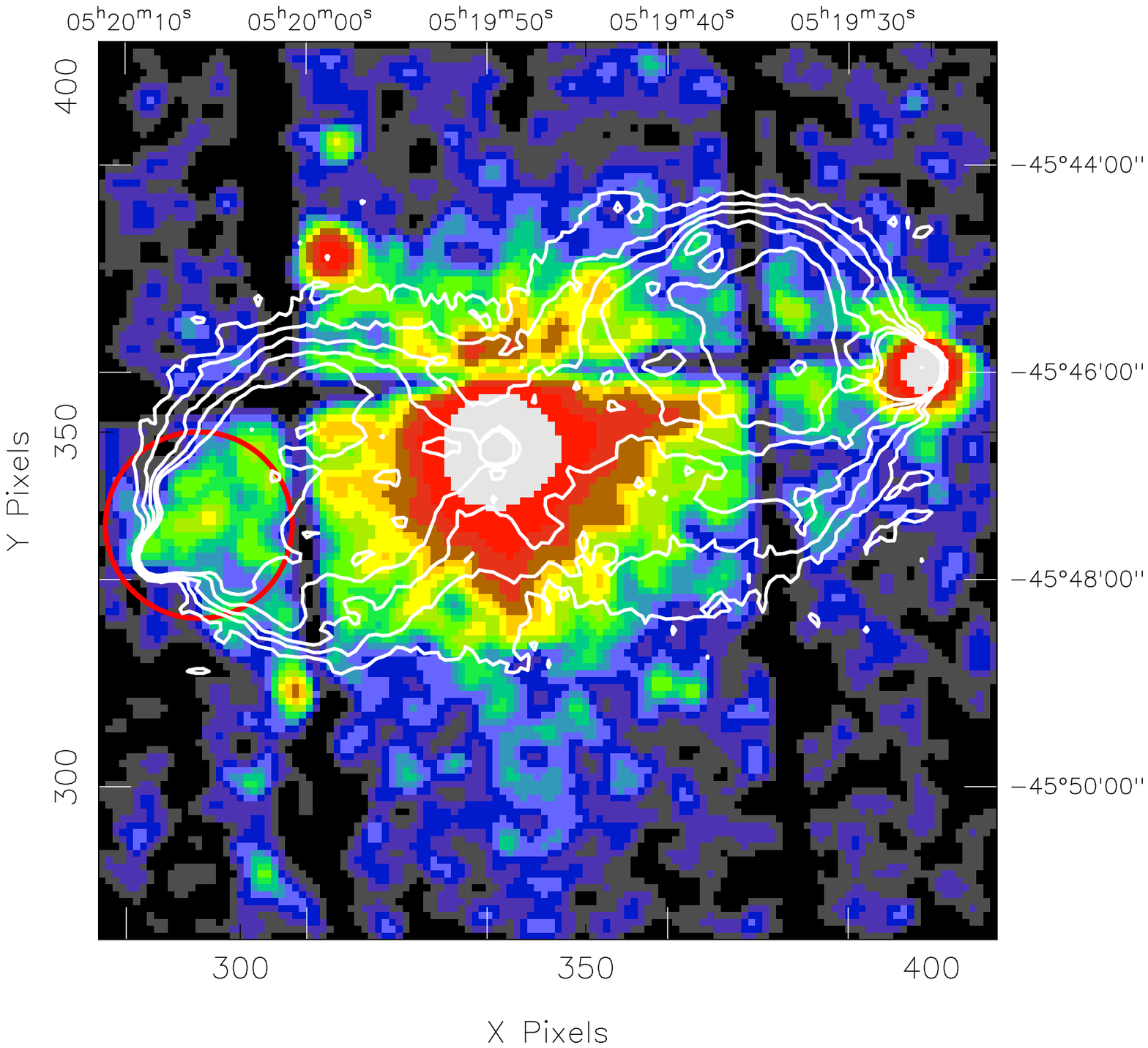,height=8.4cm,width=7.35cm,angle=-90}
}
\vspace*{-5pt}
 \caption{Top: X-ray image of the  galaxy 
Pictor A. A non-expanding jet extends
across 360000 light years towards a hot spot at least 800000
light years away from where the jet originates \cite{PG}. Bottom: 
XMM/p-n image of Pictor A in the 0.2--12 keV energy interval,
centred at the position of the leftmost spot in the upper panel,
superimposed on the radio
contours of a 1.4 GHz radio VLA map \cite{WYS}.}
 \label{Pictor}
\end{figure}

The essence of our considerations may be pictorially conveyed. The quasar
Pictor A is shown in Fig.~\ref{Pictor}. The X-ray picture in the top panel
shows one of its extremely narrow jets, which we interpret as X-ray emission
from a jet of CBs. The lower panel shows the two opposite jets, and contour
plots of their radio-emission fluence. We interpret the radio signal as the
synchrotron radiation of `cosmic-ray' electrons. Electrons and nuclei
were scattered by the CBs, which encountered
them at rest in the intergalactic medium (IGM), kicking them up to high 
energies. Thereafter, these particles diffuse in the ambient magnetic fields 
(MFs) and the electrons efficiently emit synchrotron radiation.
In applying this picture to the CRs in our Galaxy, we will simply replace
the quasar for all past Galactic and extragalactic 
core-collapse SNe, and fill in the details.

As a CB from a core-collapse SN travels through the interstellar medium
(ISM), it encounters ISM matter that has been previously ionized by the
passage of the GRB's $\gamma$ radiation. The CB's density is low enough
for individual interactions between its ionized plasma constituents and
those of the ISM to be irrelevant. The ISM ions and electrons are only
deflected by the collective effects of the CB's inner MFs,
generated by the very same ions and electrons. We shall
see in detail that this makes a CB act as a formidably efficient
{\it relativistic magnetic-racket} accelerator, which loses essentially all
of its energy to the recoiling particles: the newly born CRs. We argue
that this very simple concept explains all observed properties of 
non-solar CRs at all observed energies. 

Cosmic-ray sources other than
high-energy jets ---such as the traditional expanding SN envelopes, novae,
stellar flares, stellar winds and non-relativistic jets--- may be relevant
at low energies. Galactic high-energy CRs 
are also emitted by ordinary pulsars, by soft $\gamma$-ray repeaters,
by microquasars, and probably in the final merger of neutron stars
and black holes in binary systems.  The total CR luminosity of these objects is
smaller than the observed one by more than two orders of magnitude.
Similar considerations lead us to neglect, or to discuss
{\it cum grano salis,} the extragalactic contribution
of relativistic jets from massive black holes in AGNs, 
perhaps the most luminous potentially competitive sources. 
These topics are discussed in Section \ref{AugerAGNcorr} and
Appendix \ref{Accelerators}.

\begin{figure}
\begin{center}
\epsfig{file=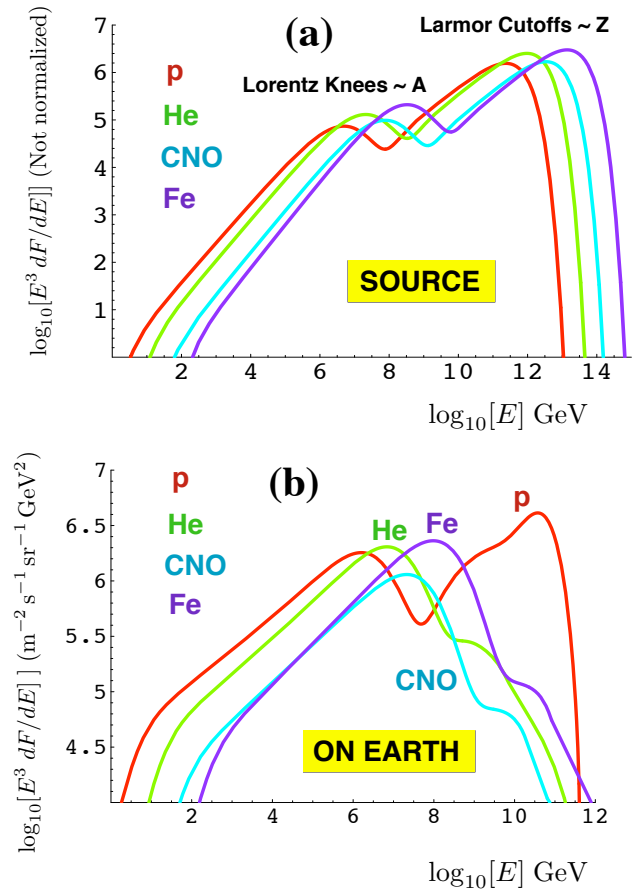, width=8.5cm,angle=-0}
\end{center}
\caption{Predicted CR spectra for the most abundant elements
and groups. The vertical scales are $E^3\,dF/dE$. (a): The source
spectra, with a common arbitrary normalization. (b): The CR spectra 
at the location of the Earth. Notice that both the horizontal and
vertical scales are different.}
\label{Groups}
\end{figure}

Our predictions for the $E^3$-weighted fluxes 
of the most abundant nuclei and `groups' of nuclei
are shown in Fig.~\ref{Groups}, 
which previews and summarizes our results. The source spectra are shown
in Fig.~\ref{Groups}a; two of their features are: `knees' at energies proportional
to the atomic number $A$; and `Larmor' cutoffs (proportional to the nuclear
charge $Z$) beyond which our CR acceleration mechanism is no longer
operative. The CR spectra arriving to our planet are shown in Fig.~\ref{Groups}b. 
The differences between these two figures ---which are 
significant and will be discussed in minute detail---
are due to the many {\it `tribulations'} a CR suffers in travelling to Earth from the
location of its source. Two examples of tribulations are: 

(1) Below a certain momentum
(some $3\times 10^9\; Z$ GeV/c) the local flux of
CRs of Galactic origin is enhanced 
by a factor proportional to their momentum-dependent Galactic
`confinement' time \cite{Swordy}: 
\begin{equation} 
\tau_{\rm conf} \propto
\left({Z\,{\rm GeV/c}\over p}\right)^{\beta_{\rm conf}}; 
\,\,\,\,\,\,\,{\beta_{\rm conf}}\sim 0.6\pm 0.10. 
\label{cintrod} 
\end{equation}
This is the origin of the differing slopes of the
lower-energy fluxes in Figs.~\ref{Groups}a,b (note their
different scales). 

(2) Extragalactic CRs other than protons are efficiently
photo-dissociated by the cosmic background infrared radiation on their
way to our Galaxy. 
This is part of the explanation for the very different
relative abundances of the elements at the higher energies in 
Figs.~\ref{Groups}a and \ref{Groups}b.

One can see in Fig.~\ref{Groups}b 
how H and He dominate up to their knees, which add up to the knee
feature of the all-particle spectrum; how the composition thereafter
becomes `heavier' and dominated by Fe up to its knee, which is the second
knee of the all-particle spectrum; and how the flux becomes once again
`lighter'  above this feature.
The UHECRs are entirely extragalactic in origin and are
dominantly protons.

Ours is also a theory of CR electrons and of their radiative contribution to
the diffuse $\gamma$ background radiation (GBR). The diffuse GBR at low 
Galactic latitudes originates mainly from $\pi^0$-generating collisions of 
CR nuclei with the ISM, followed by $\pi^0\to 2\gamma$ decay.
At high Galactic latitudes the diffuse GBR, we contend \cite{DDGBR},
is dominated by inverse-Compton radiation from CR electrons in the ISM
and the halo of galaxies (including ours) and in the IGM. 
In this sense, CR nuclei, CR electrons
and a good fraction of the diffuse GBR have the same origin \cite{Anton},
the latter radiation being a CR `secondary'. 

More specific results, to be derived in detail and in agreement with the
data, are:
\begin{itemize}

\item
The slope of the CR source spectra below the knees is predicted
to be $\beta_s\simeq 2.17$, as derived in Section \ref{belowtheknee}. 
Modified by Galactic confinement, this
corresponds to observed spectra with a slope 
$\beta_s+{\beta_{\rm conf}}\approx 2.77$,
 with ${\beta_{\rm conf}}$ as in Eq.~(\ref{cintrod}).

\item
A very slight composition dependence of the slope of the source spectra
 is expected. This can be discerned in Fig.~\ref{Groups}. It is
 discussed in Section \ref{composlopes} and shown in Fig. \ref{ArnonSlopes}. 

\item
The CR spectra at the lowest energies are affected by solar effects.
The predictions agree with the data at minimum solar activity; see
Section \ref{lowenergies} and Fig.~\ref{VeryLowEnergy}.

\item 
The spectra of the individual CR nuclei are predicted to have
`knees', scaling as the atomic weight, at energies around $ E[{\rm
knee}]\!\sim\! 3\!\times\! 10^{15}\,A\,{\rm eV}$; see Eq.~(\ref{Eknee}) and
Figs.~\ref{Groups},~\ref{ProtonGEG}. The observed and predicted
spectra for the individual elements H, He and Fe,
at energies up to their knees, are shown in
Fig.~\ref{KASKADE}.

\item
 The CR spectrum is predicted to change rather abruptly in 
slope, dominant composition (Fe to H) and dominant origin (Galactic
to extragalactic) at the `ankle' energy, 
$E[{\rm ankle}]\sim 3\times 10^{18}\;{\rm eV}$, see Section \ref{ankleregion}.

\item 
Our CR acceleration mechanism has a cutoff at the energies of Eq.~(\ref{toe}),
proportional to atomic charge and roughly coincident with the conventional 
Greisen--Zatsepi--Kuzmin (GZK) 
cutoff \cite{GZK}.  These cutoffs do not seem to be present in the AGASA
data \cite{AGASA}, but are compatible with the  HIRES 
\cite{HIRES2}, Fly's Eyes' and Auger \cite{Mello,Unger,PAC}
data, which agree well with our theory:
see Fig. \ref{UHECR}.

\item
The predicted normalization of the UHECR flux is approximate 
but `absolute', i.e.~parameter-free; see Section \ref{UHECRtext}.

\item
The prediction for the  all-particle spectrum is compared with the data
in Fig.~\ref{AllPart}.

\item
Detailed predictions for the `primary' CR abundances relative to hydrogen are 
discussed in Section  \ref{relabundances}.
They are compared with data  at 1 TeV 
in Table \ref{table1} and  illustrated in Fig.~\ref{f1}.
\item 
Data on two rough indicators of the evolution
of CR composition with energy, $\langle \ln A \rangle$ and
the depth of shower maximum $X_{\rm max}$,
are compared with predictions in  Figs.~\ref{lnAknee} and \ref{Xmax}.
\item
The confinement volume and the confinement time of CRs in the Galaxy
can be estimated theoretically. 
They agree with the estimates extracted from 
observations, as discussed in Section \ref{timevolume}.
\item
Below their respective knees, the
source spectra of CR nuclei and electrons are predicted to have the
same slope: $\beta_s=13/6$. For relatively high-energy
electrons, radiation cooling steepens
the slope to $\beta_e=\beta_s+1\sim 3.17$. 
The observed slope is $\simeq 3.2$, as shown in Fig.~\ref{CREspectrum}.
The normalization of the electron spectrum, we cannot predict.
\item
The slope of the diffuse GBR is predicted to be $(\beta_e+1)/2\simeq 2.08$. 
The observation is $\simeq 2.1$, as shown in Fig.~\ref{GBRspectrum}. 
\end{itemize}

Admittedly, the `predictions' we have referred to in the above items are
`postdictions' of existing data. Yet, the theory on which they are based
is very `predictive': only one parameter specific to the CR source will be
fitted to the hadronic CR
data.  Otherwise, only {\it priors} (items of information independent of 
the CR source) have been used as inputs, and kept at their
`central' values, or within their error brackets~\cite{cosmopriors}.

The study of GRBs, some 40 years old, is in its infancy, if compared
with the century-old study of CRs. In the GRB realm, novel and very
precise data, in particular at X-ray energies and mainly thanks to the
Swift satellite, are being gathered. The predictions of the CB model
have been precisely verified by the data having appeared since
our first posting of this paper in June 2006. This subject is very briefly
summarized in Section \ref{Swift}.

A posteriori the distinction between post- and pre-dictions, or parameters
and priors, is somewhat artificial. But there are other assets of the CR
theory presented here: it works simply and very well, and it is based on
{\it a single source of CR acceleration at all energies}.
Moreover, the underlying theory ---originally inspired by an analogy with the
relativistic ejecta of quasars and microquasars--- is part of  a unified
model of high-energy astrophysical phenomena \cite{Florence}, which also
offers simple and successful explanations of the origin and properties of
`long-duration' GRBs and X-ray flashes and their respective
afterglows (AGs) \cite{DD,AGoptical,AGradio,DDDXRF}, the natal kicks of
neutron stars \cite{DP}, 
the MFs and radio
emission from within and near galaxy clusters  \cite{DDMF}, and the X-ray
emission from  galactic clusters allegedly harbouring `cooling flows' \cite{CDD}.

The many titles and subtitles in this paper should suffice to convey its
organization. We discuss in detail or summarize in Appendices some of
the relevant background information: how a CB expands, photo-dissociation, 
the least debatable `priors' common to all theories of CRs, 
jets in astrophysics, the CB model, 
the evidence for the ejection of relativistic jets in SN explosions,
the supernova--GRB association and the 
power supply by CR accelerators other than the one we propose.

Our main point is our proposed mechanism of CR acceleration.
A reader primarily interested in it
may choose to read first Section \ref{collisionless} on `Collisionless magnetic 
rackets'. A reader primarily intrigued by the results
may choose to start with Chapter \ref{CBresults}.

\section{CB priors}
\label{CBpriors}

The `cannon' of the CB model is analogous to the ones
responsible for the ejecta of quasars and microquasars.
As an {\it ordinary core-collapse} SN implodes
into a black hole or neutron star and sheds an exploding shell,
an accretion disk or torus is hypothesized to be produced around
the newly born compact object, either by stellar material originally
close to the surface of the imploding core and left behind by the
explosion-generating outgoing shock, or by more distant stellar matter
falling back after its passage~\cite{ADR,GRB1,DD}. 
A CB is emitted, as observed in microquasars~\cite{Felix,DMR}, when part 
of the accretion disk falls abruptly onto the compact object 
\cite{GRB1,DD}. 

\begin{figure}
\centering
\epsfig{file=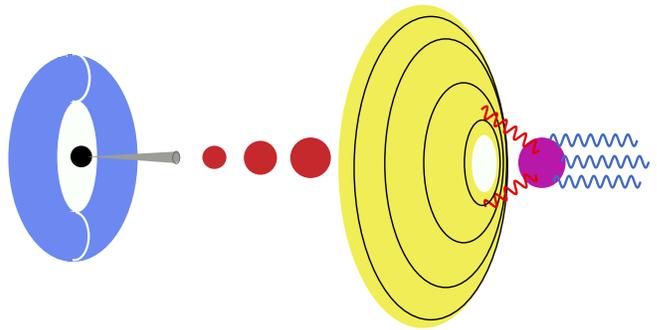,width=0.5\linewidth,angle=-90}
\vspace*{8pt}
\caption{An `artist's view'' (not to scale) of the CB model
of long-duration GRBs \cite{DD}. A core-collapse SN results in
a compact object and a fast-rotating torus of non-ejected
fallen-back material. Matter (not shown) abruptly accreting
into the central object produces
a narrowly collimated beam of CBs, of which only some of
the `northern'' ones are depicted. As these CBs move through
the `ambient light'' surrounding the star, they Compton up-scatter
its photons to GRB energies.}
\label{figCB}
\end{figure}

In the case of a core-collapse SN, the accretion torus is not fed by
a companion, it has a finite mass and can feed a limited number of
accretion episodes. Each episode corresponds to the bipolar
emission of a CB pair. A CB generates a forward cone of high-energy
photons as its constituent electrons Compton-up-scatter ambient light.
If the jet is directed close to the line of 
sight of an observer, each of its CBs generates a pulse in a GRB signal;
a bit more off axis, an  XRF is observed. The CBs, like
the matter that feeds them from the accreting torus, are made
of {\it ordinary-matter plasma}. The typical initial Lorentz factor (LF)
of a CB, $\gamma_0$, and its typical initial  baryon number, $N_{_{\rm B}}$, are
\cite{DD}:
\begin{eqnarray}
\gamma_0&\equiv& E/(M_0\,c^2)\sim{\cal{O}}(10^3),\label{typicalg}\\
N_{_{\rm B}} &\sim& 10^{50}.
\label{typicalNB}
\end{eqnarray}
The value of $M_0\sim N_{_{\rm B}}\, m_p\, c^2$
roughly corresponds to half of the mass of Mercury, a very small number
in comparison with the mass of the parent exploding star. 
An artist's view of the CB model is given in Fig.~\ref{figCB}.

The CB model of GRBs and their AGs is briefly discussed in Appendix
\ref{TheCB}.
Some of the distributions and average values of the input priors required
in our theory of CRs are specific to this model. 
They are summarized in this Section, along with the other ingredients
of the CB model  relevant to CR production.

\subsection{The distribution of initial Lorentz factors} 
\label{initial}

Let $\gamma_0$ denote the value of the LF 
of a CB as it is originally emitted by a SN and produces a GRB's
$\gamma$-ray pulse by inverse Compton scattering (ICS), before it is
slowed down by the ISM while
generating the GRB's afterglow by synchrotron radiation. An  average
value $\overline{\gamma_0}\sim 10^3$ was first estimated using
the rough hypothesis
that an asymmetry between the momenta of the diametrically
opposed jets was responsible for the `natal kick' velocity of 
neutron stars, the remnants of the core-collapse SN 
explosions of relatively light progenitors \cite{DP}.
This value of $\overline{\gamma_0}$ was confirmed by a first
study of GRBs~\cite{GRB1} within the 
CB model. It is also compatible with the roughly 1 to 1 SN--GRB
association discussed in Appendix \ref{association}.

\begin{figure}[]
\centering
 \epsfig{file=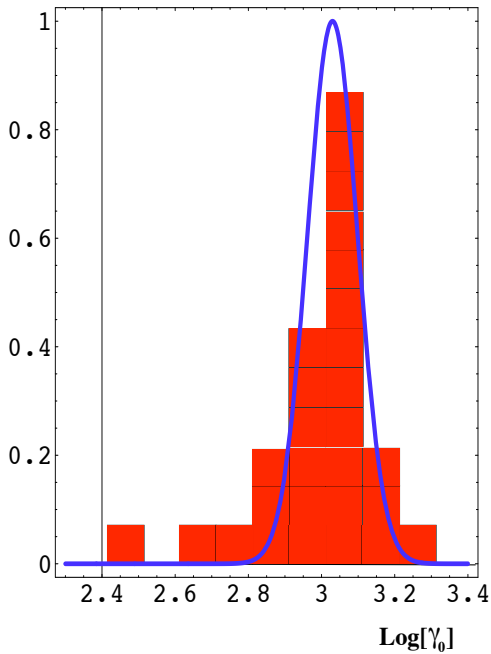,height=9.1cm,width=6.5cm}
\vspace*{-5pt}
 \caption{Distribution of $\log_{10}(\gamma_0)$ values for GRBs of known
 redshift~\cite{DD}. The continuous curve is a log-normal fit.}
 \label{DDLogGamma}
\end{figure}

A subsequent analysis of GRB afterglows (AGs) at infrared and
optical~\cite{AGoptical} as well as radio~\cite{AGradio} frequencies confirmed
$\overline{\gamma_0}\sim 10^3$ as the average initial LF. The distribution of
$\gamma_0$ values obtained from these analyses for the ensemble of GRBs of
known redshift (as of 2002)
is shown in Fig.~\ref{DDLogGamma}, constructed with the results
of Ref.~\cite{AGradio}.
The figure refers to data obtained with the selection criteria for the detection
of GRBs, which discriminates in favour of large LFs, and is the result of
fits to AGs which ---with the exception of some GRBs clearly dominated by
two CBs--- are made with the simplification of substituting an ensemble of
CBs for a single `average' one. This tends to make the extracted $\gamma_0$
distribution narrower than the `real' one, and its real average somewhat
uncertain. 

The properties of CRs depend on the `real' $\gamma_0$ distribution, which
we parametrize as:
\begin{equation}
D(\gamma_0,\overline{\gamma_0})\propto \exp 
\left[-\left({\log\gamma_0-\log\overline{\gamma_0}\over \log w}\right)^2 \right].
\label{gamdist}
\end{equation}
The distribution of Fig.~\ref{DDLogGamma} has $\overline{\gamma_0}=1070$
and $\omega=0.8$. It
 results in a good description of CR data,
but, not surprisingly, a somewhat broader distribution gives an even better
description, as discussed in Section \ref{kneeregion}. The predictions for CRs
are insensitive to the assumed form of the lower-energy tail of 
$D(\gamma_0,\bar\gamma_0)$.

\subsection{The deceleration of CBs in the ISM}
\label{Deceleration}

While a CB exits from its parent SN and emits a GRB pulse, it is
assumed~\cite{GRB1} to be expanding, in its rest system, at a speed comparable
to that of sound in a relativistic plasma ($v_s=c/\sqrt{3}$). In their voyage, CBs
continuously intercept the electrons and nuclei of the ISM, previously ionized
by the GRB's $\gamma$-rays. In seconds of (highly Lorentz- and
Doppler-foreshortened) GRB
observer's time, such an expanding CB becomes `collisionless', that is, its
radius becomes smaller than a typical interaction length between a constituent
of the CB and an ISM particle. But a CB
still interacts with the charged ISM particles it encounters, for, as we
discuss in detail in Section \ref{magfield}, it contains a
strong magnetic field.

Consider a CB of initial mass $M_0$ and initial LF $\gamma_0$.
As it travels in the ISM its LF diminishes all the way to unity.
We assume that the
ISM particles entering a CB's magnetic mesh are trapped in it
and slowly re-exit by diffusion. To a fair approximation,
a CB simply accumulates the ISM particles that it intercepts. In this case,
energy--momentum conservation implies that the CB's mass
increases as:
\begin{equation}
M_{\rm CB}\approx 
M_0\,{\gamma_0\over \beta\,\gamma},\;\;\;\;
\left[\beta\equiv {\sqrt{\gamma^2-1}/ \gamma}\right],
\label{NRmass}
\end{equation}
and, for an approximately hydrogenic ISM of local density $dn_{\rm in}$,
the LF decreases as:
\begin{equation}
{d\,\gamma\over\beta^3\,\gamma^3} \approx - \,{m_p\over M_0\,\gamma_0}\;
dn_{\rm in}(\gamma).
\label{gammadown}
\end{equation}

To compute the spectrum of the CRs produced by a CB in its voyage through
the ISM, we shall have to perform a $dn_{\rm in}$ integral over its trajectory, as the
CB decelerates from $\gamma=\gamma_0$
to $\gamma = 1$. Given Eq.~(\ref{gammadown}),
this is tantamount to integrating the CR spectra at
local values of $\gamma$ with a weight factor 
$dn_{\rm in}\propto {d\gamma/(\gamma^3\,\beta^3)}$. Notice that the CB's deceleration
law of Eq.~(\ref{gammadown}) depends explicitly on the number of ISM particles it
intercepts, but not on any CB properties other than $M_0$ and $\gamma_0$.

\subsection{The expansion of a CB}
\label{expansionCB}

We approximate a CB, in its rest system, by a sphere of radius $R(\gamma)$.
The value of $R(\gamma_0)$ is immaterial, for it becomes rapidly negligible 
as the CB initially expands at a speed $\sim c/\sqrt{3}$. The ISM particles that are
intercepted ---isotropized in the CB's inner magnetic mesh,
and re-emitted--- exert an inwards force on it that, we assume,
has as its main effect to counteract the CB's expansion. 
This expansion, in the `fast elastic' case of instantaneous re-emission,
was studied in \cite{AGoptical,AGradio}. The case of  `diffusive' re-emission 
results in a slightly better description of more recent data 
\cite{DDDSwift}. We discuss it in detail 
in Appendix \ref{ExpansionApp}  and we adopt it here.

\begin{figure}
\vspace*{.5cm}
\centering
\epsfig{file=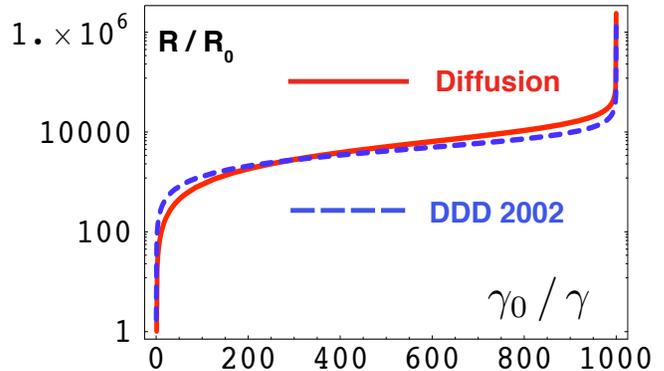,height=6cm}
\vspace*{-.2cm}
\caption{Expansion of a CB as its LF $\gamma$ diminishes
along its trajectory from an initial $\gamma_0=10^3$.
The dotted line is for the case of fast elastic CB interactions
with the ISM, discussed in \cite{AGoptical,AGradio}. 
The continuous line is for the case discussed here, in 
which the ISM gathered by the CB slowly oozes out of
it by diffusion. These are two limiting cases.}
 \label{RofGamma}
\end{figure}

The behaviour of $R(\gamma)$ is shown in Fig.~\ref{RofGamma}.
It has three distinct phases. The initial rapidly expanding quasi-inertial phase 
plays a crucial role in the description of GRB pulse shapes and is supported 
by the CB-model's correct prediction of all their other properties \cite{DD}. 
The properties of the intermediate
coasting phase are supported by the CB-model's successful description
of GRB AGs; see, e.g.~\cite{AGoptical,AGradio,DDDSwift}. The final blow-up
phase may describe the observed lobes of quasars and microquasars, such 
as the one at the right of Pictor A in Fig.~\ref{Pictor}.

A CB converts the ISM into CRs at a rate proportional to $R_{_{\rm CB}}^2$. The
initially fast-expanding phase in $R_{_{\rm CB}}(\gamma)$ has negligible effects.
The subsequent behaviour of $R_{_{\rm CB}}(\gamma)$, in the diffusive case and for
typical (or average) CB parameters, is well described by:
\begin{eqnarray}
R_{_{\rm CB}}(\gamma)&\approx&  R_0\,
\left({\gamma_0\over \beta\,\gamma}\right)^{2/3},
\nonumber\\
R_0&\sim&10^{14}\;{\rm cm}.
\label{best}
\end{eqnarray}
This behaviour gives the best description of GRB afterglows, 
as discussed in \cite{DDDSwift} and Appendix \ref{ExpansionApp}.

\subsection{The trajectories of CBs}
\label{trajectories}

How far does a CB travel before the collisions with the ISM stop it? 
The answer crucially depends on the distribution of ISM densities
that the CB encounters, and the relativistic approximation
($\beta\approx 1$) suffices to give 
it with the required precision. In the `slow' approximation ---in which
the rate at which the ISM particles enter the CB is faster than the rate at
which they exit it by diffusion--- every ISM proton intercepted by a CB
increases its mass by $\gamma\,m_p$. The mass increase per
travel length $dx$ is:
\begin{equation}
dM_{_{\rm CB}}=\pi\,R_{_{\rm CB}}^2\,\gamma\,m_p\,n_p\,dx.
\label{dM}
\end{equation}
The relation between $dM_{_{\rm CB}}$ and $d\gamma$ is,
according to Eq.~(\ref{NRmass}), 
$\gamma\,dM_{_{\rm CB}}=M_0\,\gamma_0\,d\gamma/\gamma_0$, and
$R_{_{\rm CB}}$ is given in Eq.~(\ref{best}).
Gathering this information and integrating the result in $dx$ and $d\gamma$
we obtain:
\begin{eqnarray}
x&=&x_{\rm stop}\,\left({1\over\gamma^{2/3}}-{1\over\gamma_0^{2/3}}\right),
\nonumber\\
x_{\rm stop}&\equiv&{3\,N_{_{\rm B}}\over2\pi\,R_0^2\,\gamma_0^{1/3}}\;\bar n_p=
(18\;{\rm kpc})\times
\nonumber\\
&&\!\!\!\!\!\!\!\!\!\!\!\!
\left[{N_{_{\rm B}}\over 10^{50}}\right]
\left[{10^{14}\,{\rm cm}\over R_0}\right]^2
\left[{10^{-2}\,{\rm cm^{-3}}\over\bar n_p}\right]
\left[{10^3\over\gamma_0}\right]^{1\over 3},
\label{stop}
\end{eqnarray}
where $\bar n_p$, an adequately averaged ISM density along a given CB's
trajectory, is perhaps the most uncertain of the case-by-case
varying inputs in $x_{\rm stop}$.

In the geometrically unlikely case that a CB travels in the plane
of the Galaxy and crosses its central densest regions, its reach
should be much less than the 18 kpc in Eq.~(\ref{stop}). In the opposite
extreme, if a CB exits perpendicularly to the plane of the Galaxy
from a relatively high point in its ISM density distribution, 
it can reach beyond the Galactic halo into intergalactic
space.

\begin{figure}[]
\centering
 \epsfig{file=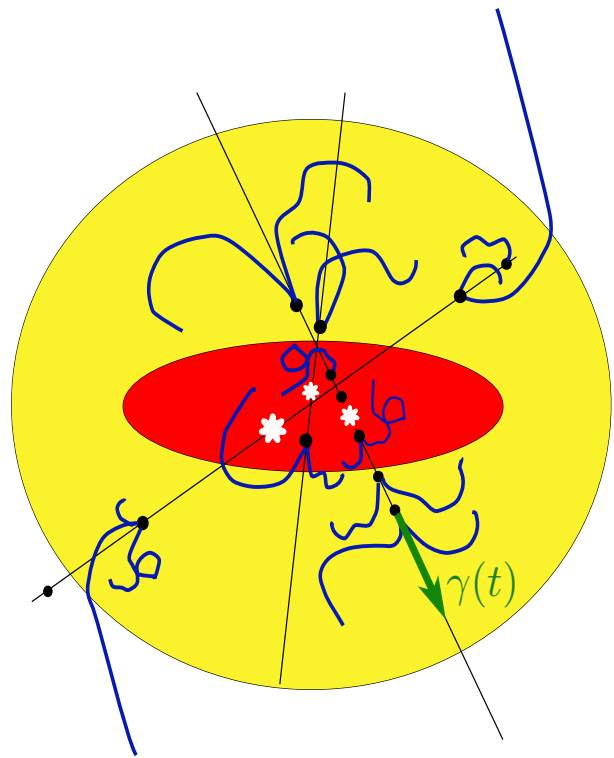,height=10cm,width=8cm}
 \caption{A schematic representation of the Galaxy and its halo. The CBs (black dots)
emitted in the explosions of a few SNe 
are shown. All along their long straight trajectories 
(thin black lines), CBs slow down by collisions with the ISM, converting it into
 forward cones of CRs (blue lines), which curve on
the magnetic fields of the Galaxy and its halo, eventually exuding into 
intergalactic space. The SN rate is such that an 
actual `snapshot' should contain many more jets of CBs.}
 \label{trip}
\end{figure}

Cannonballs typically move from the inner SN-rich realm of
the Galaxy into its halo or beyond and,  along their trajectories, they 
convert into CRs the ISM particles they encounter, absorb and re-emit,
as illustrated in Fig.~\ref{trip}. The CRs are 
forward-emitted by the fast-moving CBs, subsequently meandering 
in the Galactic MFs till they eventually escape the Galaxy.

In our Galaxy SNe occur at a rate of about twice a century. This is a much 
shorter time than it takes a CB to travel over most of its trajectory,
between a fraction of a kpc and several kpc, while still moving at
a relativistic LF. Thus the Galaxy and its halo are, at any moment, permeated
by scores of CB sources: Fig.~\ref{trip} should show many more CB trajectories.
A CB is a continuous source of CRs along its trajectory, and its
source intensity depends
on the local and previously traversed ISM density: in our theory the source
of CRs is very diffuse. Thus,
the directional anisotropy of CRs at the Earth's location
is expected to be very small and to vary little with energy, as observed 
\cite{Smith}. 

In models 
in which CRs diffuse away from point sources in the Galaxy's disk,
CR diffusion and hypothetical reacceleration mechanisms
play a crucial role, particularly for CR electrons. 
In the CB model, on the contrary, CR transport by diffusion 
should not play a significant role, and reacceleration mechanisms need 
not be invoked.

\subsection{The magnetic field within a CB}
\label{magfield}

The `collisionless' interactions of a CB and the ISM electrons and nuclei
constitute the merger of two ordinary-matter plasmas at a large relative 
LF $\gamma$. This merger should be very efficient in creating turbulent
currents and the consequent MFs within the CB, the
denser of the two plasmas \cite{DP,GRB1}. 
We assume that these MFs,
as the CB reaches a quasi-stable radius, are in `equipartition': 
their pressure (or energy density)  equals the pressure
exerted on the CB's surface by the ISM particles it re-emits
(or the energy density of the ISM particles it has temporarily phagocytized).
This results in a time-dependent magnetic-field strength~\cite{AGoptical}:
\begin{equation}
B_{_{\rm CB}}[\gamma,n_p]=3\;{\rm G}\;{\gamma\over 10^3}\;
\left({n_p\over 10^{-3}\;{\rm cm}^{-3}}\right)^{1/2}\; ,
\label{B}
\end{equation}
where $n_p$ is the ISM number density, normalized to a value characteristic of the
`superbubbles' (SB) in which most SNe and GRBs are born.  The simple
ensuing analysis of the elaborate time and frequency dependence of AGs 
---dominated by synchrotron radiation of electrons in the field of
Eq.~(\ref{B})--- is very successful~\cite{AGradio}. Thus, we adopt the result
of Eq.~(\ref{B}) in our analysis of CRs.

\subsection{Fermi acceleration within a CB}
\label{Fermiacc}

\begin{figure}[]
\vspace{.5cm}
\centering
 \epsfig{file=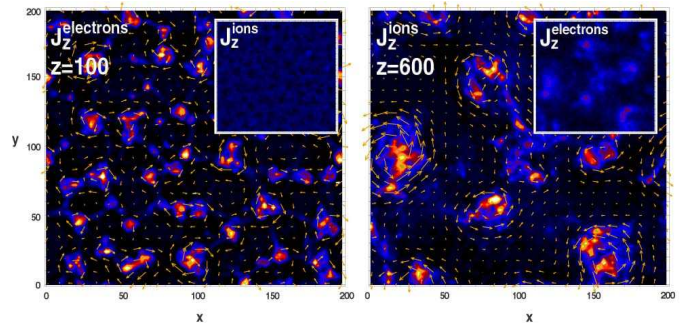,height=4.3cm,width=9cm}
 \caption{Results of a simulation of the merger of two plasmas at a relative
 LF $\gamma=3$. The left  panel shows the longitudinal electron current
density through a cut transverse to the direction of motion of the
incoming plasma, with a small inset showing
the ion current in the same plane.  The right  panel shows
the ion current deeper into the target plasma, with the small inset now  showing
the electron current instead. The arrows represent the transverse 
magnetic field~\cite{Fred}.}
 \label{DDfred}
\end{figure}

Charged particles interacting with macroscopic, turbulently moving MFs, tend
to gain energy: a `Fermi' acceleration process.  This acceleration is very efficient 
for a relativistic `injection', the case relevant to a CB, which is subject, in
its rest system, to a flow of ISM electrons and nuclei arriving with a large common LF.
A `first-principle' numerical analysis~\cite{Fred} of
the merging of two plasmas at a moderately high $\gamma$ ---based on following
each particle's individual trajectory as governed by the Lorentz force and
Maxwell's equations--- demonstrates the generation of such chaotic MFs,
and the acceleration of particles to a spectrum with a power-law tail:
\begin{equation}
{dN_{\rm ac}\over dE}\sim E^{-\beta_{\rm ac}}\;\Theta(E-\gamma\,M\,c^2),
\;\;\;\;\;\;  \beta_{\rm ac}\approx 2.2.
\label{Acc}
\end{equation}
The Heaviside $\Theta$ function is an approximate characterization of the
fact that it is much more likely for the light particles to gain than to lose
energy in their elastic collisions with the heavy `particles' (the CB's
turbulently moving collective plasma and MF domains). 
The numerical analysis~\cite{Fred} shows that this acceleration occurs
in a total {\it absence} of shocks, very much unlike what is generally
assumed for CRs accelerated in shocks produced by expanding SN shells 
\cite{GP}.
In Fig.~\ref{DDfred} we reproduce a plot of~\cite{Fred} showing the ion
and electron currents at two depths into the denser of the merging plasmas.

In our analysis of the radio, infrared, optical, UV and X-ray AGs of GRBs, we
assumed that a fraction of the ISM electrons entering a CB was accelerated
as in Eq.~(\ref{Acc}), the majority remaining unaccelerated at their
incoming LF. In the case of electrons, both populations `cool' by synchrotron
radiation in the CB's MFs. The ensuing synchrotron radiation ---the afterglow---
has a complex frequency and time dependence, which is in excellent agreement
with observations and ---assuming that the index $\beta_{\rm ac}$ of Eq.~(\ref{Acc})
is the same for electrons and nuclei--- confirms that Eqs.~(\ref{B},\ref{Acc}) 
are adequate; see Sections \ref{Swift} and \ref{GRBAGs}. The same index
governs the high-energy tail of the ``prompt" $\gamma$-ray and X-ray
spectrum of GRBs, again in agreement with observations \cite{DD}.

We assume that CR nuclei entering a CB from the ISM are also accelerated
as in Eq.~(\ref{Acc}). This acceleration cannot extend to arbitrarily high energies;
there must be a {\it Larmor cutoff,} for a CB has a finite radius and MF. 
A CB cannot significantly bend or accelerate a particle of energy larger than:
\begin{equation}
E[{\rm Larmor}]\simeq 9\times 10^{16}\;Z\;{\rm eV}\;
{B_{_{\rm CB}}[\gamma_0,n_p]\over 3\;{\rm G}}\;
{R_{_{\rm CB}}\over 10^{14}\;{\rm cm}},
\label{LarmorE}
\end{equation}
with $R_{_{\rm CB}}$ as in Eq.~(\ref{best}) and $B_{_{\rm CB}}$ as in Eq.~(\ref{B}).
This corresponds to a maximum LF in the CB's rest system:
\begin{eqnarray}
&&\gamma_{\rm max}(\gamma) = b\; \gamma^{1/3}
\label{gammamax}\\
&& b \simeq 10^5\;\gamma_0^{2/3}\;(Z/A).
\label{bb}
\end{eqnarray}
The distribution of the LFs, $\gamma_A$, of the Fermi-accelerated
nuclei that entered a CB with a Lorentz factor $\gamma$,  is:
\begin{equation}
{dN\over d\gamma_{_A}}\propto\gamma_{_A}^{-\beta_{\rm ac}}\,
\Theta(\gamma_{_A}-\gamma)\,\Theta[\gamma_{\rm max}(\gamma)-\gamma_{_A}],
\label{gammaA}
\end{equation}
where the second $\Theta$ function is a rough characterization of
the Larmor cutoff.
But for the small dependence of the coefficient $b$ on the nuclear
identity (the factor $Z/A$), the spectrum of Eq.~(\ref{gammaA}) is
universal.

\subsection{The energy of the jets of CBs}

The baryon number $N_{_{\rm B}}$ of a CB
---or, equivalently, its mass $M\simeq N_{_{\rm B}}\,m_p$---
can be roughly estimated from the fluence of the 
AG of GRBs~\cite{AGoptical,AGradio} and better constrained from the
 `spherical-equivalent' total energy and number of the $\gamma$-rays
in a single GRB pulse~\cite{DD}. The average result is $N_{_{\rm B}}\sim 10^{50}$,
cited in Eq.~(\ref{typicalNB}). 

The observed average number \cite{Quilligan} of significant pulses in a GRB's 
$\gamma$-ray light curve is $\sim 6$. The total energy of the two jets of CBs 
in a GRB event is therefore:
\begin{equation}
E[{\rm jets}]\simeq 12\,\gamma_0\,N_{_{\rm B}}\,m_p\,c^2
\simeq 2\!\times\! 10^{51}\;\rm erg.
\label{Ejets}
\end{equation}
Practically all of this energy will, in our model, be transferred to CRs.

\subsection{The CR luminosity of the Galaxy}
\label{lumGal}

In a steady state, if the low-energy rays dominating the CR luminosity
are chiefly Galactic in origin, their accelerators must
compensate for the escape of CRs from the Galaxy.
The Milky Way's luminosity in CRs must therefore satisfy: 
\begin{equation}
{L_{\rm CR} \simeq L_{p}={4\pi \over c}\int_{V_{\rm CR}} dV
 \int {dE\over \tau_{\rm conf}}
\,E\,{dF_p\over dE}}\, ,
\label{CRlum}
 \end{equation}
with $dF_p/dE$ as in Eq.~(\ref{protons}), $\tau_{\rm conf}$ as in Eq.~(\ref{cintrod}),
and ${V_{\rm CR}}$ the volume to which low-energy CRs are confined.
The coefficient $1/\tau_{\rm conf}$, for $Z=1$, converts the observed proton spectrum 
into the corresponding source spectrum. 
The conventional result of detailed models of CR production
and diffusion~\cite{SM1} is:
\begin{equation}
{L_{\rm CR} \sim 2\times 10^{41}\rm ~erg~s^{-1}}\, .
\label{them}
\end{equation}

Let $R_{\rm SN}$ be the SN rate in our Galaxy, discussed in Appendix \ref{rateSN},
and given by Eq.~(\ref{SNrate}).
The estimate of $L_{\rm CR}$ in the CB model is simply:
\begin{equation}
L_{\rm CR}{\rm [MW]}\!\approx\! 
R_{\rm SN}{\rm [MW]}\, E{\rm[jets]}\!\approx\! 1.3\times 10^{42}\,\rm erg\, s^{-1},
\label{SNEsupply}
\end{equation}
with $E{\rm [jets]}$ as in Eq.~(\ref{Ejets}). 
This estimate is uncertain by a factor of at least 2, for two reasons.
First, SNe are observed to produce roughly spherical non-relativistic ejecta,
whose kinetic energy is comparable to $E{\rm[jets]}$. The luminosity is dominated
by low-energy CRs, which may also be produced ---with debatable efficiency--- by 
these ejecta, as in the generally accepted models. This may increase the result of
Eq.~(\ref{SNEsupply}) by a factor $\leq\!2$. Second, we contend~\cite{DDMF}
that the MFs 
observed in the Milky Way and in galaxy clusters are generated by CRs, and are
in energy equipartition with them, as observed in the Galaxy \cite{Longair}, for which
\begin{equation}
\rho_{_{E}}{\rm [CR]}={4\,\pi\over c}\,\int {dF\over dE}\, E\,  dE\approx 0.5\, 
                \rm eV\, cm^{-3},
\label{rhoECR}
\end{equation}
and 
\begin{equation}
{B^2/ (8\, \pi)} \approx \rho_{_{E}}{\rm[CR]},\;\;\;\;\;\; {\rm for} \; B\sim 5\,\mu {\rm G}.
\label{rhoEB}
\end{equation}
The transfer to MFs of $\sim$ 50\% of the original CR energy may decrease the 
result of Eq.~(\ref{SNEsupply}) by a factor $\sim\!2$ \cite{DSS}.

\section{A theory of the CR source}
\label{THsource}

\subsection{Collisionless magnetic rackets}
\label{collisionless}

The essence of our theory of CRs is kinematical and trivial.
A very massive object (a CB) travelling with a Lorentz factor $\gamma$
and colliding with a light object (an ISM particle) can boost the light
object (now a CR) to extremely high energy. 

By definition, in an {\it elastic} interaction of a CB at rest 
with ISM electrons or ions of LF $\gamma$,  
the light recoiling particles retain their incoming energy.
Viewed in the system in
which the ISM is at rest,  the light recoiling particles  (of mass $m$)
have an energy spectrum extending, for large $\gamma$,
 up to $E\simeq 2\,\gamma^2\,m\,c^2$. A moving CB
is a {\it Lorentz-boost accelerator} of gorgeous efficiency:
the ISM particles it scatters
reach up to $\gamma_{_{\rm CR}}\simeq 2\,\gamma^2$, with
$\langle\gamma_{_{\rm CR}}\rangle\sim\gamma^2$ for any non-singular
scattering-angle distribution in the CB's rest system.
In a single scattering with a CB of $\gamma\sim 10^3$, and with 100\% efficiency, 
the energy of an ISM particle increases a million-fold from its value at rest.
The `accelerator' is also good at focusing: it produces a forward-collimated beam 
of CRs, the initial divergence of whose 
angular distribution is characterized by an angle $\theta\sim 1/\gamma$.

A particle with a LF $\gamma$ entering a CB at rest 
can be accelerated by elastic interactions with the CB's turbulently moving 
plasma. Viewed in the rest system of the bulk of the CB the interaction
is `inelastic': the light particle gained energy. Its LF can reach
$\gamma_{\rm max}\sim 10^7\,\gamma$; see Eqs.~(\ref{gammamax},\ref{bb}). 
Boosted by the CB's motion the spectrum of the scattered particles extends 
to $\gamma_{_{\rm CR}}\sim 2\times 10^7\,\gamma^2$, in the UHECR
 domain, for $\gamma\sim\gamma_0\sim 10^3$. 
This powerful {\it Fermi--Lorentz accelerator} completes our theory of CRs.

We have tacitly assumed in the previous paragraph that interactions
are instantaneous: a CB has the same LF when a given ISM
particle enters and leaves it; the CB has not decelerated in
the meantime via collisions with many other ISM nuclei. 
 Borrowing from the language of particles 
more elementary than CBs, we called the interactions {\it inelastic} or
{\it elastic}. In what follows we retract the cited assumption, but we keep
the italicized nomenclature to refer to our results for particles that have 
---or have not--- been Fermi-accelerated within a CB.

\subsection{Exiting a CB by diffusion}
\label{diffusiveexit}

Let $\gamma_{\rm in}$ be the LF of a given ISM proton
that entered a CB. 
Its momentum stays fixed as it is tossed around by the CB's inner 
chaotic magnetic field, or is increased by the acceleration mechanism we have
discussed. For the ISM nuclei, as opposed to electrons, 
radiative and collisional losses are negligible. We assume that
these trapped particles ooze out of the CB by diffusion, much as 
CRs do in the Galaxy. The characteristic diffusion time when
the LF [radius] of the CB has reached a value 
$\gamma$ [$R_{_{\rm CB}}(\gamma)$] is:
\begin{equation}
\tau={R_{_{\rm CB}}^2\over D}\; ,
\label{tau1}
\end{equation}
with  $D=D(\gamma_{\rm in},\gamma)$ 
a diffusion coefficient. The rate at which the 
diffusing particles are exuded by the CB is $r=\beta_{\rm in}/\tau$. 

In the CB model, the MF of the Galaxy \cite{DDMF} and that 
within a CB are both made by the same turbulence, induced by the 
injection of relativistic particles (the ISM in the case of a CB,
CRs in the case of the Galaxy). Consequently, we expect
$D$ to have the same energy dependence as observed for CRs:
$D\propto p_{\rm in}^{\beta_{\rm conf}}$, with the same 
${\beta_{\rm conf}}$ as in Eq.~(\ref{cintrod}).
In the case of a CB, the diffusion occurs in a 
MF with an energy density assumed to be in approximate
equipartition with the kinetic energy density of
the particles entering the CB at a given moment
$B\propto \gamma-1$, and $D$ should reflect the
$B$-, $A$- and $Z$-dependence of the corresponding Larmor
radius, that is: 
\begin{equation}
D\propto \left(E\over B\right)^{\beta_{\rm conf}}\propto 
\left[A\,\gamma_{\rm in}\over Z\,(\gamma-1) \right]^{\beta_{\rm conf}}.
\label{D1}
\end{equation}
The diffusion, out of a CB, of the fraction of ISM nuclei that are accelerated
within it, will be treated in an entirely analogous fashion.

\subsection{`Elastic' scattering}
\label{elasticscatt}

We have assumed that, to a good approximation, a CB ingurgitates most
of the ISM nuclei that it intercepts in its voyage, and that, within a CB, a fraction of
these nuclei keeps the energy at which they entered it. In this approximation,
and at the moment when the CB's LF has descended from $\gamma_0$
to $\gamma$, the distribution of LFs, 
$\gamma_{\rm co}$, of the collected nuclei is:
\begin{eqnarray}
{dN_{\rm co}\over d\gamma_{\rm co}} &=&
\int_{\gamma}^{\gamma_0}\,{dn_{\rm in}\over d\gamma_{\rm in}}\,
\delta(\gamma_{\rm co}-\gamma_{\rm in})\nonumber\\
&\propto&{1\over \beta_{\rm co}^3\,\gamma_{\rm co}^3} \;
\Theta(\gamma \leq \gamma_{\rm co} \leq \gamma_0),
\label{collected}
\end{eqnarray}
where we have used Eq.~(\ref{gammadown}) and an unconventional but transparent
notation for the Heaviside step function $\Theta$.

The collected particles exit the CB in its rest system at a rate $\beta_{\rm in}/\tau$,
so that the doubly differential ($\gamma$, $\gamma_{\rm co}$) oozing out rate is:
\begin{equation}
{dN_{\rm out}\over d\gamma\,d\gamma_{\rm co}}=
{dN_{\rm co}\over d\gamma_{\rm co}}\;
{dt_{_{\rm CB}}\over d\gamma}\;{\beta_{\rm co}\over\tau}\; ,
\label{dNdgdg}
\end{equation}
where 
\begin{equation}
-{dt_{_{\rm CB}}\over d\gamma}\propto {1\over \beta^4\,\gamma^4\,R_{_{\rm CB}}^2}\; ,
\label{dtCBdgamma}
\end{equation}
obtained by inserting
\begin{equation}
dn_{\rm in}(\gamma) \approx -
 \pi\,R_{_{\rm CB}}^2\,c\,n_p\,\beta\,\gamma\,dt_{_{\rm CB}} 
 \label{dninbis}
\end{equation}
into Eq.~(\ref{gammadown}).

To specify the distribution of LFs, $\gamma_{_{\rm CR}}$, of the CRs
in the ISM rest system, we must perform the corresponding boost over an
assumed isotropic distribution of exit directions in the CB's rest system:
\begin{eqnarray}
\!\!\!\!\!\!\!\!\!\!
&&{dN_{\rm out}\over d\gamma_{_{\rm CR}}\,d\gamma\,d\gamma_{\rm co}}=
\nonumber\\
\!\!\!\!\!\!\!\!\!\!
&&\int{d\cos\theta\over 2}{dN_{\rm out}\over d\gamma\,d\gamma_{\rm co}}
\delta[\gamma_{_{\rm CR}}-\gamma\,\gamma_{\rm co}\,
(1+\beta\,\beta_{\rm co}\,\cos\theta)].
\label{Lboost}
\end{eqnarray}
The condition $|\cos\theta|\leq 1$ introduces two constraints which, solved
for $\gamma_{\rm co}$, read:
\begin{eqnarray}
\!\!\!\!\!\gamma_{\rm co}&\leq& T(\gamma,\gamma_{_{\rm CR}})\equiv 
\gamma\,\gamma_{_{\rm CR}} (1+\beta\,\beta_{_{\rm CR}}),
\nonumber\\
\!\!\!\!\!\gamma_{\rm co}&\geq& B(\gamma,\gamma_{_{\rm CR}})\equiv 
\gamma\,\gamma_{_{\rm CR}} (1-\beta\,\beta_{_{\rm CR}}).
\label{constraints}
\end{eqnarray}

To compute the CR flux $dF/d\gamma_{_{\rm CR}}$, we must integrate over
$\gamma$ and $\gamma_{\rm co}$. Collecting all the results of this section
and using Eqs.~(\ref{best}), (\ref{tau1}) and  (\ref{D1}), 
we obtain:
\begin{eqnarray}
&&{dF_{\rm elast}\over d\gamma_{_{\rm CR}}}\propto n_{_A}
\beta_{_{\rm CR}}\left({A\over Z}\right)^{\beta_{\rm conf}}
\int_1^{\gamma_0}{d\gamma\over (\beta\,\gamma)^{7/3}}
\,{G[\gamma,\gamma_{_{\rm CR}}]\over (\gamma-1)^{\beta_{\rm conf}}}\; ,
\nonumber\\
&&G[\gamma,\gamma_{_{\rm CR}}]\equiv
\int_{\rm max(\gamma,B)}^{\rm min(\gamma_0,T)}
{\beta_{\rm co}\,d\gamma_{\rm co}\over 
(\beta_{\rm co}\,\gamma_{\rm co})^{4-{\beta_{\rm conf}}}}\; ,
\label{RFlux}
\end{eqnarray}
where we have introduced the factor $n_{_A}=n(A,Z)$ of proportionality to the
number-density of intercepted ISM nuclear species, thereby specifying the full
$A$- and $Z$-dependence of the result.
Except for the overall factor $(A/Z)^{\beta_{\rm conf}}$, 
Eq.~(\ref{RFlux}) is very insensitive to 
${\beta_{\rm conf}}$
(over most of their extension, the integrals are powers and the powers
of $\gamma^{\beta_{\rm conf}}$ and $\gamma_{\rm co}^{-{\beta_{\rm conf}}}$ 
simply cancel).
It is also, down to $\gamma\sim 2$, well approximated by its very simple,
relativistic and analytical version:
\begin{eqnarray}
{dF_{\rm elast}\over d\gamma_{_{\rm CR}}}&\propto& n_{_A}
\left({A\over Z}\right)^{\beta_{\rm conf}}
\int_1^{\gamma_0}{d\gamma\over \gamma^{7/3}}
\,{G[\gamma,\gamma_{_{\rm CR}}]}\; ,\nonumber\\
G[\gamma,\gamma_{_{\rm CR}}]&\equiv&
\int_{\rm max[\gamma,\gamma_{_{\rm CR}}/(2\,\gamma)]}
^{\rm min[\gamma_0,2\,\gamma\,\gamma_{_{\rm CR}}]}
{d\gamma_{\rm co}\over \gamma_{\rm co}^{4}}\; ,
\label{NRFlux}
\end{eqnarray}
from which we have eliminated the weak dependence 
of the integrand on ${\beta_{\rm conf}}$.
Notice that the function $dF_{\rm elast}/d\gamma_{_{\rm CR}}$
depends only on the priors $n_{_A}$, ${\beta_{\rm conf}}$, 
and $\gamma_0$, but not on any
parameter specific to the mechanism of CR acceleration.

The flux of Eq.~(\ref{NRFlux}) has an abrupt upper limit at
$\gamma_{_{\rm CR}}\simeq 2\,\gamma_0^2$. The initial LFs of CBs
peak at $\gamma_0\sim 10^3$ and have a distribution extending
 up to  $\gamma_0\sim 1.5\times 10^3$, as in Fig.~\ref{DDLogGamma}. 
 Thus, the spectrum of a nucleus elastically scattered by CBs should end 
 at a {\it knee} energy
 \cite{DP}:
 \begin{equation}
 E_{{\rm knee}}(A) = 2\,\gamma_0^2\,A\,m_p
 \sim (2\;{\rm to}\;4)\times 10^{15}\,A\;{\rm eV.}
 \label{Eknee}
 \end{equation}
In our comparisons of theory and data, the distribution in Eq.~(\ref{NRFlux})
is convoluted with distributions of $\gamma_0$ values described by
Eq.~(\ref{gamdist}).
 
 \subsection{`Inelastic' scattering}
\label{inelastscatt}

A fraction of the ISM nuclei impinging on a CB is Fermi-accelerated within it.
We assume this process to be fast on the scale of a CB's slow-down time.
At a fixed LF of the CB, the spectral shape of the 
accelerated nuclei, in the CB's rest system, is that of Eq.~(\ref{gammaA}), 
independent of the particle's species and proportional to the number density
of intercepted ISM particles $n_{_A}$. We assume that a fixed, 
$A$-independent fraction of $n_{_A}$ is thus accelerated, so that
---in the CB's rest system--- their instantaneous distribution
of LFs, $\gamma_{_{\rm ac}}$,  is of the form:
\begin{equation}
{dN^{\rm inst}\over d\gamma\,d\gamma_{_{\rm ac}}}\propto
{n_{_A}\over \gamma^{4-\beta_{\rm ac}}}\,
{1\over \gamma_{_{\rm ac}}^{\beta_{\rm ac}}}\,
\Theta(\gamma_{_{\rm ac}}-\gamma)\,
\Theta(b\,\gamma^{1/3}-\gamma_{_{\rm ac}}),
\label{gammaac}
\end{equation}
where, for the typical reference parameters of Eq.~(\ref{bb}), $b\sim 10^7$. 
In analogy with the `elastic' case, we assume that
these particles keep the energy to which they were fastly
accelerated  within the CB. At 
the moment when the CB's LF, $\gamma_{_{\rm CB}}$,
has descended from $\gamma_0$
to $\gamma$, the distribution of LFs, 
$\gamma_{\rm {ac}}$, of the accumulated and accelerated particles is:
\begin{eqnarray}
{dN_{_{\rm ac}}\over d\gamma_{_{\rm ac}}}&\propto&
\int_\gamma^{\gamma_0}d\gamma_{_{\rm CB}}
{dN^{\rm inst}\over d\gamma_{_{\rm CB}}\,d\gamma_{_{\rm ac}}}\propto
{1\over \gamma_{_{\rm ac}}^{\beta_{\rm ac}}}\,F,\nonumber\\
F&\equiv&-{1\over\gamma_{_{\rm CB}}^{3-\beta_{\rm ac}}}
\Big|_D^U\;\;\Theta(U-D),\nonumber\\
U&\equiv&{\rm min}[\gamma_0,\gamma_{_{\rm ac}}],\nonumber\\
D&\equiv&{\rm max}[\gamma,(\gamma_{_{\rm ac}}/b)^3].
\label{Nacc}
\end{eqnarray}

The accelerated particles exit the CB by diffusion, as in Eq.~(\ref{dNdgdg}),
and are Lorentz-boosted by the CB's motion, as in Eq.~(\ref{Lboost}).
The accelerated contribution to the CR spectrum is important only for
energies above the knees, and the relativistic ($\beta\sim 1$) approximation
is good, except in some of the integration limits, wherein
factors such as $1-\beta$ do appear. The final result for this 
`Fermi-accelerated' or `inelastic' contribution
to the flux is:
\begin{equation}
{dF_{\rm inel}\over d\gamma_{_{\rm CR}}}\propto n_{_A}
\left({A\over Z}\right)^{\beta_{\rm conf}}
\int_1^{\gamma_0}{d\gamma\over \gamma^{{\beta_{\rm conf}}+7/3}}\,
\int_{C}^{S}
{F\,d\gamma_{\rm ac}\over \gamma_{\rm ac}^{\beta_{\rm ac}+1-{\beta_{\rm conf}}}}\, ,
\label{Inelastic}
\end{equation}
where $F$ is defined in Eq.~(\ref{Nacc}) and
\begin{eqnarray}
C&\equiv&
{\rm min}[b\,\gamma_0^{1/3},T(\gamma,\gamma_{_{\rm CR}})]\nonumber\\
S&\equiv&
{\rm max}[\gamma,B(\gamma,\gamma_{_{\rm CR}})],
\label{Somelimits}
\end{eqnarray}
with $T$ and $B$ as in Eq.~(\ref{constraints}).
Once again, except for the overall factor $(A/Z)^{\beta_{\rm conf}}$, 
Eq.~(\ref{Inelastic}) is very insensitive to ${\beta_{\rm conf}}$. As for the elastic case,
the function $dF_{\rm inel}/d\gamma_{_{\rm CR}}$
depends only on the priors $n_{_A}$, ${\beta_{\rm conf}}$, $\beta_{\rm ac}$
and $\gamma_0$, but not on any parameter not previously constrained.
In our comparisons of theory and data, the distribution in Eq.~(\ref{Inelastic})
will be convoluted with distributions of $\gamma_0$ values described by
Eq.~(\ref{gamdist}).

The  flux of Eqs.~(\ref{Inelastic}, \ref{Somelimits}) cuts off at a 
maximum energy:
nuclei exiting a CB after having been accelerated within it have
energies extending up to $E_{{\rm end}}=2\,\gamma_0\,E[{\rm Larmor}]$, 
with $E[{\rm Larmor}]$ as in Eq.~(\ref{LarmorE}), that is:
\begin{equation}
E_{{\rm end}}(Z)\sim(2\;{\rm to}\;6)\times 10^{20}\,Z\;{\rm eV}.
\label{toe}
\end{equation}
These `end-points''  scale as $Z$, unlike the knees, which scale
like $A$. The predictions in
Eq.~(\ref{toe}) will not be easy to test, for three reasons: 
the end-point energies are in the same ball-park as the GZK cutoff;
for $A>1$ the ultra-high energy flux is strongly suppressed by photo-dissociation; 
and the extraction of relative CR abundances at very high energies is a 
very difficult task.

\subsection{The complete spectrum}

\begin{figure}
\begin{center}
\epsfig{file=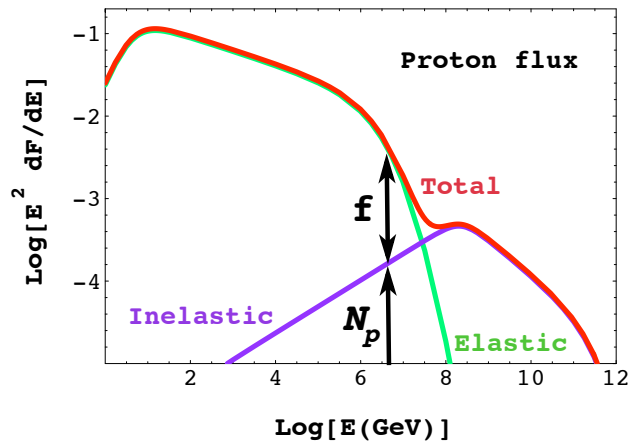, width=8.5cm,angle=-0}
\end{center}
\caption{Elastic and inelastic contributions to the proton source spectrum.
 The vertical scale is 
$E^{2}\,dF/dE$. }
\label{DDelinel}
\end{figure}

The complete source spectrum of each CR nucleus is the sum of an elastic and
an inelastic contribution. This sum and its addends are illustrated, for protons, 
in Fig.~\ref{DDelinel}. The figure shows an elastic flux larger than the inelastic one
by a factor $f\simeq 10$ at the nominal position of the proton's knee. 
This ratio $f$ is the only required input  for which we have
no `prior' information. It is the only parameter we need to choose in an
unpredetermined range. We assume $f$ to be the same for all nuclei, 
in accordance with the purely
`kinematical' character of the acceleration by  `magnetic racket' CBs. 

The other parameter in Fig.~\ref{DDelinel}, ${N_p}$, is the normalization of the 
proton inelastic flux at the nominal position of the proton's knee. 
Albeit within large error bars, $N_p$ will be determined from the predicted
luminosity of Eq.~(\ref{SNEsupply}), in the way discussed in Sections
\ref{EGnorm}, \ref{ankleregion}. The abundances of 
the other elements relative to protons ---or, equivalently, the normalization of their 
fluxes--- are predicted, as  discussed in Section \ref{relabundances}. Thus, the 
ensemble of source fluxes in Fig.~\ref{Groups}a has been constructed with just one 
fit parameter:  $f$.

Notice in Fig.~\ref{DDelinel} the different  shapes of the elastic and
inelastic contributions, implying that the fraction of accelerated
nuclei is small, as in the results of the numerical analysis of the
relativistic merging of plasmas \cite{Fred}.

Many salient features of the source fluxes of CRs ---the pronounced
knees in the individual-element spectra, the differential changes
of slope, and a maximum energy for proton acceleration--- survive
unscathed the many tribulations transmogrifying the source spectra into
the observed ones. To discuss the comparison of predictions and data
we must first summarize these tribulations.

\section{Tribulations of a Cosmic Ray}
\label{tribulations}

On its way from its source to the Earth's upper atmosphere, a CR 
is influenced by the ambient magnetic fields, radiation and matter, 
which it encounters. Extragalactic CRs are also affected by cosmological
redshift ($z$) and the dependence of their source strength on the 
star-formation rate as a function of  `look-back time'. In this section 
we list the tribulations of CRs ---which are discussed in detail in 
several Appendices---
and we  summarize the choices we make for 
the priors that are not very well understood either observationally or 
theoretically. 
Three types of interactions constitute a CR {\it tribulation}:
\begin{itemize}
\item{} {\it Interactions with magnetic fields} that
permeate galaxies and clusters and, presumably, the IGM.

The fluxes of CRs of Galactic origin are, below their free-escape energy,
enhanced by a factor proportional to their
confinement time. At higher energies they escape the Galaxy practically
unhindered.

Extragalactic CRs entering the Galaxy must overcome the effect of its 
exuding magnetic wind.

\item{}{\it Interactions with radiation,} significant for
CRs of extragalactic origin. The best studied one is $\pi$
photoproduction by nuclei on the cosmic microwave background radiation,
the GZK effect \cite{GZK}.

Pair ($e^+ e^-$) production is akin to the GZK effect.

The photo-dissociation of extragalactic CR nuclei, mainly on the cosmic
infrared background radiation, is also extremely relevant. 

\item{}{\it Interactions with the ISM} are fairly well understood for relatively 
low-energy CRs of Galactic origin. Their spallation gives rise
to `secondary' stable and unstable isotopes in the CR flux. 
\end{itemize}

Of the above items, three need be discussed here:

\subsection{Magnetic confinement and escape; the ankles}

The Galaxy's MF, whose typical value is of ${\cal{O}}(5)\,\mu$G,
as in Eq.~(\ref{rhoEB}),
varies on scales ranging up to a 
 `coherence length' of ${\cal{O}}(1)$ kpc. 
The MF in the Galaxy's halo is not well charted; its typical
value is similar. The Larmor radius  of a 
CR of charge $e\,Z$ and momentum $p(E)$ is:
\begin{eqnarray}
&&R_L\approx 0.65\;{\rm kpc}\;{5\;\mu{\rm G}\over B}\;
{p(E)\over E_{\rm ankle}(Z)}\, , \label{LarmorR}\\
&&E_{\rm ankle}(Z)\equiv Z\times (3\times 10^{18} \;\rm eV).
\label{Larmor}
\end{eqnarray}
A CR of energy $E\ge E_{\rm ankle}$ cannot be significantly bent in
the Galaxy. For $Z=1$, Eq.~(\ref{Larmor}) coincides with the `ankle' in
the CR flux, see Eqs.~(\ref{flux}). 

At $E\sim E_{\rm ankle}(Z)$, Galactic CR nuclei
undergo a random walk process of moderate
deflections on the Galactic MF domains. Their cumulative deflection
angle has a Gaussian distribution, analogous to the one describing the multiple
deflection of high-energy muons in matter \cite{PDG}. The escapees are the CRs
deflected by less than an angle of order one radian. We need a rough
description of the corresponding confinement and escape probabilies, which we 
characterize by:
\begin{equation}
P_{\rm conf}=1-P_{\rm esc}
=\exp\left[-\left({E\over E_{\rm ankle}(Z)}\right)^2\right]\, .
\label{Escape}
\end{equation}

The Galactic CR flux is modulated by the momentum dependence of the CR
confinement time, $\tau_{\rm conf}$, in the disk and halo of the
galaxy, affecting the different species in the same way, at fixed $p/Z$.
Confinement effects are not well understood \cite{Swordy,Ptuskin},
but observations of astrophysical
and solar plasmas indicate that \cite{Swordy}: 
\begin{eqnarray} 
\tau_{\rm conf}&\sim& K \left({Z\,{\rm GeV/c}\over p}\right)^{\beta_{\rm conf}},
\label{galconf}\\
K&\sim& 2\times 10^7\,{\rm y,}
\label{K}\\
{\beta_{\rm conf}}&\sim& 0.6\pm 0.1. 
\label{c} 
\end{eqnarray}
Measurements of the relative abundances of secondary CR isotopes 
 \cite{Swordy}
agree with the functional form of Eq.~(\ref{galconf}).
The observed ratios of unstable CR isotopes \cite{Connell}
result in  $K$ as in Eq.~(\ref{K}), but the method is well known to be biased
towards low-$K$ values \cite{Longair}.
Our theory results in a somewhat larger predicted value of $K$, as discussed in
Sections~\ref{timevolume}, \ref{electrons}. 

The spectrum of observable CRs of Galactic origin,
$dF_{\rm Gal}/dE$, is their source
spectrum, $dF_s/dE$, modified by confinement \cite{Axford} 
so that $dF\!\propto\! P_{\rm conf}\,\tau_{\rm conf}\,dF_s$, or:
\begin{eqnarray}
{dF_{\rm Gal}\over dE}&=& C(E,Z)
\;{dF_s\over dE}\, ,\nonumber\\
C(E,Z) &\equiv& P_{\rm conf}
{\left[{E_{\rm ankle}(Z)\over p(E)}\right]}^{\beta_{\rm conf}}\, ,
\label{ConfSpec1}
\end{eqnarray}
with $E_L$ and $P_{\rm conf}$ as in Eqs.~(\ref{Larmor}, \ref{Escape}).
Note that $C(E,Z)$ does not depend on the magnitude of the factor
$K$ in Eq.~(\ref{K}).

\subsection{CR penetration into the Galaxy}

In a steady-state situation, the CR flux escaping a galaxy has the 
energy dependence of the {\it source} flux, not the confinement-modified
flux. In our CR theory, the extragalactic flux arriving to our Galaxy is
simply the CR flux exiting other galaxies, modified by the tribulations
of an intergalactic journey. How do these extragalactic
CRs penetrate our Galaxy?

The penetration of Galactic CRs into the solar system is hindered 
by the `wind' of solar CRs and MFs. Analogously, we proceed to argue,
the penetration of extragalactic CRs into the Galaxy is hindered by 
the `wind' of Galactic CRs and MFs. 
The Galaxy certainly exudes a wind of CRs: in a steady state the outgoing 
Galactic flux is that of the sum of Galactic sources. The question is whether
the Galaxy also has an accompanying MF `wind'.

The Galactic CR- and MF-energy densities are known to be approximately
coincident, a strong hint of an intimate 
relationship. In the CB model CRs are the dominant 
source of MFs in galaxies, clusters and the IGM (this is a tenable statement 
for two reasons: CR sources are kiloparsecs-long CB trajectories,
as opposed to SN shells in star-formation regions, and the Galactic CR
luminosity is almost one order of magnitude bigger
than in the conventional view). For all these systems, the simple hypothesis 
of rough energy--density equipartition between CRs and MFs results in 
correct predictions for the intensity of the latter \cite{DDMF}. 

If the interaction between CRs and the ambient medium 
results in turbulent currents whose MFs end up storing some 50\% 
of the energy density, we expect a large
fraction of the {\it momentum} of CRs to be transferred to the MFs.
This would imply the existence of a hefty Galactic MF wind. The expanding shells
of SNe ---and the superbubbles that ensembles of SNe  generate--- 
should also carry in their motion a Galactic MF wind.

Knowing little about the magnetic wind of the Galaxy, we cannot ascertain the
probability $P_{\rm pen}(E,Z)$ that an extragalactic CR penetrates
it. The flux of such CRs at the Sun's location is renormalized by a factor
$C'(E,Z)\propto P_{\rm pen}\,\tau_{\rm conf}$,
the extragalactic source analogue to $C(E,Z)$ in Eq.~(\ref{ConfSpec1}).
In the absence of a wind the Galaxy would act as a diffusive magnetic
`trap' and, for a steady-state external flux, $P_{\rm pen}=1$.
At energies above the ankle, $P_{\rm pen}$ and $C'$ must be close to
unity. At smaller energies $P_{\rm pen}$ must decrease in a 
manner reminiscent of the quenching of low-energy CRs by
the Sun's wind. 

We have faced our ignorance on $C'$ by trying very many different
ansatzes. The features of the source spectrum (slopes, knees,
ankle) are very `robust' and survive unscathed the choice of a reasonable $C'$.
This is true even for the extreme `no-wind' possibility: $P_{\rm pen}=1$ 
at all energies. However, the overall description of the data is much more 
satisfactory if $P_{\rm pen}<1$ below the ankle or, at least, below the knees.
To illustrate this, we shall report results for two very different cases:
\begin{eqnarray}
{\rm (a)}\;\; C'(E,Z)&=&1 \label{conf-a}\\
{\rm (b)}\;\; C'(E,Z)&=& {\left[{E_{\rm ankle}(Z)\over p(E)}\right]}^{-{\beta_{\rm conf}}}
\;\;{\rm for}\;\;E<E_{\rm ankle}(Z)\nonumber\\
&=&1\;\;\;\;\;\;\;\;\;\;\;\;\;\;\;\;\;\;{\rm for}\;\; E>E_{\rm ankle}(Z).
\label{conf-b}
\end{eqnarray}
Case (a) corresponds to a Galactic wind that quenches the entrance
of extragalactic CRs by as much as the Galactic confinement enhances their
flux, once they are in. Case (b), with ${\beta_{\rm conf}}$ as in Eq.~(\ref{c}), 
corresponds to a wind that is `twice as repellent' as in case (a).

\subsection{Photo-dissociation}

At energies higher than a few $10^8$ GeV,
CR nuclei of extragalactic origin interact with the
cosmic background radiation (CBR) and are photo-dissociated:
one or a few nucleons per collision are stripped off. The important
CBR wavelength domain extends from the ultraviolet to the far infrared,
corresponding to centre-of-mass energies at which the giant
dipole resonance lies. Computing the effects of photo-dissociation
for a given CR source spectrum and composition, 
given present and past radiation densities, and given cross sections
and lifetimes for parent and daughter nuclei, would be straightforward 
but extremely lengthy. For our current purposes it suffices to estimate 
the effect, which we do in Appendix \ref{pd}, summarized below.

Independent of atomic number, the approximate energy at which 
the mean photo-dissociation time of a nucleus travelling in the
current CBR coincides with the age of the Universe is:
\begin{equation} 
E_{\rm PhD}\simeq 7\times 10^{17}\;\,\rm eV,
\label{PhDisCutNA}
\end{equation}

The photo-dissociation effect on the extragalactic 
CR flux of (arrival) energy $E$ and (departure) atomic weight $A$ is
well approximated by an attenuation factor:
\begin{equation}
A_{\rm PhD}(E,A)\approx  
{1\over\sqrt{ 1+[3.15\,E/ (n(A)\,E_{\rm PhD})]^2}},
\label{PhDisCut2NA}
\end{equation}
where $n(A)$ is an estimate of the average number of successive 
photo-dissociations required to reduce to $A/2$ the largest fragment of
an original nucleus $A$.

\subsection{The processed fluxes}
\label{versus1}

\begin{figure}[]
\centering
\vbox{
\epsfig{file=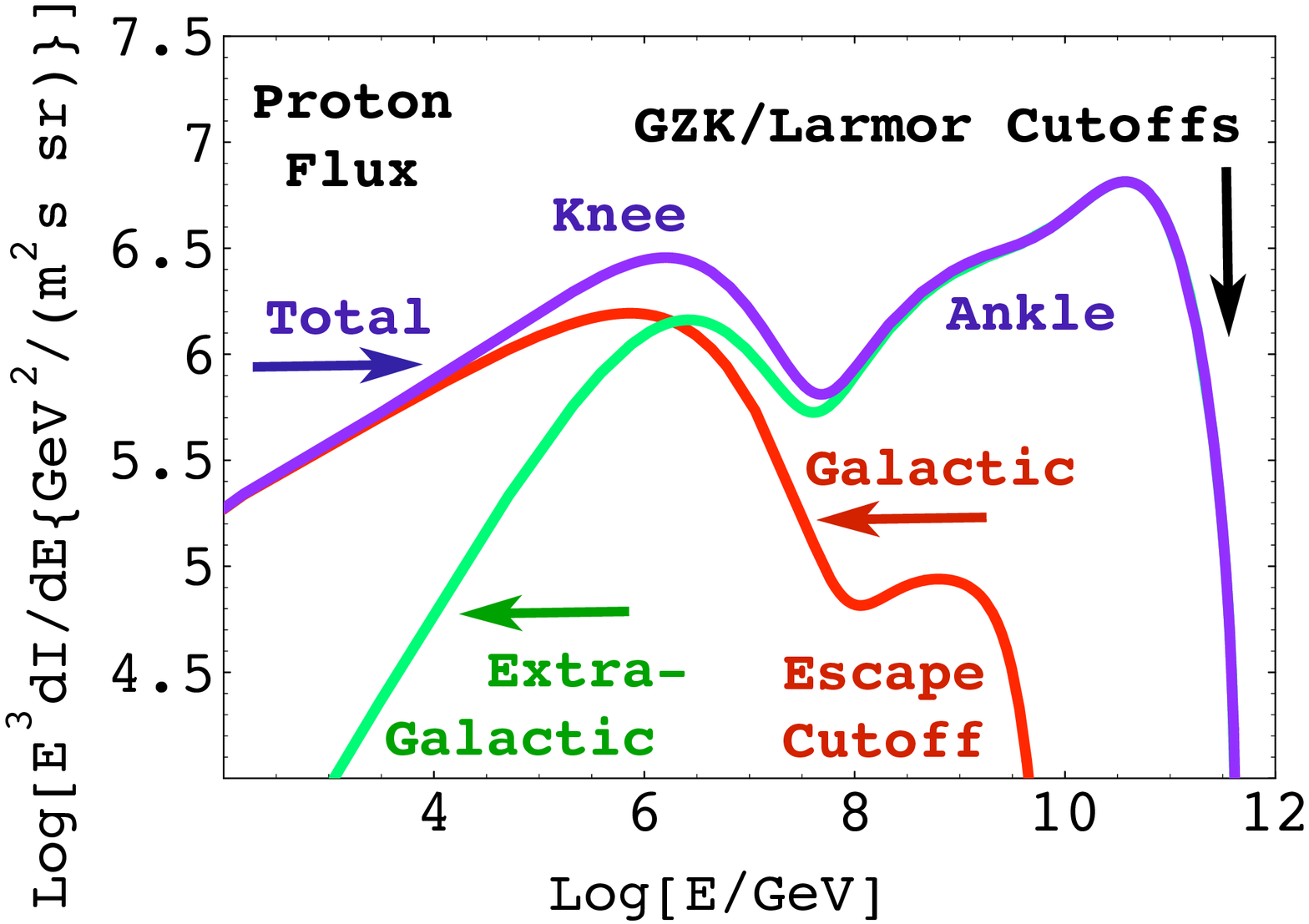, width=8.5cm,angle=-0}
}
\vbox{
\epsfig{file=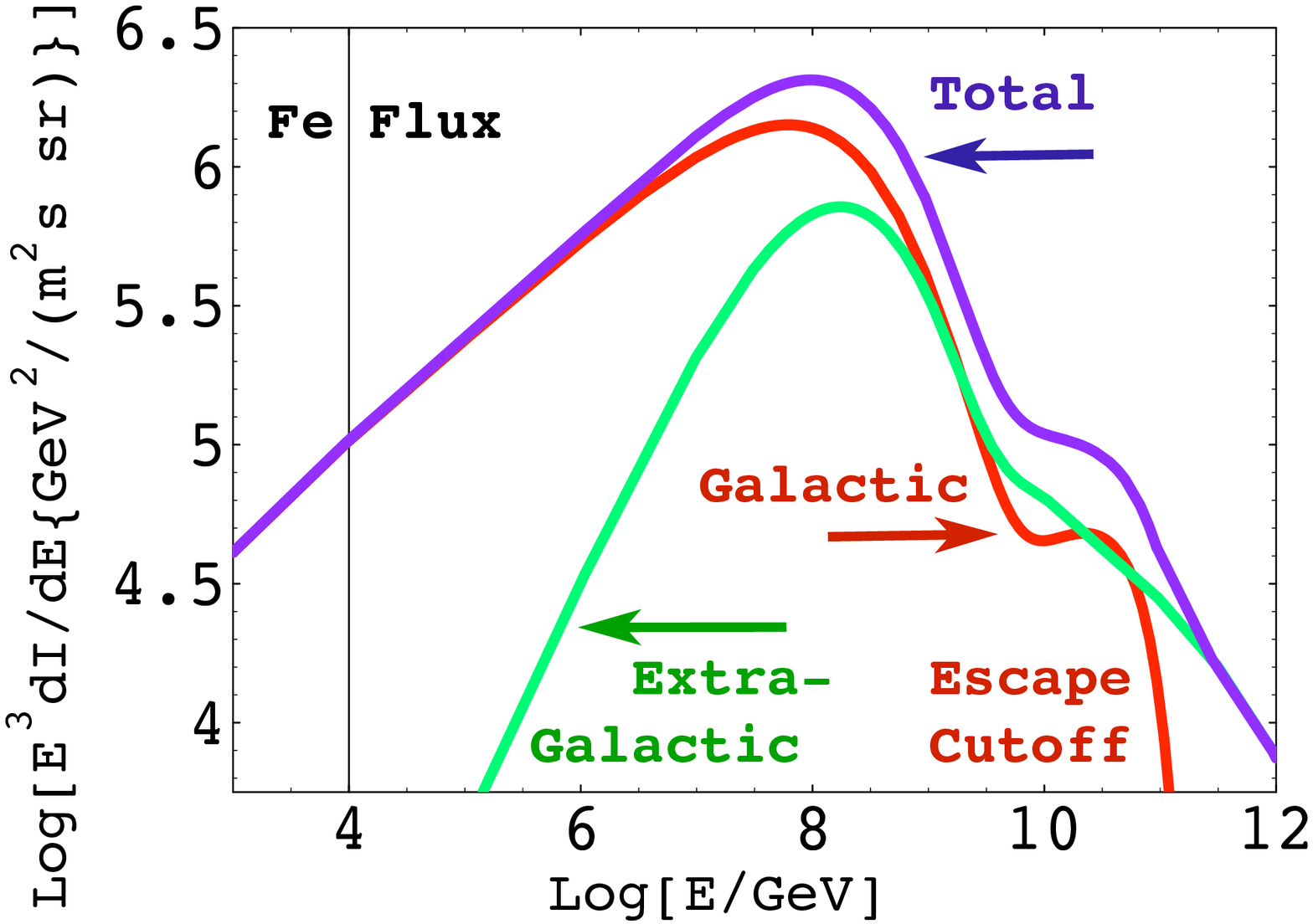, width=8.5cm,angle=-0}
}
\vspace*{-5pt}
 \caption{Features of the CR spectra for protons and Fe.}
 \label{ProtonGEG}
\end{figure}

The Galactic and extragalactic  fluxes at the Earth's location are 
affected by the tribulations we just discussed. To illustrate this point we have split
the proton and Fe fluxes of Fig.~\ref{Groups}b into their Galactic and extragalactic 
contributions, and we report the result in Fig.~\ref{ProtonGEG}.
The cutoffs in the Galactic fluxes are due to CR escape, parametrized as in
Eq.~(\ref{ConfSpec1}). The extragalactic fluxes are suppressed below
the ankle by the Galactic penetrability effect of Eq.~(\ref{conf-a}) or (\ref{conf-b})
and redshifted as discussed in Appendix \ref{Redshift}.
The high-energy flux of extragalactic Fe is attenuated by photo-dissociation,
parametrized by Eq.~(\ref{PhDisCut2NA}). The ultra-high energy proton flux 
is almost exclusively extragalactic in origin. Its shape at the
highest energies is governed by the acceleration end-point of Eq.~(\ref{toe}),
the GZK cutoff of Eq.~(\ref{GZK2}) and the pair-production suppression
of Eq.~(\ref{PairProd}).

\section{Detailed CB-model Results}
\label{CBresults}

\subsection{The index below the knee}
\label{belowtheknee}

The elastic contribution to the CR flux dominates below the knee,
as can be seen in Fig.~\ref{DDelinel}. For a large range of energies
($\sim$ ten to a million times $A\,m_p\,c^2$),
this source flux is very well approximated by a power law:
\begin{equation}
{dF_{\rm elast}\over d\gamma_{_{\rm CR}}}\propto \gamma_{_{\rm CR}}^{-\beta_s}.
\label{powerlaw}
\end{equation}
The value of $\beta_s$ can be trivially extracted from Eq.~(\ref{NRFlux}).
It is:
\begin{equation}
\beta_s={13\over 6}\approx 2.17.
\label{beautyfulindex}
\end{equation}
The observed spectrum should be steeper, in accordance
with the Galactic confinement effect of
Eq.~(\ref{cintrod}). The predicted index is: 
\begin{equation}
\beta_{\rm th}=\beta_s+{\beta_{\rm conf}}\approx 2.77\pm 0.10, 
\label{betapred}
\end{equation}
in agreement with the observed value, $2.73\pm 0.03$ for protons, 
reported in Eq.~(\ref{protons}); or $\sim\! 2.7$ for the all-particle flux,
as in Eq.~(\ref{flux}). Above the knee and
over the range, illustrated in Fig.~\ref{DDelinel},
 in which the inelastic contribution $dF_{\rm inel}/d\gamma_{_{\rm CR}}$ is well
described by a power law, $\gamma_{_{\rm CR}}^{-\beta_s'}$,
 its slope is steeper than that in Eq.~(\ref{powerlaw}):
$\beta_s'\approx \beta_s+0.3$.

The prediction of the spectral index is gratifying: simple, analytical, and almost
exclusively based on trivial kinematics. It is, moreover, very insensitive to many
assumptions, e.g.~any non-singular non-isotropic angular distribution of particles
elastically scattered by the CB in its rest system gives the same result for $\beta_s$
as the isotropic distribution we used here.

\subsection{Relative abundances}
\label{relabundances}

It is customary to present results on the composition of CRs at a fixed
energy per nucleus $E_{_A}=1$ TeV, as opposed to a fixed LF.
This chosen energy is relativistic ($E_{_A}\simeq p_{_A}$), it is below the 
corresponding knees for all $A$, and it is in the domain wherein the source 
fluxes are dominantly elastic and
are very well approximated by the power-law in Eq.~(\ref{powerlaw}),
with the index $\beta_s$ of Eq.~(\ref{beautyfulindex}). Up to a common
species-independent factor, then:
\begin{equation}
{dF_{\rm source}\over d\gamma_{_{\rm CR}}}
\propto n_{_{A}}\,\left({A\over Z}\right)^{\beta_{\rm conf}}
\;\gamma_{_{\rm CR}}^{-\beta_s},
\label{composource}
\end{equation}
where we have taken into account the species dependence of the
source flux, as in Eq.~(\ref{NRFlux}). Change variables
($E_{_A}\!\propto\! A\,\gamma$) in Eq.~(\ref{composource}) and
modify the result by the multiplicative confinement factor, 
$(Z/E)^{\beta_{\rm conf}}$, of Eq.~(\ref{galconf})
to obtain the prediction for the observed fluxes:
\begin{equation}
{dF_{\rm obs}\over dE_{_A}}\propto \bar{n}_{_A}\,A^{\beta_{\rm th}-1}
\,E_{_A}^{-\beta_{\rm th}},\,\,\,\,\,\,\,\, \beta_{\rm th}-1\sim 1.77,
\label{compo}
\end{equation}
with $\bar{n}_{_A}$ an average ISM nuclear abundance and  $\beta_{\rm th}$
from Eq.~(\ref{betapred}). At  fixed energy the predictions for the CR abundances 
$X_{_{\rm CR}}$ relative to protons are:
\begin{eqnarray}
X_{_{\rm CR}}(A)&\approx& X_{\rm amb}(A) \; A^{1.77},\nonumber\\
X_{\rm amb}(A) &\equiv& {\bar{n}_{_A}/ \bar{n}_p},
\label{composimple}
\end{eqnarray}
where $X_{\rm amb}$ are the ambient `target' abundances relative to hydrogen.

\begin{figure}
\begin{center}
\epsfig{file=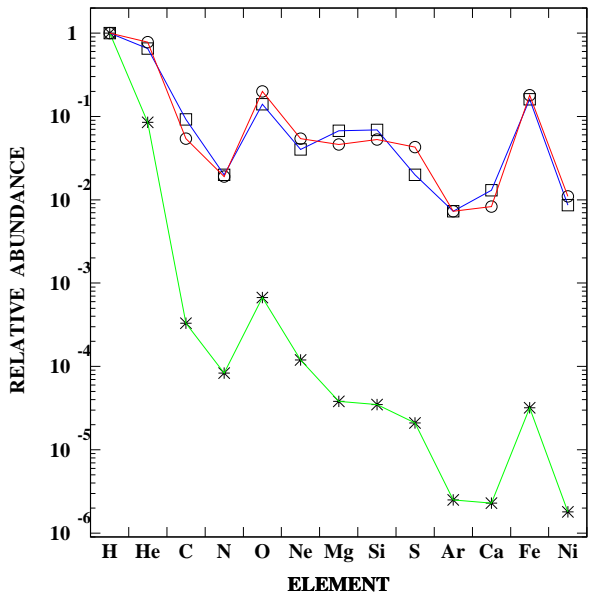, width=8.8cm,angle=0}
\end{center}
\vspace{-.5cm}
\caption{The relative abundances of primary CR nuclei, from H to Ni
around 1 TeV \cite{WBM}. The
	stars (joined by green lines) are solar-ISM abundances \cite{GS}.
The circles (joined in red) are the predictions, with input superbubble
abundances. The squares (joined in black) are the CR observations.}
\label{f1}
\end{figure}

Cannonballs produce CRs while travelling in the large `metallicity'
environments of a SN-rich domain and the enclosing superbubble (SB). 
Let $X_{_{\rm SB}}$ be
the abundances in these domains, relative to H.
 Only late in their voyage do CBs reach regions wherein
the relative abundances of the ISM are solar-like, 
$X_{\rm amb}\! \simeq \! X_\odot$. For He the observations yield 
$ X_{_{\rm SB}}\! \simeq \! X_\odot$. 
For the intermediate elements ranging from C to Ne, 
$ X_{_{\rm SB}}\! \simeq \! 2\, X_\odot$ \cite{HL}. The abundances of heavier metals 
 in SBs  are poorly known. They should be close to those of old SNRs, 
also not well measured.
One exception is SNR W49B,  recently observed with XMM-newton \cite{Miceli}.
The best-fitted spectral parameters have resulted in
$ X_{_{\rm SNR}}/ X_\odot$ values, $3.3\pm 0.2$ for Si, 3.7 +0.1/$-$0.2 
for S, 4.2 +0.3/$-$0.4
for Ar, $6.4\pm 0.5$ for Ca, 6 +0.1/$-$0.2
for Fe and 10 +4/$-$1
for Ni. We use these  values in the
predictions of Eq.~(\ref{compo}), reported in Fig.~\ref{f1} and Table
\ref{table1}, even though the mean abundances in SBs may
differ from those measured in a given SNR.

The results of Fig.~\ref{f1} are for the most abundant, dominantly primary
CRs. We have suppressed the error bars of the input $X_{_{\rm SB}}$
values: even the size of the errors is debatable. Yet, the results are
satisfactory. In spite of its simplicity, Eq.~(\ref{compo})
snugly reproduces the large enhancements in the heavy CR abundances
relative to hydrogen, with respect to solar or SB abundances. 

In Table \ref{table1} we report in detail the abundances of the
primary and secondary (mainly odd-$Z$) elements.
The predictions of Eq.~(\ref{compo}) are slight overestimates of the CR  
observations for intermediate and heavier primaries. For the CR
secondaries, the predictions are always underestimates. When
elements are added in groups of primaries and their most
abundant secondaries, the agreement between theory and observation is 
even better. All this is to be expected: we have not considered the nuclear 
spallations depleting primaries and making secondaries.
\begin{table}
\caption{Solar abundances and CR abundances at 1 TeV, both relative to 
hydrogen.
\label{table1}}
\begin{tabular}{lcccc}\colrule
& Z & $X_\odot$\footnote{Solar=ISM abundances
\cite{GS}.} & $X_{_{\rm CR}}$\footnote{CR abundances relative to hydrogen
at 1 TeV \cite{WBM}.} &  $A^{1.73}\, X_{_{\rm SB}}$\\
\colrule

H  & 1 &  1                   & 1   &  1  \\
He & 2 & $7.5\times 10^{-2}$ & $6.5\times 10^{-1}$
& $8.2\times 10^{-1}$\\

\colrule
C  & 6 & $3.3\times 10^{-4}$ & $9.2\times 10^{-2}$
& $4.9\times 10^{-2}$\\

N  & 7 & $8.3\times 10^{-5}$ & $2.0\times 10^{-2}$
& $1.6\times 10^{-2}$\\

O  & 8 & $6.7\times 10^{-4}$ & $1.4\times 10^{-1}$
& $1.6\times 10^{-1}$\\

Ne & 10& $1.2\times 10^{-4}$ & $3.8\times 10^{-2}$
& $ 4.3\times 10^{-2}$\\

\colrule

C--Ne &   &   &  $2.9\times 10^{-1}$
& $2.7\times 10^{-1}$\\

\colrule
Na & 11& $2.2\times 10^{-6}$ & $6.5\times 10^{-3}$
& $1.7\times 10^{-3}$\\

Mg & 12& $3.8\times 10^{-5}$ & $6.7\times 10^{-2}$
& $3.2\times 10^{-2}$\\

Al & 13& $3.0\times 10^{-6}$ & $1.0\times 10^{-2}$
& $3.1\times 10^{-3}$\\

Si & 14& $3.5\times 10^{-5}$ & $6.9\times 10^{-2}$
& $3.7\times 10^{-2}$\\

P  & 15& $2.8\times 10^{-7}$ & $2.3\times 10^{-3}$
& $3.7\times 10^{-4}$\\

S  & 16& $2.1\times 10^{-5}$ & $2.0\times 10^{-2}$
& $3.2\times 10^{-2}$\\

Cl & 17& $3.2\times 10^{-7}$ & $2.6\times 10^{-3}$
& $6.0\times 10^{-4}$\\

Ar & 18& $2.5\times 10^{-6}$ & $7.3\times 10^{-3}$
& $6.2\times 10^{-3}$\\

K &  19& $1.3\times 10^{-7}$ & $4.7\times 10^{-3}$
& $3.8\times 10^{-4}$\\

Ca & 20& $2.3\times 10^{-6}$ & $1.3\times 10^{-2}$
& $8.7 \times 10^{-3}$\\
\colrule

Na--Ca &   &   &  $2.0\times 10^{-1}$
& $1.3\times 10^{-1}$\\

\colrule

Sc & 21& $1.4\times 10^{-9}$ & $2.6\times 10^{-3}$
& $6.5\times 10^{-6}$\\

Ti & 22& $1.0\times 10^{-7}$ & $9.8\times 10^{-3}$
& $5.2\times 10^{-4}$\\

V  & 23& $1.0\times 10^{-8}$ & $5.5\times 10^{-3}$
& $5.8\times 10^{-5}$\\

Cr & 24& $4.7\times 10^{-7}$ & $1.2\times 10^{-2}$
& $2.8\times 10^{-3}$\\

Mn & 25& $2.5\times 10^{-7}$ & $1.2\times 10^{-2}$
& $1.6 \times 10^{-3}$\\

Fe & 26& $3.2\times 10^{-5}$ & $1.6\times 10^{-1}$
& $2.0\times 10^{-1}$\\

Co & 27& $8.3\times 10^{-8}$ & $6.5\times 10^{-4}$

& $6.2\times 10^{-4}$\\

Ni & 28& $1.8\times 10^{-6}$ & $8.6\times 10^{-3}$
& $2.0\times 10^{-2}$\\

\colrule
Sc--Ni &   &   &  $2.1\times 10^{-1}$
& $2.2\times 10^{-1}$\\
\colrule
\end{tabular}
\\
\end{table}
 
\subsection{Composition dependence of the spectral slopes}
\label{composlopes}

At each local value of its decelerating LF, $\gamma$, a CB exudes 
elastically scattered CRs with LFs $\gamma_{_A}$ in the range 
$1\leq\gamma_{_A}\leq 2\,\gamma^2$, as well as internally pre-accelerated 
CRs in the range $1\leq\gamma_{_A}\leq 2\,b\,\gamma^2$; see Section 
\ref{collisionless}. The higher-energy CRs must have been gathered by a CB 
from the ISM when $\gamma$ is close to $\gamma_0$ and the CB is close to 
its place of origin, where the abundance of the elements is that of a 
star-forming region or its surrounding SB. The lower-energy CRs are 
generated all along the CB's trajectory and pile-up from its low-$\gamma$ 
end, a point at which a CB is typically travelling in a `normal' ISM, with 
a composition close to that of the solar neighbourhood. This complicated 
effect may be approximated by a composition-dependence of the spectral 
slopes, $\beta_{_A}$, of the flux of the different nuclei.

To illustrate this point, consider the CR flux below the knee, dominated
by elastically scattered CRs. Adopt an extreme simplifying ansatz: that CRs 
with  $\gamma\sim 2\, \gamma_0^2$ are accelerated within SBs,
whereas CRs with $\gamma\sim 1$
are accelerated in the ISM further away from the SBs.
Since the abundance of Fe nuclei in the SB is $\sim 6$ times 
larger 
than their abundance in the average ISM, the flux of CR Fe nuclei with 
$\gamma\sim 2\, \gamma_0^2$ is enhanced by a factor $\sim 6$ 
relative to their flux at $\gamma\sim 1$. This is equivalent to a change
$\Delta \beta_{\rm Fe}$ in the slope of the CR Fe flux which 
satisfies
\begin{equation}
[2\, \times \gamma_0^2]^{\Delta\beta_{\rm Fe}}\approx
 [3\times 10^6]^{\Delta\beta_{\rm Fe}} =6,
\label{betaFe}
\end{equation}
or $\Delta \beta_{\rm Fe}\approx 0.12$.
The predicted slope of the Fe spectrum below the knee is
then $\beta_{\rm Fe}=\beta_{\rm th}-\Delta\beta_{\rm Fe}=2.65$, with 
$\beta_{\rm th}$
as in Eq.~(\ref{betapred}), in good agreement with the observed
$\beta_{\rm Fe}=2.60\pm 0.10$ \cite{WBM}.

\begin{figure}
\begin{center}
\epsfig{file=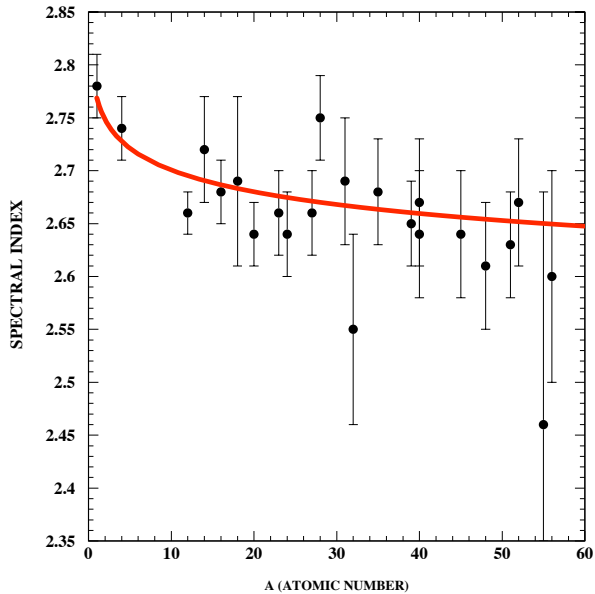, width=8.8cm,angle=-0}
\end{center}
\vspace{-0.5cm}
\caption{A rough prediction of the $A$-dependence of the spectral indices
around 1 TeV,
shown for some of the most abundant primary elements \cite{WBM}.
The spectral slopes of H and He are those measured by AMS \cite{AMS}.}
\label{ArnonSlopes}
\end{figure}

The above exercise can be redone for the rest of the elements, with the
result that, to a good approximation, $\beta_{_A}\simeq \beta_{\rm th}-0.03\,\ln(A)$
\cite{DAD}. The predicted and observed slopes below the knees are shown
in Fig.~\ref{ArnonSlopes}. 

\subsection{The normalization of the extragalactic flux}
\label{EGnorm}

The cleanest place of choice to discuss the normalization of the CR
spectrum is at ---or slightly above--- the ankle.  At such an energy CRs
are extragalactic and their spectrum is insensitive to Galactic
MFs and winds, to the GZK and CB-acceleration cutoffs
of Eqs.~(\ref{GZK2}, \ref{toe}), to the effects of photo-dissociation above the
energy of Eq.~(\ref{PhDisCutNA}), to the distribution of LFs of
Eq.~(\ref{gamdist}), to the `elastic-scattering' contributions ending at
the much lower energies of Eq.~(\ref{Eknee}), and to the detailed
composition of CRs, for only H is abundant at that energy.

Let $\epsilon_{_{\rm U}}$ be the energy-integrated,
average current energy flux of  CRs in intergalactic space, accumulated 
over look-back time $t(z)$. Up to the highest energies ---at which effects such
as the GZK cutoff are relevant--- there is nearly no energy loss 
except for the redshift effect, and:
\begin{equation}
\epsilon_{_{\rm U}}={c\over 4\,\pi}\,R_{\rm SN}{\rm[U]}
\int dz\,{dt\over dz}\,{E[{\rm jets}]\over 1+z}\,{R_{\rm SF}(z)\over R_{\rm SF}(0)},
\label{AccEnergy}
\end{equation}
where $dt/dz$ is the time to redshift relation specified in Appendix \ref{Redshift};
$R_{\rm SN}\rm{[U]}$ is the average current SN 
rate per unit universal volume, given by Eq.~(\ref{SNunivrate});  $E[{\rm jets}]$, as in Eq.~(\ref{Ejets}), is the
average jet energy per SN; and $R_{\rm SF}(z)$ is the star-formation rate 
reviewed in Appendix \ref{star}. 

The extragalactic flux has the energy distribution of Eq.~(\ref{z}). Its 
normalization is specified by the constraint:
\begin{equation}
\int_{E_{\rm min}}E\;{dF{\rm[EG]}\over dE} = \epsilon_{_{\rm U}},
\label{Econstraint}
\end{equation}
which allows us to compute $dF{\rm[EG]}/dE$
 at any energy, using the observed (or fitted)
 CR flux and the adopted 
correction for Galactic confinement, Eq.~(\ref{ConfSpec1}).
The result is proportional to  $E_{\rm min}^{-0.2}$: insensitive to $E_{\rm min}$.
For $E_{\rm min}=1$ GeV
and our predicted indices, the result at $E=E$[ankle]  is:
\begin{equation}
E^3\,{dF{\rm[EG]}\over dE}\Big|_{E{\rm [ankle]}} \sim 
10^{24}\; \rm eV^2\, m^{-2}\, s^{-1}\, sr^{-1}.
\label{EGatankle}
\end{equation}
An extrapolation from $E\sim E_{\rm min}$ ---where most of the CR 
flux and energy reside--- to $E=E$[ankle] would seem to be inordinately
sensitive to the adopted spectral indices. But the theory fits the data
over this large domain! The result of Eq.~(\ref{EGatankle}) is a gratifying 
number, as we proceed to discuss.

\subsection{Questions of presentation}

We shall see in the next Subsection that the result of Eq.~(\ref{EGatankle})
allows us to predict the shape and normalization of the UHECR  flux.
The prediction of the normalization has an uncertainty that reflects
the combined uncertainties of various inputs, such as 
the fraction of core-collapse SNe that generates GRBs (to which we 
dedicate Appendix \ref{association}), the error in the value
of the prior $E[{\rm jets}]$, the uncertainty in the distribution of $\gamma_0$
values and in the star formation rate 
$R_{\rm SF}(z)$ at $z\sim 1$ to 2, the redshifts dominantly contributing to 
the integral in Eq.~(\ref{AccEnergy}). The nominal error on each of these
quantities is a factor of 2 or more and hard to ascertain with precision. 
The combined error in the prediction is larger than that of the normalization of
the UHECR flux  (a statistical error of a factor of about 2, if we 
restrict ourselves to measurements made with a single technique, such as 
the fluorescence of the CR showers).

We could choose to present our prediction for the UHECR flux as a
wide band reflecting the uncertainty in the normalization. Alternatively,
we could use the CR data to constrain the priors to a multidimensional 
domain narrower
than the prior one. We opt for a third possibility: to choose a 
normalization ---within its predicted domain--- that compares well
with the UHECR observations, making the result for the spectrum `look better'.
We make the same choice elsewhere, e.g.~the CR abundances relative
to protons are correctly predicted within a factor of order 2, yet, 
we shall fix the overall normalization of the corresponding spectra
to compare well with the normalization of the observations. None of
the above affects the results for the shapes of the spectra.

In all of our results, a comprehensive best fit of all parameters and priors may 
make the comparison with data `look even better'. Such an effort would be
premature: the observations of the CR flux are still a fluid issue,
our detailed choices do not all indisputably follow from first
principles. 

We choose to present our results for the CR spectra in 
order of descending energy.

\subsection{The UHECR spectrum}
\label{UHECRtext}

\begin{figure}
\centering
\epsfig{file=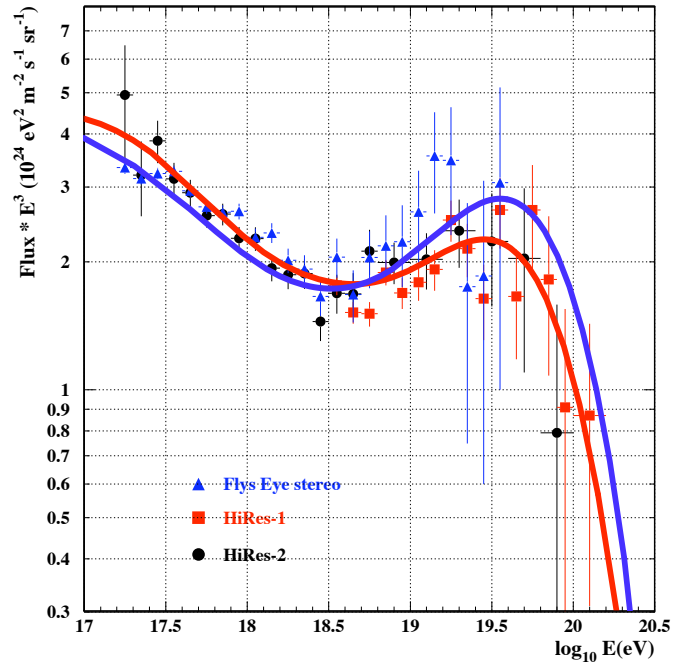, height=9.5cm,  width=9.5cm,angle=-0}
\caption{Predicted and observed \cite{HIRES1} UHECR spectrum. The 
vertical scale is 
$E^3\,dF/dE$. The normalization and shape of the spectrum above the
ankle do not involve any fit parameters, only choices of priors within
their predetermined domains.}
\label{UHECR}
\end{figure}

Our prediction for the UHECR all-particle spectrum is
shown in Fig.~\ref{UHECR}. At $E=E$[ankle] the extragalactic
contribution of Eq.~(\ref{EGatankle}) is about 1/2 of the observations
reported in the figure. Its normalization, at this energy or above it, is
approximate but `absolute', in the sense discussed in the previous subsection.
The shape of the flux above the ankle is entirely predicted; it is
the shape of the redshifted flux of Eq.~(\ref{z}).

The two curves of Fig.~\ref{UHECR}  correspond to the two choices of penetrability
 of extragalactic CRs to the Galaxy:
the blue curve rising higher uses Eq.~(\ref{conf-b}) with ${\beta_{\rm conf}}=0.55$, 
the red curve uses Eq.~(\ref{conf-b}) with ${\beta_{\rm conf}}=0.5$. The curves have 
slightly different central values and widths of the $\gamma_0$
distributions of Eq.~(\ref{gamdist}), both within the corresponding prior domains: 
$\bar \gamma_0=1200$ (1300), $w=0.4$ (0.5) for the blue (red) lines.
In the two curves in Fig.~\ref{UHECR} the shape of the high-energy end-point 
and the height of the hump reflect not only the GZK cutoff of Eq.~(\ref{GZK2}), 
but also the acceleration-cutoff energy for protons, which has been
properly smeared with the corresponding $\gamma_0$ distribution.

\subsection{The ankle region and the flux normalization}
\label{ankleregion}

Above the ankle, the CR flux is dominated by protons of extragalactic origin
belonging to the high-energy `inelastic' part of their source spectrum.
The overall normalization of this spectrum is the quantity $N_p$ illustrated
in Fig.~\ref{DDelinel}, whose approximate predicted value is implied by 
Eq.~(\ref{EGatankle}). The refinement of this prediction to agree with the 
data shown in Fig.~\ref{UHECR} narrows down the value of $N_p$ to better 
than a factor of 2.

To fit the flux of protons below the proton knee, we will have to fix our only
free parameter: the elastic-to-inelastic ratio $f$  of Fig.~\ref{DDelinel}.
In our theory, $f$ is species-independent and the relative CR
abundances are predicted. Hence, once $N_p$ and $f$ are fixed, the spectrum
of CRs of all nuclear species is fixed. In particular, the Fe flux is predicted.
The knee of the Fe flux dominates the all-particle spectrum just below the ankle.
At the ankle, its contribution to the total
flux of Fig.~\ref{UHECR} is about 50\%. So, the ankle is indeed the energy above
which the extragalactic flux takes over \cite{Cocconi}.

The ankle may be defined as the energy at which CR {\it protons} 
are no longer expected to be confined to the Galaxy, as in 
Eqs.~(\ref{LarmorR}, \ref{Larmor}). The ankle happens to occur at this energy, 
but it is not the end-point of a dominantly Galactic {\it proton} flux. 
It is, however, the starting point of a dominantly extragalactic proton flux.
This is not the only `ankle coincidence'. The shape of the CR flux, at the
ankle and just above it, is  partly due to the effect of $e^+ e^-$ 
production on the extragalactic proton flux, illustrated in Fig.~\ref{PhDisGZK}.
The energy at which this attenuating effect is maximal coincides with 
$E_{\rm ankle}(Z\!=\!1)$,
but has nothing to do with CR confinement in the Galaxy. In galaxies unlike ours 
these coincidences need not take place. This prediction may be particularly
difficult to test.

\subsection{The all-particle spectrum}
\label{all-part}

\begin{figure}
\vspace {0.5cm}
\begin{center}
\epsfig{file=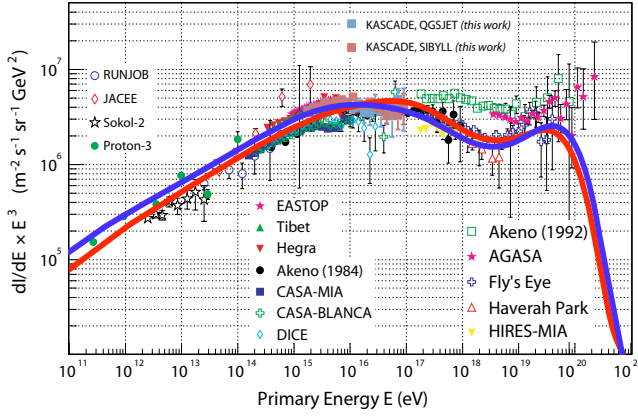, width=8.5cm,angle=-0}
\end{center}
\caption{Fitted and observed  all-particle CR spectrum. The vertical scale is 
$E^3\,dF/dE$. Some of the
UHECR data in this figure disagree with the ones in Fig.~\ref{UHECR}.
The color-coded lines
correspond to the same choices as in Section \ref{UHECRtext}
and Fig.~\ref{UHECR}.}
\label{AllPart}
\end{figure}

Our results for the all-particle spectrum are shown in  
Fig.~\ref{AllPart}. The normalization of this plot is fixed by the
parameter $f$ (fit to proton data at the knee), the combination of
priors $N_p$ (adjusted within its pre-established domain) and
the predicted relative abundances of the CR elements.
The shape of the theoretical curves is thereby fixed.
Naturally, their tilt and the sharpness of the ankle are sensitive
to the chosen value of ${\beta_{\rm conf}}$, which appears in the exponential of
an energy dependence that extends over many decades.
The colour-coded lines
correspond to the same choices as in Section \ref{UHECRtext}
and Fig.~\ref{UHECR}.

\subsection{The knee region}
\label{kneeregion}

\begin{figure}[]
\centering
\vskip .5cm
\vbox{\epsfig{file=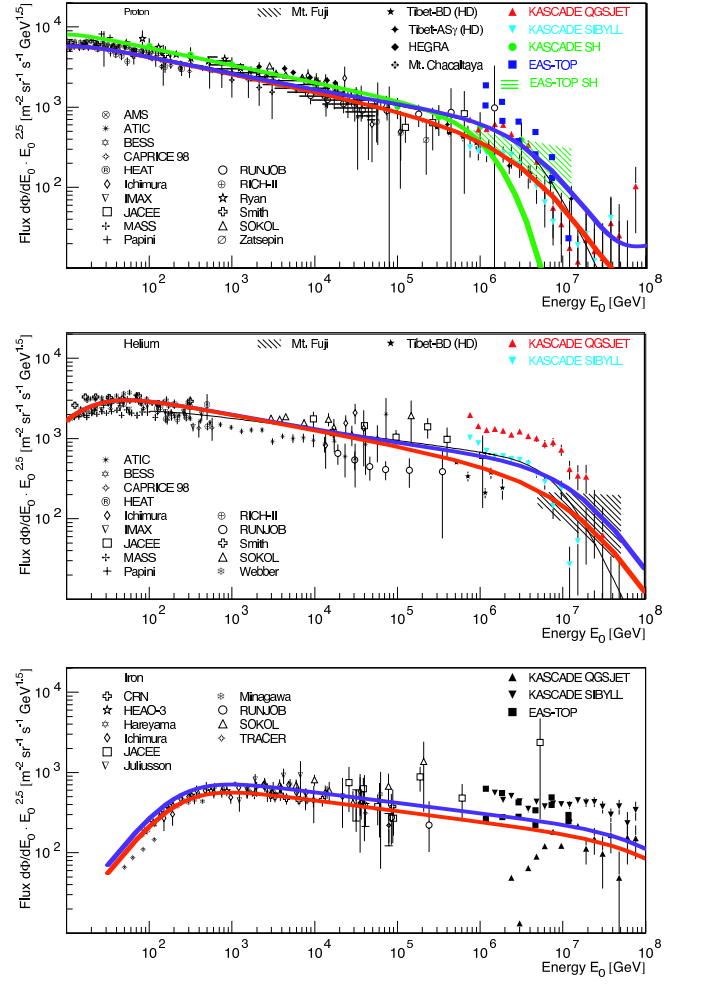,width=8.6cm}}
\caption{Measurements of individual-element CR spectra in
the `knee' region \cite{Hoer}. The vertical scales are $E^{2.5}\,dF/dE$.
Top: protons; Middle: $\alpha$ particles; Bottom: iron nuclei. The colour-coded lines
correspond to the same choices as in Section \ref{UHECRtext}
and Fig.~\ref{UHECR}.
The green line in the proton entry has $w=0.8$ for the width of the 
$\gamma_0$ distribution, as in Fig.~\ref{DDLogGamma}. 
The compilation of data was kindly provided to us by K.H. Kampert.}
\label{KASKADE}
\end{figure}

There are recent data from the KASKADE collaboration attempting
to disentangle the spectra of individual elements or groups in the
knee region. The data are preliminary in that
their dependence on the Monte Carlo programs used to simulate
hadronic showers is still unsatisfactorily large. Our predictions
for the spectra of  H, He and Fe are shown in Fig.~\ref{KASKADE}.
The red and blue lines
correspond to the same choices as in Section \ref{UHECRtext}
and Fig.~\ref{UHECR}.
The green line in the proton entry has $w=0.8$ for the width of the 
$\gamma_0$ distribution, as in Fig.~\ref{DDLogGamma}, the 
red and green lines correspond to distributions about $2\sigma$ and $3\sigma$
wider and, within the large systematic uncertainties of the data,
seem to be `better'.

At the highest energies, the blue line in the H figure curves up,
as the corresponding inelastic contribution begins to dominate.
Since the elastic and accelerated distributions are additive, the
theory predicts not only a knee ---at the point where the elastic
contribution is rapidly cut off--- but rather a `kneecap', ending
at the point at which the inelastic contribution takes over. This
is more clearly visible in Fig.~\ref{DDelinel}.

\subsection{The low-energy spectra}
\label{lowenergies}

\begin{figure}
\vspace {-0.5cm}
\begin{center}
\epsfig{file=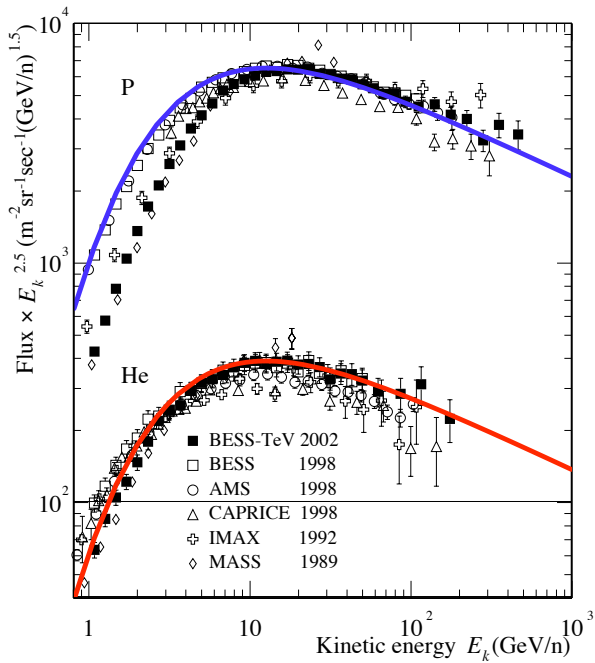, width=8.5cm,angle=-0}
\end{center}
\caption{The very low energy fluxes of protons and $\alpha$ particles
at various times in a solar cycle. The 1998 data are close to solar-minimum
time.}
\label{VeryLowEnergy}
\end{figure}

The lower the energy, the easier the sieving of CRs into individual
elements and their isotopes. In Fig.~\ref{VeryLowEnergy} we show
the weighted spectra $E_k^{2.5}\,dF/d\,E_k$
of protons and $\alpha$ particles, as functions of $E_k$, the 
kinetic energy per nucleon. The figure shows data taken at
various times in the 11-year solar cycle. The most intense
fluxes correspond to data taken close to a solar-minimum time.
The theoretical curves do not include an attempt to model the
effects of the solar wind. They should agree best with the
solar-minimum data, as they do, particularly for protons.

The theoretical spectra, dominated by the elastic contribution to
the CR spectrum, are given by Eq.~(\ref{NRFlux}). The data in
Fig.~\ref{VeryLowEnergy} are well below the elastic cutoff
at $\gamma_{_{\rm CR}}\simeq 2\,\gamma_0^2$, meaning that the result is 
independent of the chosen $\gamma_0$ distribution. Thus, the shape 
of the theoretical source spectra is, in this energy domain, parameter-free.
At the lowest energies shown in Fig.~\ref{VeryLowEnergy}, the differences
between the exact result of Eq.~(\ref{RFlux}) and its non-relativistic
approximation of Eq.~(\ref{NRFlux}) are at the 20\% level. At these
energies CRs are confined for very long times and their interactions 
with the ISM ---which we have not corrected for--- result in similar
corrections. Uncertainties of the same order are also introduced by our 
neglect of solar-neighbourhood effects. The results of Fig.~\ref{VeryLowEnergy}
may look better than they should.

The curves in Fig.~\ref{VeryLowEnergy} are sensitive to the chosen value
of the confinement exponent ${\beta_{\rm conf}}$ of Eq.~(\ref{cintrod}), which
governs the overall `tilt' of the curves. In this figure we have chosen
${\beta_{\rm conf}}=0.6$, reflecting a general tendency of the data to be better described
by slightly higher values of ${\beta_{\rm conf}}$ at low energies (recall that the 
results for the all-particle spectrum and the knee region
have either ${\beta_{\rm conf}}=0.5$ or ${\beta_{\rm conf}}=0.55$). 
We could have chosen to present
all results with an input ${\beta_{\rm conf}}$ reflecting the errors in this prior; see
Eq.~(\ref{cintrod}). To some extent this is purely a question of cosmetics
in the presentation; one reason is that the addition of best fits to
the data on individual elements differs from a best fit to the
all-particle spectrum!

\subsection{Rough measures of CR composition}

\begin{figure}
\vspace {0.5cm}
\hbox{\hskip 0.7cm\epsfig{file=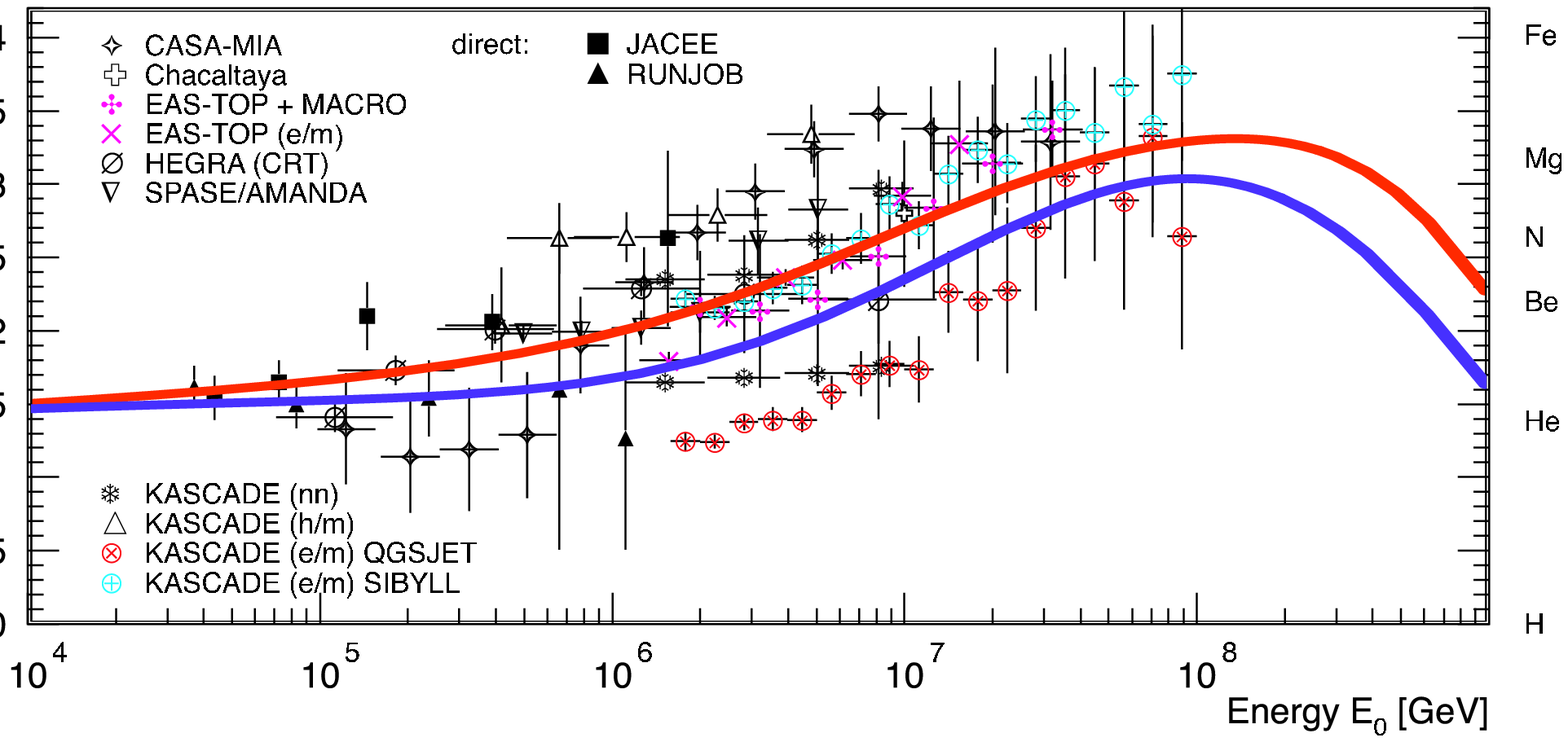, width=8cm,angle=-0}}
\caption{Mean logarithmic mass of high energy CRs. Data points were
         compiled  by Hoerandel \cite{Hoer} from
         experiments measuring electrons, muons, and hadrons at
         ground level. The colour-coded lines correspond to the same choices 
as in Section \ref{UHECRtext} and Fig.~\ref{UHECR}.
The compilation of data was kindly provided to us by K.H. Kampert.}
\label{lnAknee}
\end{figure}

\begin{figure}
\begin{center}
\epsfig{file=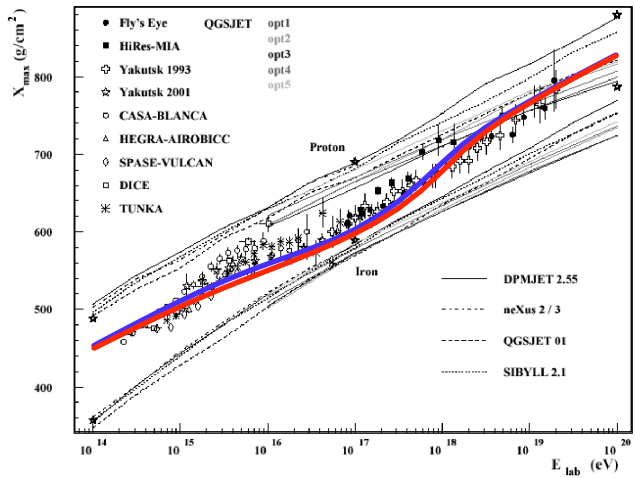, width=10.cm,angle=-0}
\end{center}
\vspace{-7cm}
\caption{The depth of shower maximum as a function of energy. The data are from a
compilation in Ref.~\cite{ZKP}. The color-coded lines
correspond to the same choices as in Section \ref{UHECRtext}
and Fig.~(\ref{UHECR}).}
\label{Xmax}
\end{figure}

The evolution of the CR composition as a function of energy is
often presented in terms of two quantities: the mean logarithmic
atomic weight $\langle \ln A \rangle$ and the depth into the
atmosphere of the `maximum' of the CR-generated particle shower,
$X_{\rm max}$. The predicted   $\langle \ln A(E) \rangle$ is
compared with relatively low-energy
data in Fig.~\ref{lnAknee}. The predicted $X_{\rm max}(E)$,
constructed with a simplified method described by Wijmans
\cite{Wigmans}, is shown in Fig.~\ref{Xmax}.

\begin{figure}
\vspace {0.5cm}
\begin{center}
\epsfig{file=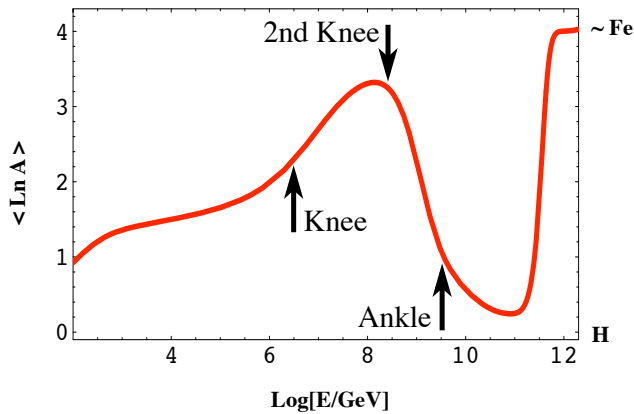, width=8.5cm,angle=-0}
\end{center}
\caption{The prediction for $\langle \ln A(E) \rangle$ at all energies.}
\label{lnA}
\end{figure}

The predicted $\langle \ln A(E) \rangle$ at all energies, 
shown in Fig.~\ref{lnA}, shows how at very high energies
the flux is once more Fe-dominated: lighter elements have
reached their acceleration and Galactic-escape cutoffs.
Naturally, this prediction is very sensitive to the assumed
details of Galactic escape and extragalactic photo-dissociation.

\subsection{The confinement time and volume}
\label{timevolume}

The adopted form of the confinement-time function $\tau_{\rm conf}(E)$
of Eq.~(\ref{c}) implies a constraint that may be re-expressed as a prediction
of the coefficient $K$ in Eq.~(\ref{c}). Similarly, 
the CB-model value of the Milky-Way's luminosity, Eq.~(\ref{SNEsupply}),
and its expression Eq.~(\ref{CRlum}) in terms of the confinement volume,
$V_{\rm CR}$, can be used to predict the value of the latter.

The largest coherent magnetic-field domains in the Galaxy have sizes 
of ${\cal{O}}(1)$ kpc  \cite{Beck}.
The light-travel time in such a domain must approximately correspond to the
confinement time for protons of energy $E=E(\rm ankle)$. With this constraint, 
using Eq.~(\ref{c}) for $Z=1$, we can
make a very rough estimate of the coefficient $K$ in the expression for 
$\tau_{\rm conf}$. The result is $K\sim 2\times 10^8\,{\rm y}$, one order of
magnitude larger than the value quoted in Eq.~(\ref{K}), which is also fairly 
uncertain \cite{DDGBR} and is known to be an underestimate \cite{Longair}.
In discussing the spectrum of CR electrons in Section \ref{electrons} we shall see that 
in our theory there is another way of estimating $K$, whose result is also
$K\sim 2\times 10^8 \,{\rm y}$.

Approximate Eq.~(\ref{CRlum}) with an assumed fairly uniform 
CR flux in the Galaxy, the expectation in our theory, thereby defining an
effective confinement volume:
\begin{equation}
V_{_{\rm CR}}\sim L_p\;
\left({4\pi \over c}\, \int {dE\over \tau_{\rm conf}}\,E\,{dF_p\over dE}\right)^{-1} .
\label{CRlum2}
 \end{equation} 
For the predicted luminosity of the Galaxy, Eq.~(\ref{SNEsupply}),
and the observed  proton luminosity, the result is
$V_{_{\rm CR}}\!\approx\! 1.6 \!\times\! 10^{69}$  cm$^3$. This is in 
agreement with the volume, $V_{_{\rm CR}}\!=\!(\pi)^{3/2}\, 
\rho_e^2\, h_e \!=\! 1.6\times 10^{69}\, {\rm cm^3}$,
of a Galactic CR halo of 35 kpc radius and 8 kpc 
height above the disk, inferred from our study of the GBR \cite{DDGBR,DDDGBR2},
summarized in Section \ref{theGBR}. 
This volume is consistent with our estimate of the confinement time 
of Galactic CRs.
The volume $V_{_{\rm CR}}\!\approx\! 6.6\!\times\! 10^{68}$ cm$^3$,
obtained by Strong et al.~\cite{Str4}
in an elaborate leaky box model of the Galaxy,
is smaller by a factor $\approx 2.5$ than our estimate, reflecting the 
shorter confinement time of CRs in leaky box 
models, and the higher value adopted in \cite{Str4} for the extragalactic GBR.

\section{Discussion}

We have presented a specific version of the theory, implying several concrete
assumptions and choices. In this Section we further discuss 
these choices, as well as the `robustness' of the
predictions, i.e.~their relative independence of our chosen inputs. The conclusion
is that all of the general properties of the results, listed in our Conclusions, are robust.

\subsection{The CB-model priors}

\subsubsection{The CR luminosity}

We have argued in Sections \ref{lumGal}, \ref{UHECRtext} and \ref{all-part} 
that our predicted CR luminosity agrees with observations. This means
that no other sources of non-solar CRs need be invoked.
Yet, the agreement of theory and observations is `within large errors'.
The question arises of whether or not other sources of CRs may be relevant.

We devote Appendix \ref{Accelerators} to the discussion of the relative
CR luminosity of other sources: pulsars, soft $\gamma$-ray repeaters,
neutron-star mergers, and micro-quasars. Our
conclusion is that their putative contribution is in every case negligible.
We dedicate Section \ref{GBRAGN} to the contibution of AGNs
to the GBR, which is very relevant, in our theory, to their putative
contribution to UHECRs, discussed in Sections \ref{CRAGN}
and \ref{AugerAGNcorr}.

The only remaining conventional candidate for a source of CRs is
the non-relativistic expanding shells of SNe. Literally thousands of
papers have been written on this subject. Many of them recognize
that the theory is not supported by observations of the flux of $\gamma$-rays 
that nuclear CRs impinging on the local ISM would generate
via $\pi^0$ production and subsequent decay, nor by the near-isotropy
of the CR flux at our location in the solar circle, external to the domain
where most massive stars (potential SNe) lie. One prominent
example of a discussion of these points
is the 1957 review by Philip Morrison \cite{Morrison}.
A much later and very incisive example is the commentary by Rainer Plaga
in \cite{Plaga}. Yet, having received so much attention in spite of
its well-known flaws and limitations, this conventional SNR theory of relatively
low-energy CRs is unlikely to be abandoned. We choose not to attempt
to discuss the subject in detail, except for the important questions of the
total CR luminosity, which has triggered controversy, and of recent results 
by the HESS and MAGIC collaborations, which have triggered great interest,
and are discussed in Section \ref{gammaray}.

The standard result of Eq.~(\ref{them}) is a factor $\sim\! 6$ smaller than our
Eq.~(\ref{SNEsupply}).  Dogiel, Sch\"onfelder and Strong \cite{DSS} criticized
our original result for the luminosity~\cite{DDLum}, which was somewhat larger
(we used at the time a rough estimate of $E{\rm[jets]}$ based on the natal
kicks of neutron stars). Their critique is phrased entirely within models in
which the CRs are produced in the central realms of the Galaxy
(interior to the solar circle) 
and diffuse  to the rest of the Galaxy and its halo. 
The result is inapplicable to the 
CB model, wherein CRs  are made much more
uniformly: over the entire trajectories of CBs, which extend all the way to
the halo and beyond, as discussed in Section \ref{trajectories}. 

In this paper we have adopted a 1 to 1 association between GRBs and
core-collapse SNe. As discussed in Appendix \ref{association} this is
subject to `cosmological' and CB-model uncertainties, both of ${\cal{O}}(2)$.
Within errors, it may be that a subclass of SNe, e.g.~those of Type Ib/c,
is responsible for the majority of long-duration GRBs. In that case,
our estimates of the CR luminosity may be correspondingly reduced.

\subsubsection{The deceleration of a CB and the index below the knee}

In deriving Eqs.~(\ref{NRmass},\ref{gammadown}) we have worked in the
approximation wherein the diffusive rate of escape of ISM particles from a CB 
is slower than their incoming rate. To study a large range of possibilities,
consider the opposite extreme, in which the escape is instantaneous
at all values of the CB's
LF $\gamma$. Let $a$ be the ratio between the average energy of a nucleus exiting
a CB in its rest system and the energy at which the nucleus entered, so that
$\langle \gamma_{\rm out}\rangle\equiv a \, \gamma$. For elastic scattering, $a=1$;
for nuclei phagocytized by the CB, $a=0$; and for those Fermi-accelerated within
the CB, $a\!>\! 1$. Let $\bar a$ be the mean value in the average over these
processes. Energy--momentum conservation implies
 a CB's deceleration law:
\begin{equation}
{d\gamma\over\beta^k\,\gamma^k}\simeq-{m_p\over M_0}\,{\gamma_0^{\bar a-1}}\; 
dn_{p}\; ,\;\;\;\;\;\;
k\equiv3-\bar a\, ,
\label{decelerate}
\end{equation}
and a CB's inertial mass evolving as:
\begin{equation}
M=M_0\; \left({\beta\,\gamma\over\gamma_0}\right)^{2-k}.
\label{mass}
\end{equation}

For $a>0$, Eqs.~(\ref{decelerate},\ref{mass}),  imply a slightly different deceleration
law than Eqs.~(\ref{NRmass},\ref{gammadown}), and a smaller $\beta_s$ than that
of Eq.~(\ref{beautyfulindex}). A related uncertainty is the one introduced by the
adopted form of $R_{\rm CB}(\gamma)$, which affects $\beta_s$ via
Eqs.~(\ref{tau1}), (\ref{dtCBdgamma}). We argue in detail in Appendix
\ref{ExpansionApp} that this is the right choice, 
and is supported by GRB X-ray AG data, but our 
theoretical arguments are admittedly over-simplifications of the extremely 
complex problem of the CB--ISM collisional process.
Yet another source of uncertainty in the prediction of $\beta_{\rm th}$ 
relates to the fact that cosmic MFs are in rough energy equipartition
with CRs, both in the Galaxy and in larger systems~\cite{DDMF}. The transfer
of as much as 50\% of their initial energy from CRs to the MFs they generate 
may affect the slope of the CR spectrum, if the transfer efficiency is not
energy-independent. In spite of all these caveats, since the errors in 
the uncertainty ${\beta_{\rm conf}}$ of 
Eq.~(\ref{cintrod}) are large, the prediction for the observed $\beta_{\rm th}$ of
Eq.~(\ref{betapred}) is still  quite satisfactory. 

The prediction of the spectral slope of Eq.~(\ref{beautyfulindex}) is  insensitive 
to all other details of its derivation. An example: we have assumed the re-emission 
of ISM particles in the CB's rest system to be isotropic. We may have assumed
the distribution to be that of scattering by a hard ball (modelling a CB's
highly magnetized surface), by a monopole
(modelling an electrically charged CB) or a dipole (modelling a CB's longest-range
MF component). These distributions do not affect $\beta_s$, though
they give slightly different shapes to the elastic flux close to the knee.

\subsubsection{Location of the Galactic CR sources}
\label{location}

We have adopted here a version of our theory wherein the rate at which
CRs exit a CB is much
slower than that at which they enter it, as in our recent study
of X-ray AGs \cite{DDDSwift}. In previous analyses of GRB AGs
\cite{AGoptical,AGradio}, as well as in our first results on CRs \cite{Florence},
we studied the  `fast' opposite limit: the ISM particles intercepted by a CB 
are instantaneously scattered. The results in the two limits, for
AGs and CRs, are very similar: we have no convincing way to opt for one
or the other limit. 

Let $\vec r$ be the vector position of a point in the Galaxy relative to
its centre. Let $n(\vec r)$ be the ISM density.
To sketch a point, consider the rough approximation wherein
SNe occur only close to the Galactic centre.
In the `fast' limit, the CR source is distributed in proportion to
$n(\vec r)$ times the density of CB trajectories, that is $n(\vec r)/r^2$.
In the extreme `slow' limit  \cite{DP}
 the sources of CRs are located at the points where CBs end their voyage;
 a prediction of their distribution would require a very detailed modelling
 of the distribution $n(\vec r)$ in the entire Galaxy.
 
 There is yet another source of uncertainty in the precise distribution of the
 Galactic CR sources. The narrow conical beams of CRs 
 produced by the decelerating CBs may propagate collectively, sweeping up the 
 MF in front of them until the energy density in the beam becomes smaller than that of the field. Thereafter the CRs would begin to diffuse in the ambient MF as individual particles. 
Such a mechanism may effectively remove the source further away
from the SNe firing the CBs, and contribute to the explanation of
the high isotropy of Galactic CRs at all energies.

\subsubsection{Fermi acceleration within a CB}

We have been very specific in choosing the Fermi-accelerated spectrum
of Eq.~(\ref{gammaA}). Its abrupt threshold $\Theta(\gamma_A-\gamma)$
may be substituted by a much smoother function describing how unlikely
it may be to `Fermi-decelerate' a fraction of the particles tossed around
by moving MFs. Once processed through the CB-deceleration
integrals in Eq.~(\ref{Inelastic}),
 no significant changes occur in the predicted CR flux, except that the
 `little knee' shown in Fig.~\ref{DDelinel} at $E\sim 10^8$ GeV becomes less
 pronounced (depending on the specific choices of this and other priors, one 
 may obtain smoother spectra, such as those in Fig.~\ref{Groups}, or move the 
 little knee to an energy at which it looks like a `rotula' before the steepening 
 at the knee). The results are even more insensitive to the abrupt Larmor cutoff
$\Theta(\gamma_{\rm max}-\gamma_A)$ of the assumed input spectrum.
The prediction of the UHECR flux would only be affected if the radius of a
CB, or the MF within it,
were smaller than the ones of Eqs.~(\ref{best}), or (\ref{B}),
by more than one order of magnitude.

In the calculations we have presented, we used 
$\beta_{\rm ac}=2.2$, as in Eq.~(\ref{Acc}). This affects the slope 
of the accelerated flux via Eq.~(\ref{Inelastic}). We have also worked
out the results for $\beta_{\rm ac}=2.2\pm 0.2$. For the upper value,
they `look even better' than the results we have presented. For the
lower value they are similar, for slightly different chosen values of 
the confinement exponent ${\beta_{\rm conf}}$ and the width of the $\gamma_0$
distribution. This relative insensitivity is good
news: for the mechanism accelerating particles within a CB, we
have relied on `first-principle' numerical analyses \cite{Fred}, but so far
their results are very limited in their study of the parameter space:
the electron to proton mass ratio is unrealistic, the LF values are
much smaller than $10^3$, the density contrast between the two
merging plasmas is low, radiative effects (which are important for
electrons) are neglected, and the merging plasmas are semi-infinite in
extent (modelled with a finite transverse size and periodic boundary
conditions).

On the positive side, the value of $\beta_{\rm ac}$ that we have
adopted, if common ---as we assume--- to nuclei and electrons, is
strongly supported by the observations of the ``prompt"
ICS-generated spectrum of GRBs \cite{DD} and of the 
SR-generated spectrum of their AGs, discussed in some detail
in Section \ref{Swift}.

\subsection{The non-CB-model priors}

\subsubsection{The relative abundances in the ISM}

\begin{figure}
\begin{center}
\epsfig{file=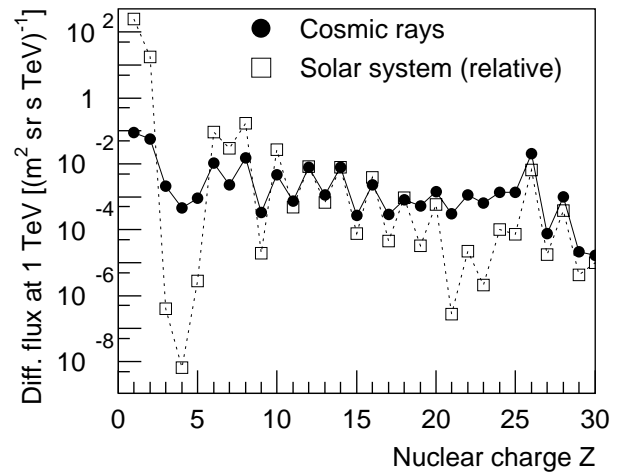, width=8cm,angle=0}
\end{center}
\vspace{-.5cm}
\caption{Composition of CRs with $Z\!\le\!28$ 
at 1~TeV~\cite{wiebel,pg}.}
 \label{f1standard}
\end{figure}

We contend that CBs accelerate the target ISM particles
to CR energies, mainly during
their voyage through the superbubble domains enclosing most SNe. 
As we discussed in Section \ref{relabundances}, the relative abundances
in these domains are poorly known, resulting in large errors in the CR
abundances predicted in Eq.~(\ref{composimple}). The errors
are not large enough to invalidate the comparison between different
mechanisms of CR acceleration, a simple task in the
analysis of the CR composition:

In the conventional theory of CR acceleration by the
non-relativistic ejecta of SNe,  the CR composition directly
reflects the relative abundances of the ISM, in this case the medium
surrounding the SN shell. The comparison
 in Fig.~\ref{f1standard} is often used to illustrate
how the abundance of secondary  CRs is
enhanced relative to their ambient (solar or interstellar) abundances 
and how the abundances of the primary CRs follow the pattern
of ambient abundances. The second conventional
claim is not quite correct: the figure
demonstrates that the statement is up to three orders of magnitude wrong
in the comparison of Fe to H. Contrarywise, for $A\!=\! 56$ in Eq.~(\ref{composimple}),
$X_{_{\rm CR}}\!=\! 1242\,X_{\rm amb}$. This explains the difference
between Figs.~\ref{f1} and \ref{f1standard} regarding the abundances of
primaries.

\subsubsection{Confinement in and penetration into the Galaxy}

In implementing the effects of Galactic MFs on the confinement of
CRs we have used the results of 
Eqs.~(\ref{galconf}), (\ref{ConfSpec1}), which are based on observational 
results for relatively low energies, and applied them up to the
Galactic escape energies of Eqs.~(\ref{Larmor},\ref{ConfSpec1}).
It is quite conceivable that the confinement exponent ${\beta_{\rm conf}}$ is
not a fixed number over all of this range, and we have commented
on how a ${\beta_{\rm conf}}$ that slightly decreases with energy improves the
presentation of the results. There is some theoretical
understanding of the value of ${\beta_{\rm conf}}$. A turbulent Kraichnan spectrum of 
magnetic inhomogeneities
yields ${\beta_{\rm conf}}=0.5$ \cite{Ptuskin}, compatible with the low-energy observations,
Eq.~(\ref{c}). But not enough is known about the MF of the Galaxy to
opt for a particular spectrum of inhomogeneities, or a transition energy
from one case to another.

Fits to CR data employing our fixed source spectra and a value of
${\beta_{\rm conf}}$ that slowly diminishes with energy provide very satisfactory
results. Moreover, they increase the domain of acceptable possibilities for
the description of Galactic penetrability by extragalactic CRs, which
we have narrowed down in this paper to the choices of 
Eqs.~(\ref{conf-a},\ref{conf-b}). We do not report the various possibilities
we have explored, given our ignorance of the
detailed properties of Galactic fields and winds.

\subsubsection{Interactions with matter and radiation}

The interactions of CRs with radiation and the ISM, listed in Section
\ref{tribulations} and discussed in various Appendices, are well understood.
The exception is the photo-dissociation of extragalactic nuclei which,
though well understood in principle, has not been studied in full
detail. We have estimated it in a manner that should be sufficient for our current
purposes.
The main impact of a careful study would be to gain confidence in
the predictions of our theory (or any other sufficiently specific theory of
UHECR sources) concerning the relative abundances at very high
energy.

\section{Cosmic-Ray electrons}
\label{electrons}

In this Section and the next, we give a simplified description of
CREs and the GBR in the context of the CB model. More
details can be found in Refs. \cite{DDGBR} and \cite{DDDGBR2}. 

Electrons and nuclei are accelerated by the `magnetic-racket'
CBs in the same manner. The functional form of their source spectra is
therefore the same, approximately $dF_s/d\gamma\propto\! \gamma^{-2.17}$, 
in the range $10\!<\!\gamma\!<\!10^6$; see Eqs.~(\ref{powerlaw}, \ref{beautyfulindex}). 
Electrons lose energy much more
efficiently than protons in their interactions with radiation, MFs and the ISM.
Moreover, the Larmor radii of nuclei and electrons ($\propto\! m/Z$) are
enormously different. We have no way to relate the normalization of the
CREs to that of CR nuclei.

The energy-loss rates  $b\equiv -dE/dt\propto E^{\alpha}$ of the various 
mechanisms via which electrons lose energy have different energy dependences. 
For Coulomb losses $\alpha=0$; for bremsstrahlung $\alpha=1$; for ICS
and SR losses at the relevant energies, $\alpha=2$. The time evolution
of an assumed approximately
uniform flux of electrons, $dF_e/dE$, with a source density 
$dF_s/dE$, is governed by \cite{GandS}:
\begin{equation} 
{d\over dt}{dF_e\over dE} =
 {d \over dE} \left[ {dF_e\over dE}\, {\Sigma_i} \,b_i \right] 
 -{1\over \tau_{\rm conf}^e}\,{dF_e\over dE}
+ R\,{dF_s\over dE}
\label{CRe}
\end{equation}
where the term involving $\tau_{\rm conf}^e(E)$ represents the `energy loss'
by diffusion in the Galaxy's MF and $R$ in the CRE injection rate. We assume 
$\tau_{\rm conf}^e$ to have the same energy dependence as the corresponding
one for nuclei, that is:
\begin{eqnarray}
\tau_{\rm conf}^e&=&K_e\,\left({{\rm GeV}\over E}\right)^{\beta_{\rm conf}},
\nonumber\\
{\beta_{\rm conf}}&=&0.6\pm 0.1,
\label{electronconf}
\end{eqnarray}
where we have used Eqs.~(\ref{galconf}, \ref{c}). The term involving
 $\tau_{\rm conf}^e$ in Eq.~(\ref{CRe}) can be
formally treated as an additional energy loss with a rate:
\begin{equation}
b_{\rm conf}\propto E^{1+{\beta_{\rm conf}}}\approx E^{1.6}.
\label{bconf}
\end{equation}

Let $U$ be the energy density of real and virtual photons permeating
the medium through which CREs move. In the solar neighbourhood,
starlight and the CBR have similar energy densities: 
$U_\star\!\approx\! 0.26$ eV cm$^{-3}$, and $U_0\!\approx\! 0.24$ eV 
cm$^{-3}$, at the current CBR temperature of ${T_0\!=\!2.728}$ K \cite{MandF}.
The energy density in the virtual (MF) photons is
$U_{_{\rm B}}=B^2/(8\,\pi)\approx 0.62$ eV cm$^{-3}$, for $B\!\sim\! 
5\,\mu$G.
Electrons lose energy
by ICS on real photons and by SR in the field of virtual
ones,  basically the same processes. The radiative energy-loss rate is:
\begin{equation}
b_\gamma ( E ) = {4\over 3}\,c\, \sigma_{_{\rm T}} \,E^2\,  \Sigma_j\,U_j,
\label{Radloss}
\end{equation}
where $\sigma_{_{\rm T}}\!\approx\! 0.65\!\times\! 10^{-24}\, {\rm cm^{-2}}$ is  
Thomson's cross-section.

\begin{figure}[]
\centering
 \epsfig{file=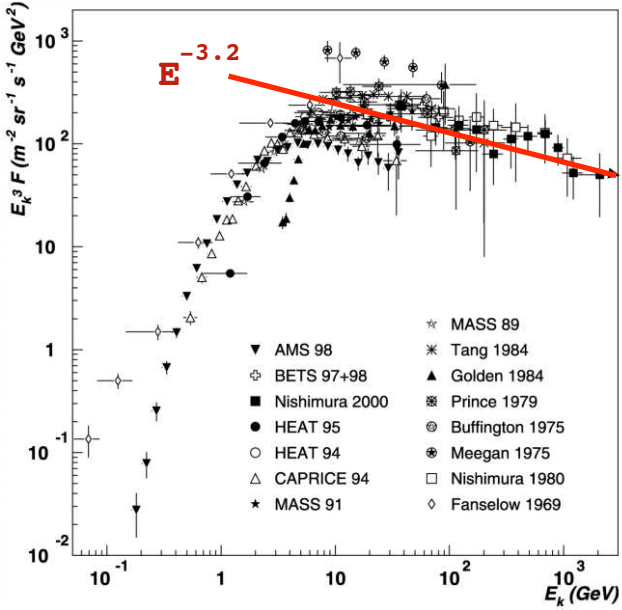,width=8.4cm}
\caption{The CRE spectrum, summarized in  \cite{Sapinski}.
The line is the central result of a power-fit to the higher-energy data; 
its slope is $3.2\pm0.1$ \cite{Sreek}.}
 \label{CREspectrum}
\end{figure}

At sufficiently high energy, the radiative energy loss of Eq.~(\ref{Radloss})
must dominate the others, since it has the fastest growth with energy.
In this domain, and in a steady-state situation, the solution to Eq.~(\ref{CRe}) 
for an input $dF_s/dE\propto E^{-\beta_s}$ is:
\begin{equation}
{dF_e\over dE}\propto E^{-\beta_e};\;\;\;\;\;\;\; \beta_e=\beta_s+1\approx 3.17,
\label{espectr}
\end{equation}
where we have used the predicted $\beta_s$ of Eq.~(\ref{beautyfulindex}).
This result is in  agreement with the observed slope of
the CRE spectrum,  see Fig.~\ref{CREspectrum}.
The best-fitted value above $E\!\sim\!6$ GeV is
$\beta_{\rm obs}=3.2\pm 0.10$, and the
fit is excellent if all the experiments are recalibrated to yield the same 
flux at high energy \cite{AMS, DDGBR}.
The radiative loss rate of Eq.~(\ref{Radloss}) corresponds to a 
cooling time:
\begin{equation}
\tau_\gamma\equiv{E\over b_\gamma}\approx 
(2.85\times 10^8\,{\rm y})\;\left[{E\over {\rm GeV}}\right]^{-1}.
\label{Radcool}
\end{equation}

Close inspection of Fig.~\ref{CREspectrum} results in the conclusion that,
for the most precise observations (the AMS experiment, \cite{AMS}),
and at $E_{_{\rm eq}}\!\sim\! 2.5$ GeV,
the data are a factor $\sim\! 2$ below the extrapolation to low energies
of the predicted or best-fitted higher-energy behaviour of Eq.~(\ref{espectr}).
The `diffusive' energy-loss rate of Eq.~(\ref{bconf}) has the fastest 
growth with energy, after the radiative loss of Eq.~(\ref{Radloss}).
If we interpret $E_{_{\rm eq}}$ as the energy at which the corresponding
characteristic times are equal, that is, if we equate 
$\tau_{\rm conf}^e$ in Eq.~(\ref{electronconf}) to 
$\tau_\gamma$ in Eq.~(\ref{Radcool}) at $E=E_{_{\rm eq}}$, we obtain: 
\begin{equation}
K_e\sim 2\times 10^8 \, \rm y,
\label{Ke}
\end{equation}
as an estimate of the CR confinement time at $E \sim\! 1$ GeV.

\section{The GBR}
\label{theGBR}

\begin{figure}[]
\centering
 \epsfig{file=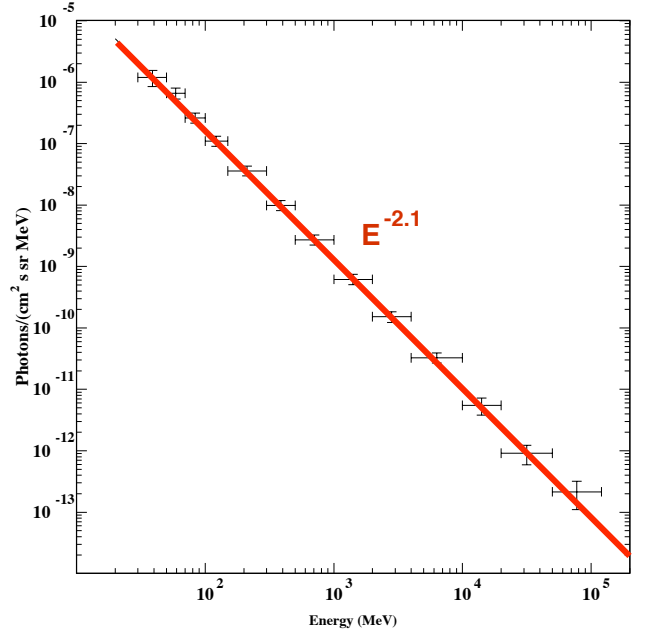,width=9.2cm}
\caption{The GBR spectrum, measured by EGRET \cite{Sreek}.
The line is the central result of a power-law fit of
slope  $2.10\pm0.03$.}
 \label{GBRspectrum}
\end{figure}

The existence
of an isotropic, diffuse gamma background radiation (GBR)
was first suggested by
data from the SAS 2 satellite \cite{TF}.
The EGRET instrument on the Compton Gamma Ray Observatory
confirmed this finding \cite{Sreek}.  
We call ``the GBR'' the diffuse emission observed by EGRET
by masking the galactic plane at latitudes
$\rm{|b|\le 10^o}$, as well as the galactic centre
at $\rm{|b|\le 30^o}$ for longitudes $\rm{|l|\le 40^o}$,
and by extrapolating to zero column density, to eliminate the $\pi^0$
and bremsstrahlung contributions to the observed radiation and
to tame the model-dependence of the results.
Outside this `mask', the GBR flux integrated over all directions in the 
observed energy range of ${\rm 30}$ MeV to ${\rm  120~GeV}$, shown in
Fig.~\ref{GBRspectrum}, is well described by a power law 
$dF_\gamma/dE\!\propto\! E^{-2.10\pm 0.03}$ \cite{Sreek}.

The EGRET GBR data show a significant deviation from isotropy, correlated with the
structure of the Galaxy and our position relative to its centre \cite{Anton,DDDGBR2}.
Contrariwise, the GBR's spectral index is uncannily directionally uniform.
These facts suggest a GBR which is partially local, as opposed
to dominantly cosmological, and a common origin for the
Galactic and extragalactic contributions.

In \cite{DDGBR} we have analized the directional and spectral properties of the
EGRET data and concluded that:
\begin{itemize}
\item{} The GBR is produced by ICS of CREs on starlight and the CBR.
\item{} The GBR has comparable contributions from CREs in a Galactic halo
of dimensions akin to the CR confinement volume discussed in Section \ref{timevolume}
(a directional and local source) and from other galaxies 
(an isotropic cosmological component).
\item{} The contribution of active galactic nuclei (AGNs) to the GBR
is at most comparable to that of the ensemble of external galaxies.
\end{itemize}

The first item implies that the GBR is a CR `secondary': it behoves us to
include it in a detailed discussion of CRs. To the third item, we
dedicate Section \ref{GBRAGN} below.
The GBR's spectral index, which we
discuss next, is the same for the local and the cosmological contributions.

\subsection{The GBR index}
\label{GBRindex}

Consider the ICS of high energy electrons on starlight, of typical
energy $\epsilon_\star\!\sim\! 1$ eV, and on the CBR,
whose mean current energyy is 
$\epsilon_0\!\approx\! 2.7\,k\,T_0\!\approx\! 0.64$ meV.
The mean energy ${E_\gamma}$ of the Compton upscattered photons is:
\begin{equation}
{E_\gamma(\epsilon_i) \approx  {4\over 3}\,\left({E_e\over m_e\,
c^2}\right)^2\,\epsilon_i} \, ,
\label{loss}
\end{equation}
with $\epsilon_i\!=\!\epsilon_\star$ or $\epsilon_0$.

The ICS photon spectrum originating in our galaxy is the sum of CBR
and SL contributions.
The ICS final-photon spectrum --a cumbersome
convolution \cite{Felten} of a CR power spectrum with a
photon thermal distribution-- can be approximated very simply.
Using again the index ``i'' to label the CBR and SL fluxes:
\begin{eqnarray}
&&{dF^i_\gamma\over dE_\gamma}
\propto{dE^i_e\over dE_\gamma}~
\left[ {dF_e\over dE_e}
\right]_{E_e=
{E}_e^i}\; ,\nonumber\\
&&{E}_e^i\equiv m_ec^2
\sqrt{{3\, E_\gamma\over 4\,\epsilon_i}}\; ,
\label{ICSCBR}
\end{eqnarray}
where ${E_e^i}$ is obtained from Eqs.~(\ref{loss}) by inverting
${E_\gamma(\epsilon_i)}$.
Introducing the CR-electron flux of 
Eq.~(\ref{espectr}) into Eqs.~(\ref{ICSCBR}), we obtain:
\begin{eqnarray}
&&{dF^i_\gamma\over dE}\propto E^{-\beta_\gamma},\nonumber\\
&&\beta_\gamma={\beta_e-1\over 2}\simeq 2.08.
\label{ICSphotpred}
\end{eqnarray}
The predicted index agrees with the measured one, 
${2.10\pm 0.03}$ \cite{Sreek}. 
Given Eq.~(\ref{loss}), CREs of energy $E_{\rm CBR}\!\geq\! 96$ GeV 
produce the GBR above 30 MeV by ICS of the 
current ($z\!=\! 0$) CBR; CREs with energy $E_\star\!\geq\! 2.4$ GeV
suffice for ICS on starlight. 
For electrons of energy $E_i$, the radiation cooling times are 
$\tau_{\rm rad}(i)\! =\! 3\, m_e^2\, c^3/(4\,  \sigma_{_{\rm T}} \, E_i\,  U_i)$,
so that locally $\tau_{\rm rad}(\star)\!\sim\!6\times10^8$ y, and
globally $\tau_{\rm rad}({\rm CBR})\!\sim\!1.3\times10^7/(1+z)^4$ y.
These numbers are much shorter than a Hubble time.  
ICS of CBR photons dominates the production
of the extragalactic GBR, as we argue next.

\subsection{The GBR flux and its directional-dependence}
\label{GBRflux}

Adopt the standard cosmology and the usual notation
($H_0,\,\Omega,\, \Omega_M,\,\Omega_\Lambda$) for its parameters,
specified in Appendix \ref{Redshift}. 
For a  Galactic magnetic field $B\!\sim\!3\,\mu$G,
$U_{_{B}}\!\sim\!U_0$; synchotron cooling and emission
are locally relevant \cite{DDGBR, DDDGBR2}. 
In our model, CBs transfer their kinetic
energy to CRs all along their trajectories, which extend from the SN-rich
inner galaxies to their halos and beyond. In galactic halos and  
galaxy clusters, $B\!<\!3\, \mu$G, and in the IGM, $B\sim 50\,$nG 
\cite{DDMF}.  In both places starlight is irrelevant, and
ICS of the CBR, whose energy density increases with $z$ like $(1+z)^4$,
dominates over synchrotron losses on the MFs. Thus, we 
calculate the intensity of the extragalactic GBR from the conclusion that 
the kinetic energy of CREs in the Universe 
with a lifetime shorter than the Hubble time 
has been converted by ICS of the CBR to 
$\gamma$-rays with the predicted spectrum of 
Eq.~(\ref{ICSphotpred}).
  
The main accelerators of high-energy CREs are SNe and the AGNs
to be separately discussed in the next subsection.
Other putative sources are
negligible, as discussed in Appendix \ref{Accelerators} for
CRs in general.

The SN rate, $R_{_{\rm SN}}(z)$, is proportional to the star-formation rate,
$R_{_{\rm SF}}(z)$, discussed in
detail in Appendix \ref{star}.
Let $f_{_{\rm SN}}$ be the fraction of the luminosity in CREs out of the total 
luminosity $L_{_{\rm CR}}$ in CRs generated by SNe. 
The CB model does not currently imply a prediction for
 $f_{_{\rm SN}}$, which we assume to be equal
to the ratio of the Milky Way's luminosity in CREs  
to its total luminosity in CRs, i.e.~$f_{_{\rm SN}}\!\sim\! 1/40$.

Given our inferred 100\% ICS conversion of CRE energy to photon energy, 
the GBR spectrum satisfies  \cite{DDDGBR2}:
\begin{equation}
\!\!\int {dF_\gamma\over dE}\,E\,dE\!\approx\!
{c\, f_{_{\rm SN}}\,L_{\rm CR}{\rm [MW]}\over 4 \pi\,H_0\, R_{_{\rm SF}}(0)}
\int\! {dz\,(1+z)^{-\beta_s}\, 
R_{_{\rm SF}}(z)\over \sqrt{\Omega_M\, (1+z)^3+\Omega_\Lambda}},
\label{SNcont}
\end{equation}
where $L_{\rm CR}{\rm [MW]}$ is specified in Eq.~(\ref{SNEsupply}).
Thus normalized, the spectrum of the contribution to the
GBR from extragalactic SNe is estimated to be:
\begin{equation}
 {dF_\gamma\over dE}\simeq 0.9\times 10^{-3} 
 \left[{E\over {\rm MeV}} \right]^{-2.08}
{1\over \rm cm^{2}~s~sr~MeV}\; .
\label{SNgam}
\end{equation}

The GBR contains a considerable Galactic foreground due to ICS of CBR, 
starlight and sunlight photons by Galactic CREs. The 
convolution of a CRE power-law spectrum with a photon thermal 
distribution \cite{Felten} can be approximated very simply \cite{DDGBR,DDDGBR2}. 
Using the index $i$ to label the CBR, starlight and sunlight fluxes, we 
have:
\begin{equation}
{dF_\gamma\over dE_\gamma}
\simeq N_i(b,l)~\sigma_{_{\rm T}}\,
 {dF_\gamma^i\over dE_\gamma},
\label{ICSGal}
\end{equation}
where $N_i(b,l)$ is the column density of the radiation field 
weighted by the distribution of CREs in the direction $(b,\, l)$, 
and the $dF_\gamma^i/dE_\gamma$ are as in  Eq.~(\ref{ICSCBR}). 
The distribution of the
non-solar starlight is approximated as $\propto 1/r^2$, with $r$ the
distance to the Galactic centre, and the CREs are assumed to be
distributed as a Gaussian  ``CR halo" \cite{DDGBR}.
Naturally, the results depend crucially on the size and shape
of this halo. We use a Gaussian distribution with a scale length of
$\rho_e\!=\! 35$ kpc in the Galactic disk and a scale height of
$h_e\!=\!8$ kpc perpendicular to the disk \cite{DDDGBR2}.
The justification for this choice of $h_e$ is the following.
The radio emission of ``edge-on" galaxies  --interpreted as synchrotron 
radiation by electrons on their magnetic fields-- offers direct 
observational evidence for CREs well above galactic disks, see,
e.g.~Ref. \cite{Duric1998}. For the particularly well observed case of NGC 
5755, the exponential scale height of the synchrotron radiation is ${\cal{O}} 
(4)$ kpc. If their energies are in equipartition, CRs and MFs
should have similar distributions, and the Gaussian scale height,
${ h_e}$, of the electrons ought to be roughly twice that of the 
synchrotron intensity, which reflects the convolution of the electron- and 
magnetic-field distributions. The inferred value is ${h_e}\sim 8$ kpc.

Due to Feynman scaling, the GBR from $\pi^0$ production and decay in 
hadronic CR collisions in the ISM and IGM has the same power-law index as 
that of CRs \cite{Dar1995}, i.e.~$-2.77$ in the ISM of galaxies and 
$-2.17$ in the IGM inside and outside galaxy clusters. This 
contribution to the extragalactic GBR is much smaller than that of CREs.

\begin{figure}
\begin{center}
\epsfig{file=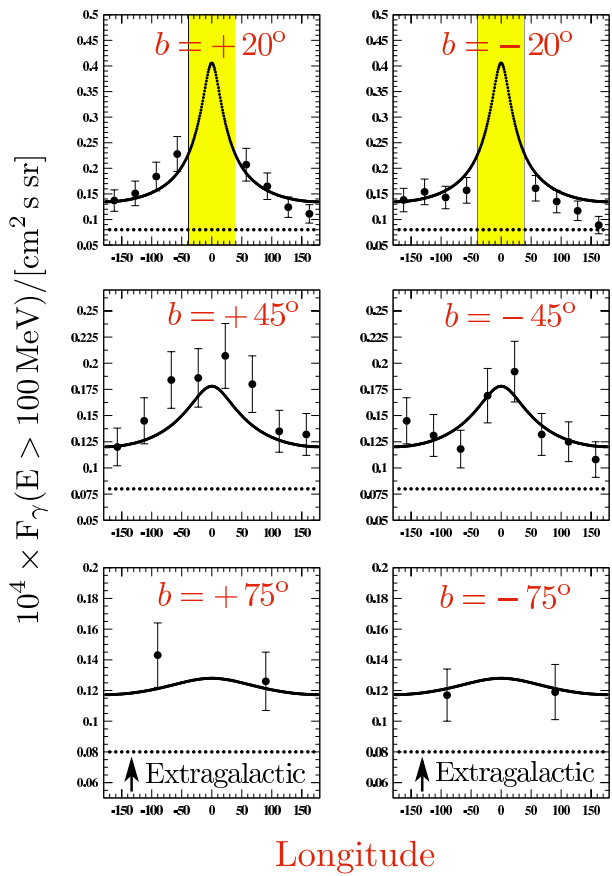,width=8cm}
\caption{The flux of GBR photons above 100 MeV: comparison
between EGRET data and our model
for ${h_e\!=\!8}$ kpc, ${\rm \rho_e\!=35\!}$ kpc, as functions
of longitude $l$ at  fixed latitudes $b$. The shaded domain
is EGRET's mask. 
Notice that the vertical scales do not start at zero. }
\vspace*{-0.5cm}
\label{GBRfits}
\end{center}
\end{figure}

In Fig.~\ref{GBRfits} we compare the observed GBR with our predictions, as functions 
of Galactic coordinates. The prediction is a sum of a $(b,\, l)$-dependent 
Galactic foreground produced by ICS of the CBR, starlight and sunlight, 
and a uniform extragalactic GBR. The result has $\chi^2/{\rm dof}\!=\!0.85$, a 
vast improvement over the constant GBR fit by EGRET, for which 
$\chi^2/{\rm dof}\!=\!2.6$. The ratios of $l$-integrated extragalactic to galactic fluxes 
are $\sim\!0.5$, 0.9, 1.5, for $\vert b\vert=20^{\rm o}$, $45^{\rm o}$, $75^{\rm o}$. 
The `foreground' component of the $\gamma$ `background' 
is $\sim\!50\%$ of the total radiation.

We conclude that the GBR can be explained by 
standard physics, namely, ICS of CBR and starlight by CREs from 
core-collapse SNe. At $E_\gamma\! >\! 100$ 
GeV, most of the extragalactic GBR is absorbed by pair production 
on the Cosmic Microwave Background (CMB) \cite{Salamon} and the diffuse 
GBR reduces to the Galactic foreground. This suppression, and a
decisive determination of the angular dependence in Fig.~2, should be 
observable by GLAST.

\subsection{The contribution of AGNs to the GBR}
\label{GBRAGN}

Active galactic nuclei, powered by mass accretion onto
massive black holes, eject powerful relativistic jets,
mainly during their $\sim 10^8$ y bright phase around $z\!=\! 2.5$. The
CBs of these jets 
should generate CR electrons and nuclei in precisely the
same way as the CBs from SNe do. The kinetic power of AGN jets has 
been estimated from their radio lobes, assuming
equipartition between CR energy and MF energy, and a ratio of
CR electrons to CR nuclei similar to that observed in our Galaxy. 
Kronberg et al.~\cite{Kronberg}, for instance, have estimated that `giant'
extragalactic radio sources, powered by accretion onto massive black holes
($M>10^8\, M_\odot$), inject $10^{61}$ to $10^{62}$ erg into radio lobes.
Such powerful CR sources may contribute to the observed spectrum
of CRs. The most reliable estimate of the contribution to the CR flux,
discussed in the next Section, may be the estimate
that parallels the study of the relative contribution
of external galaxies and AGNs to the GBR, which we
discuss next \cite{DDDGBR2}.

In search of an upper 
bound, we assume that the kinetic energy release in relativistic jets is 
the maximal energy release from mass accretion onto a Kerr black hole 
($\approx\! 42\%$ of its mass), and that this energy is equipartitioned 
between magnetic fields and cosmic rays with a fraction 
$f_{_{\rm AGN}}$  of the CR 
energy carried by electrons. These CREs also cool rapidly by ICS of the 
CBR. The energy of CREs whose radiative cooling rate, $\tau_{\rm rad}(z)$, 
is larger than the cosmic expansion rate, $H(z)$, is converted to 
$\gamma$-rays. Their energy is redshifted by $1+z$ by the cosmic 
expansion. Using a black hole density, 
$\rho_{_{\rm BH}}(z=0)\!\sim\!2\times 10^5\, M_\odot\,{\rm Mpc^{-3}}$ 
in the current Universe \cite{Tremaine}, and the CB-model injection 
spectral index, $\beta_s=13/6$ of Eq.~(\ref{beautyfulindex}), 
we estimate a contribution from AGNs to the 
extragalactic GBR flux:

\begin{equation}
{dF_\gamma\over dE} <
{2.4\times 10^{-3}\, c\, f_{_{\rm AGN}}\, \rho_{_{\rm BH}}\, c^2\over 4\,\pi\, 
 {\rm MeV}} 
\left[{E\over {\rm MeV}} \right]^{-2.08}\, ,
\label{AGNs1gam}
\end{equation} 
where, again looking for a bound, we have neglected the
cosmological redshift of the GBR energy.
With the above priors, Eq.~(\ref{AGNs1gam}) corresponds to a spectrum:
\begin{equation}
{dF_\gamma\over dE}\simeq 
4.0\times 10^{-4} 
\left[{E\over {\rm MeV}} \right]^{-2.08}~
{f_{_{\rm AGN}}/f_{_{\rm SN}}\over \rm cm^{2}~s~sr~MeV}\, .
\label{AGNs2gam}
\end{equation} 

For an assumed $f_{_{\rm AGN}}\!\sim\!f_{_{\rm SN}}$, Eq.~(\ref{AGNs2gam})
bounds the contribution of AGNs to $<\! 44$\% of that of extragalactic
SNe, Eq.~(\ref{SNgam}). It is clear from the derivation of this result
that it is at best `an estimate of an upper bound'. On the other hand,
the successful study of the GBR in \cite{DDDGBR2} ---wherein the AGN
contribution is neglected--- implies that the
bound cannot be significantly violated.

In \cite{DDDGBR2} we discussed the contribution of various established 
point sources  to the GBR, concluding
that they cannot explain its origin. 
The fact that all the AGNs detected by EGRET 
are blazars led various authors to suggest that these objects are the
main sources of the GBR \cite{Bignami,Chiang}. 
Subsequent studies have shown that at
most 25\% of the GBR can result from unresolved blazars \cite{Mukherjee},
i.e.~AGNs pointing close to our direction.
The contribution from
the much more abundant AGNs not pointing to us, is much more important
\cite{DDDGBR2}.

\section{The contribution of AGNs to UHECRs}
\label{CRAGN}

A comparison between the contributions of AGNs and extragalactic SNe
to the UHECR flux can be made along the same lines as in Section 
\ref{GBRAGN}. Assuming that the source spectra of these contributions
have the same energy dependence, the end result is the same: the AGN 
flux cannot be much bigger than $\sim\!44$\% of the extragalactic
core-collapse SN flux. This is 
because the tribulations (discussed in Section \ref{GBRAGN})
that CRs of either source suffer on their way are the same. In particular,
because of the effect of the GZK cutoff, UHECRs reach us from a
look-back time (or redshift) much smaller than that of the extragalactic
contribution to the GBR, but that `finite-volume' effect  drops from the
ratio of the fluxes, in the
approximation in which the matter of the universe is uniformly distributed.

\section{Recent Data}
\label{recent}

We call `recent' the observations that have been published after the
first posting of the current paper in June 2006. On the realm of GRBs,
on which our theory of CRs rests, enormous progress has been made 
in the observations, particularly in the X-ray domain, with the help
of the Swift satellite \cite{Swift}, amongst others. Concerning the 
CR domain, the novelties include data from HIRES \cite{HIRESGZK}
on the GZK cutoff,
from the Tibet AS$\gamma$ collaboration on CRs 
in the knee region \cite{Amenomori,TibetpHe}, from the Pierre Auger collaboration 
on UHECRs \cite{Mello,Unger,PAC}, and from the HESS and MAGIC collaborations on
$\gamma$-ray astronomy \cite{Ribo}. We comment on these 
data in view of our original results.

\subsection{GRBs in the ``Swift era"}
\label{Swift}

Various satellites are currently contributing 
to a wealth of new data on GRBs and XRFs. Swift, a true technological
jewel, is one of them.
With nominal celerity, Swift has filled a gap in GRB data: the very `prompt'
X-ray and optical radiations, and very precise measurements of X-ray
afterglows at later times.
The recent data fully corroborate the CB model of GRBs and XRFs.

\begin{figure}[]
\vskip -.4cm
\centering
\vbox{
 \epsfig{file=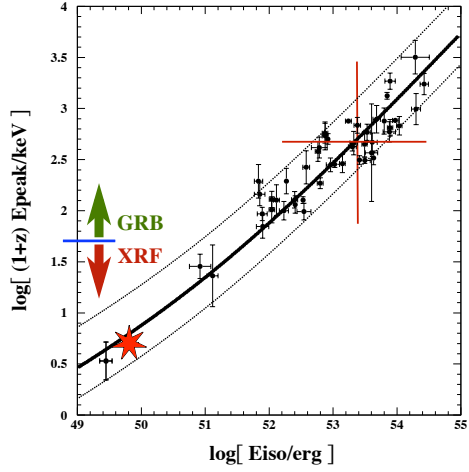,width=6.8cm}
}
\vbox{
 \epsfig{file=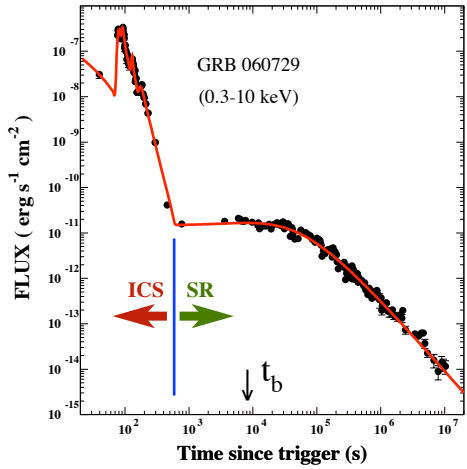,width=6.8cm}
}
\vbox{
 \epsfig{file=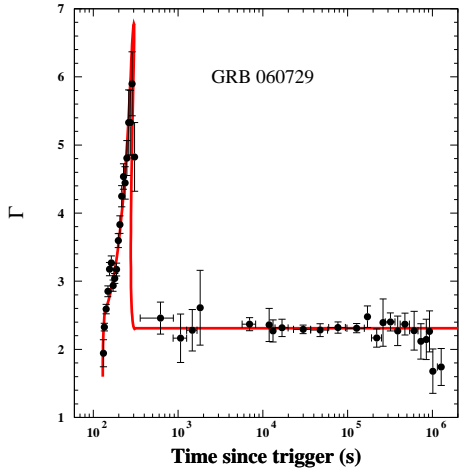,width=6.8cm}
}
 \caption{Top (a). The correlation between $(1+z)\,E_p$ and $E_{\rm iso}$.
 The ``official" distinction in terms of $E_p$ between GRBs and XRFs
  is shown. The red cross marks the average expectations for GRBs.
  The red star is XRF 060218.
 Middle (b). The X-ray light curve of GRB 060729. The transition from
 Compton to synchrotron dominance is at the start of the flat ``plateau".
 Botton (c). Time evolution of the X-ray spectral index $\Gamma$
 ($dN_{\gamma}/dE\!\approx\!E^{-\Gamma}$) of the same
 GRB, with the transition occurring at the same time.}
 \label{SWIFTGRBs}
\end{figure}

A very simple result concerns the predicted correlations
between ``prompt" GRB observables \cite{DDOldCorr}. These follow directly 
from the kinematics underlying the
assumption that a GRB's $\gamma$-rays are produced by
ICS of `ambient-light' photons, see Section \ref{initial}
and Appendix \ref{TheCB}. 
Consider the
``peak energy", $E_p$, of GRBs and XRFs and their ``isotropic-equivalent"
energy, $E_{\rm iso}$. For a point-like source 
$(1+z)\,E_p\!\propto\!\gamma\,\delta$ and $E_{\rm iso}\!\propto\!\delta^3$,
with $\delta$ the Doppler factor of Eq.~\ref{Doppler}, which varies
extremely rapidly as a function of the observer's viewing angle $\theta$.
If that large and inevitable case-by-case variation is the dominant one, we
expect $(1+z)\,E_p\!\propto\! E_{\rm iso}^{1/3}$. A CB initially
expands in its rest system at a speed of ${\cal{O}}(c)$, so that in its
motion it traces a cone of aperture of  ${\cal{O}}(1/\gamma)$. By ICS,
its interacting electrons emit radiation also within a forward
angle of ${\cal{O}}(1/\gamma)$. For energetic GRBs,
viewed at angles of ${\cal{O}}(1/\gamma)$, the volume-averaged
Doppler factor is $\delta\!\propto\!\gamma$, so that the case-by-case
variation of $\gamma$ results in the expectation 
$(1+z)\,E_p\!\propto\!E_{\rm iso}^{2/3}$. The 
transition from a 1/3 to a 2/3 slope is precisely what is seen in
Fig.~\ref{SWIFTGRBs}a. 
The crossed lines are the expectation for
a ``typical" GRB \cite{DD}. The predicted correlations are also verified
by the data for a handful of other pairs of observables \cite{DDDCorrelations}.

Swift has established the predicted \cite{AGoptical,AGradio}
{\it canonical behaviour} 
of the X-ray and optical AGs of a 
large fraction of GRBs.
The X-ray fluence decreases very fast after
the `prompt' peaks of the GRB. 
It subsequently turns into a `plateau'. After a time of
${\cal{O}}(1$d), the fluence bends (has a `break', in the usual parlance)
steepening to a power-decline. 
In Fig.~\ref{SWIFTGRBs}b, this is shown for a Swift GRB
\cite{DDDBreaks}, and compared
with the CB-model expectation. The early peaks are produced by 
ICS by a succession of CBs, that become `weaker' as the accreting
material that generates them is exhausted. The plateau is due to
the dominant CB, as it coasts in the ISM and synchrotron-radiates
the energy of the ISM electrons that it intercepts. The end decline
reflects the deceleration of the CB in the ISM.

\begin{figure}[]
\vskip -.4cm
\centering
\vbox{
 \epsfig{file=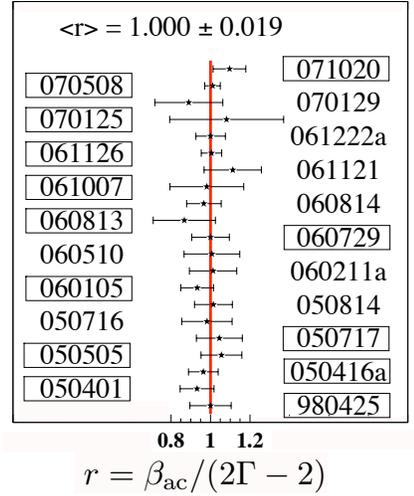,width=7cm}
}
\vskip .5cm
\vbox{\hskip -.5mm
 \epsfig{file=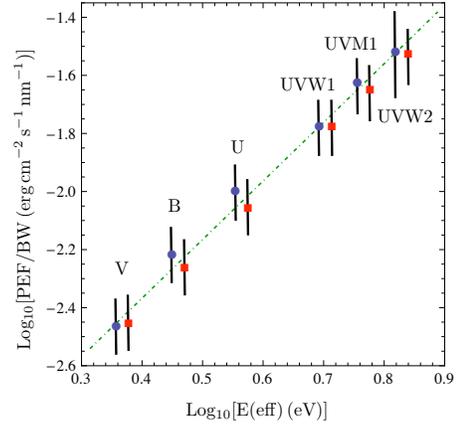,width=6.1cm}
}
\vskip -2.8cm
\vbox{
 \epsfig{file=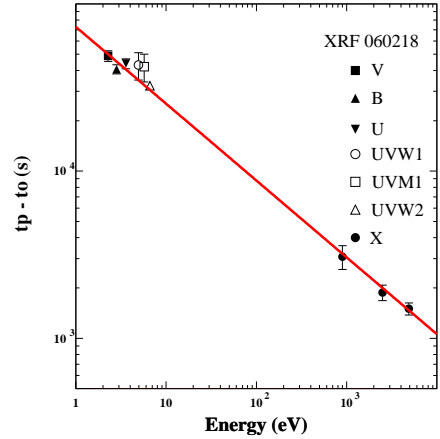,width=6.4cm}
}
\vspace*{-3pt}
 \caption{Top (a). Test of the relation between the ``index" $\beta_{\rm ac}$ 
 governing
 the whole spectral evolution of the X-ray synchrotron AG, and the index
 of its spectrum, Eq.~(\ref{Asymptotic}). 
 Middle (b). Test of the ``peak-energy fluxes" in UV and
 optical intervals predicted from the same observable in the X-ray domain,
 for the single-peak XRF 060218. Data and predictions have been slightly
 shifted for clarity. Bottom (c). Test of the correlation between peak times
 and energy for the same XRF, in a wide domain. The line is the approximate
 ``$E\, t^2$ law", $t_0$ is the energy-independent pulse start-time.
 }
 \label{rand218}
\end{figure}

The rapid transition from ICS to synchrotron-radiation dominance
in the X-ray light curves is accompanied by an abrupt change of spectral
behaviour \cite{AGoptical, DD}. This is illustrated 
in Fig.~\ref{SWIFTGRBs}c where the radiation's spectral index,
$\Gamma$, is shown and compared with the model's expectation
\cite{DDDDecline}. The time dependence of the AG's flux and
its spectral index are related for individual GRBs as in Eq.~(\ref{Asymptotic}).
The test of the predicted relation is shown in 
Fig.~\ref{rand218}a.

As explained in Appendix \ref{promptphase},
the ambient light around a just-exploded SN
has a ``thin bremsstrahlung"
spectrum $dN_\gamma/ dE_i\!\propto\!(1/E_i)\,{\rm Exp}[-E_i/T_i]$, with
$T_i$ a pseudo-temperature. The electrons in a CB boost
this light by ICS to a GRB spectrum with the same shape and 
a higher final $T_f\!\propto\!\gamma\,\delta\,T_i$. 
On occasion a GRB or an XRF is dominated by a single CB,
and its analysis is particularly simple. One example
is XRF 060218, the star in Fig.~\ref{SWIFTGRBs}a. 
For  this XRF one may use the observed X-ray `peak energy flux'
of its single-peak X-ray light curve to predict the corresponding fluxes 
of its (much broader) UV and optical peaks, also produced
via ICS of ambient light by the CB \cite{DDDXrays, DNature}. 
This is done in Fig.~\ref{rand218}b.
The ambient light sampled by a CB becomes increasingly 
radially-directed with distance from the SN, $r$, so that the incident
angle of photons, $\theta_i$, on the CB's electrons obeys
$\langle 1+\cos\theta_i\rangle \!\to\! 1/r^2$; and 
$T_f\!\propto\!\langle 1+\cos\theta_i\rangle \! \to \! 1/r^2$.
In the ``prompt" phase, while a CB has not 
significantly decelerated, $r\!\propto\! t$, the observer's time.
To a fair approximation, then, the time-energy correlation of the
spectrum is such that $E\,dN/(dE\,dt)\!\propto\!F(E\,t^2)$.
This `$E\,t^2$ law' is tested in Fig.~\ref{rand218}c for the
``peak times" of the XRF's pulse at frequencies ranging from X-rays
to optical \cite{DDDXrays}.

The Swift data display a panoply of X-ray 
AG shapes, ranging from some having a long 
and very flat plateau to others well approximated at all times by 
a single  power law. In the CB model this is the 
prediction of Eq.~\ref{Fnux}. A GRB pointing close to the
observer ($\theta\,\gamma_0\!<\! 1$) is very luminous,
its radiation being highly Doppler boosted, see Eq.~\ref{dist}.
Its synchrotron radiation diminishes uniformly 
as $\gamma(t)$ and $\delta(t)$ decrease, resulting in an approximately
power-law AG. A GRB pointing away from the
observer's direction ($\theta\,\gamma_0\gg 1$) is relatively
under-luminous (and is often classified as an XRF)
because its beam is forward-collimated within an
angle $1/\gamma$. As $\gamma$ diminishes the beam
opens up to reach the observer, resulting in a plateau
or even an increased radiation. Finally $\gamma(t)$
and $\delta(t)$ become small enough for the AG
to tend to a power law. Studied in detail, these
simple facts reproduce the entire panoply of
observed light-curve shapes \cite{DDDBreaks}.

The successful analysis of ``Swift-era" data that we have briefly illustrated
is entirely based on predictions made before the launch
of Swift. From the CB-model's point of view the data of this
era has taught us two things. First, during the fast decline of the
X-ray flux, and sometimes later, one can occasionally see the
effects of late and ``weak" CBs, the dying pangs of their
accretion-governed ``engine". Second, the bremsstrahlung
and line-emission phase cited in Appendix \ref{GRBAGs}
may be generally subdominant: a simplification.

\subsection{Tibet AS$\gamma$}
\label{Tibet}

\begin{figure}[]
\vskip -4.4cm
\centering
\vbox{
 \epsfig{file=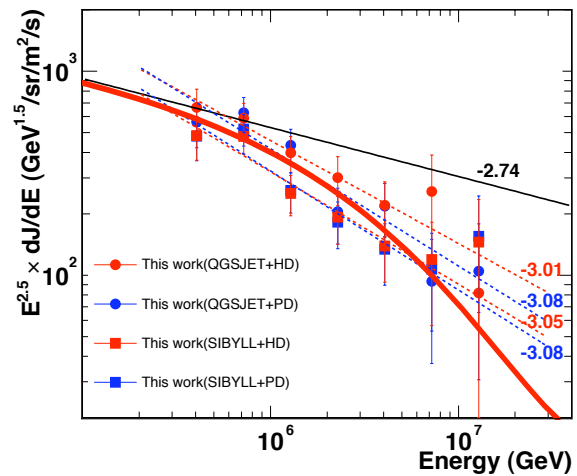,width=9cm}
}
\vskip -4.4cm
\vbox{
 \epsfig{file=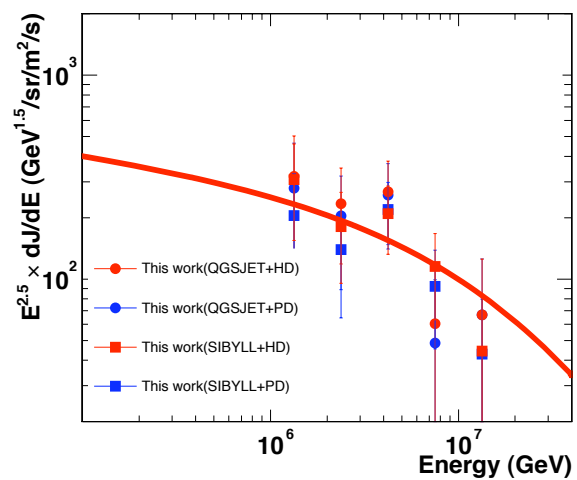,width=9cm}
}
\vspace*{-5pt}
 \caption{The proton (Top) and He (Bottom) spectra around their `knees',
 obtained by the Tibet AS$\gamma$ collaboration
 \cite{Amenomori}. The (red) curves are those of Fig.~\ref{KASKADE}. }
 \label{TibetFig}
\end{figure}

This group has studied the all-particle spectrum in the energy
range of $\sim\! 10^5$ to $\sim\! 10^8$ GeV, finding it to be
compatible with measurements from previous experiments \cite{Amenomori}.
It has also analyzed in detail the p and He spectra around their
knees \cite{TibetpHe}. Their results are shown in Fig.~\ref{TibetFig},
superimposed on the (red) curves of Fig.~\ref{KASKADE}.

\subsection{HIRES and Auger}
\label{Auger}

The HIRES collaboration has announced the observation of the 
GZK cutoff \cite{HIRESGZK}.
The Pierre Auger observatory in Argentina has improved the precision
of the data on UHECRs \cite{Mello},
studied their composition \cite{Unger}, and reportedly located some of their
sources \cite{PAC}. 

\begin{figure}[]
\centering
\epsfig{file=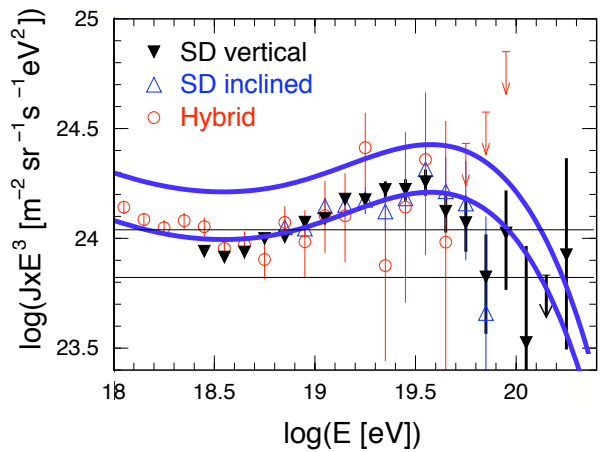,width=8cm}
 \caption{The Auger UHECR spectrum \cite{Mello}.
 The upper curve is the same as the (blue) curve 
 in Fig.~\ref{UHECR}. The lower curve is the same one,
but shifted down in flux by a factor $\sim\!1.7$.}
 \label{Augerspec}
\end{figure}

The UHECR all-particle spectrum of Auger is shown in Fig.~\ref{Augerspec}.
The upper curve in this figure is the same as the (blue) curve in Fig.~\ref{UHECR};
if shifted down in flux by a factor $\sim\!1.7$, it results in the lower curve.
This is the overall factor by which the Northern-hemisphere data of
Fig.~\ref{UHECR} exceed the Southern-hemisphere ones.
This difference may be due to the well-known calibration difficulties.
It may also be real, for the Northern sky is, within a radius of the
order of the distance to the Virgo Cluster, more densely populated
that the Southern sky. If CRs above the Ankle are mainly extragalactic,
and if the MFs they cross are insufficient to scramble 
completely their arrival directions,
one would expect a higher Northern flux. 

\begin{figure}[]
\centering
\epsfig{file=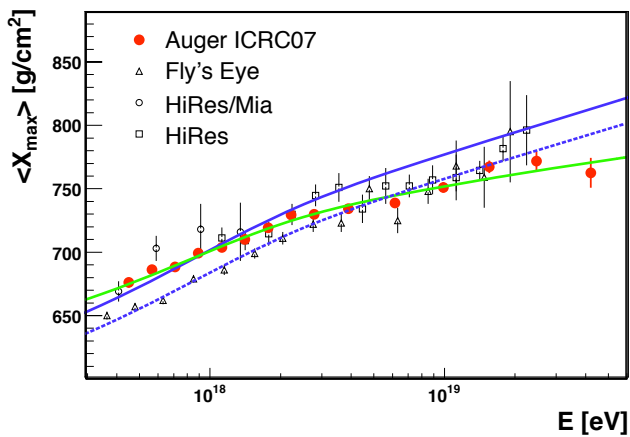,width=9cm}
 \caption{The trend of $X_{\rm max}(E)$ at very high energies.
 Data from Auger are compared with those of previous experiments
 \cite{Unger}. The blue line is that in Fig.~\ref{Xmax}, the others
 are commented in the text.}
 \label{AugerXmax}
\end{figure}

The Auger data on $X_{\rm max}(E)$ \cite{Unger} are compared
with previous observations in  Fig.~\ref{AugerXmax}.
The blue line is that of Fig.~\ref{Xmax}, in
which one can ascertain the dependence of the 
expectations on various Monte-Carlo (MC) simulations. The dashed
line, which agrees with the Fly's eye results, is the same blue
curve, rescaled down in $X_{\rm max}$ by 2.3\%, well
within the MC uncertainties. The green line, which agrees with the 
Auger data, is the blue line, shifted in $dX_{\rm max}/d\,{\rm Log}_{10}(E)$
by 3.3\% per decade of energy, not larger than the difference
in slope from some MCs to others. These considerations
illustrate how difficult it is to extract
the CR composition at large $E$. Note 
that {\it changes} in composition are easier to ascertain, since
{\it changes} in the slope of $X_{\rm max}(E)$ clearly
reflect them. The Auger results have a clear
change of slope at the Ankle, as expected in our model. Note also
that the MC-to-MC spread in $dX_{\rm max}/d\,{\rm Log}_{10}(E)$
is much smaller than that of $X_{\rm max}$ itself. It might be
useful to study $dX_{\rm max}/d\,{\rm Log}_{10}(E)$,
which would signal compositional changes in a much-reduced
MC-dependent way.

\subsection{Auger's directional correlation with AGNs}
\label{AugerAGNcorr}

The Auger collaboration has reported \cite{PAC} a
correlation between arrival directions of UHECRs and nearby AGNs
from the V\'eron catalog \cite{Veron}. The effect is maximized
for events with $E\!>\! 57$ EeV $\!=\!5.7\times 10^{10}$ GeV, and a 
$3.1^{\rm o}$ aperture around AGNs with $z\!<\! 0.018$, corresponding
to a distance $D\!<\! 75$ Mpc along a straight trajectory, smaller than
the corresponding distance, $D\!=\! 200$ Mpc, to the GZK ``horizon", 
defined in Ref.~\cite{PAC} by a 10\% attenuation of protons with $E_p\!>\! 60$ EeV.
In the optimized sample some 20 events are correlated
and $\sim\! 1/3$ as many are not.
The collaboration cautiously warns that AGNs could be mere tracers
of the density distribution of matter.

As we saw in Section \ref{CRAGN}, if AGNs accelerate CRs the
way that core-collapse SNe allegedly do, an AGN contribution
to UHECRs could be significant. That is a correct statement for the
angularly-integrated flux, but not for the flux from 20
arrival directions of $3.1^{\rm o}$ aperture, which ought to be,
on pure geometrical grounds, 
significantly smaller. Neither do we expect the matter-density 
enhancement traced by AGNs to correspond to
a sufficient number of SNe, close to AGNs, and emitting
jets pointing close to our direction. 

Other difficulties in {\it our} understanding of the Auger data
are more model-independent. The V\'eron catalog is not
directionally uniform in its coverage and sensitivity, unlike
the Auger coverage within its field
of view. A $3.1^{\rm o}$ deviation is of the order of magnitude
of that inflicted on UHECRs by the MF of the Galaxy, it would
be surprising if extragalactic CRs did not encounter other 
MFs with similar or larger effects. The Auger correlation is purely
directional, not investigated case-by-case for the possible effects
of AGN distance, luminosity, jet direction and radio loudness.
The effect of distance is obvious, the correlation with
luminosity is very plausible. Concerning jet-direction,
one has to understand
how the UHECRs from AGNs could be fairly isotropically emitted,
given that AGNs produce extremely collimated jets, and that
they are seen in $\gamma$-rays as very luminous blazars
when the jets are pointing in our direction.
The proton- and electron-acceleration efficiencies of
CR sources are presumably correlated. The radio loudness
is a measure of the electron-acceleration efficiency,
since the radio signal, as in Fig.~\ref{Pictor}b, is due
to synchrotron-radiating electrons. Finally, the number of 
UHECR Auger events is roughly the same in each of
five equal-exposure domains, but the AGNs of the
V\'eron catalog, which cluster
along the super-galactic plane, are differently distributed.

For the above reasons we do not attempt to adapt at the moment our
model to the Auger findings, limiting ourselves to the few comments
that follow.

The CBs of AGNs are much more massive than those of 
SNe, but their Lorentz factors, estimated from their superluminal motion 
\cite{gammaAGN}, 
are much smaller, $\gamma\!=\!{\cal{O}}(10)$. Their observed SR was used to estimate
a field $B\!=\!{\cal{O}}(1\,\rm mG)$ \cite{MFAGN}. This is precisely
consistent with the equipartition estimate,
$B\!\sim\!\gamma\,\sqrt{8\, \pi\, n\, m_p\, c^2}$, for $\gamma\!=\! 10$ and 
the mean IGM baryon density, $n\!\sim\! 2.3 \times 10^{-7}\, {\rm cm^{-3}}$
determined from the observations of WMAP \cite{WMAPdens}.

The CBs of an AGN may `inelastically' accelerate CRs 
to  well above the GZK limit and collimate them forward
in a cone of aperture $\theta\!\sim\!1/\gamma$. In view of the
Auger results, we are interested in a potentially more isotropic
source, the end `lobe' of an AGN jet being the obvious choice
\cite{LobeCRs}. These lobes have radii $R_L$ of a few kpc. They are
steadily energized by the incoming CBs. Traveling in a medium 
swept up by previous CBs, a jet may deposit in its lobe half of
the energy, $E_{\rm AGN}\!=\!{\cal{O}}(10^{60}\,\rm erg)$, emitted
by the central black hole during the AGN's active life.
In equipartition, $E_{\rm AGN}\!=\!(2/3)\,B_L^2\,R_L^3$, 
corresponding to a field $B_L\!=\!{\cal{O}}(1\,\rm mG)$.
The larmor-limit energy for  the acceleration of
a proton in a lobe is then
$E_{\rm max}\sim e\, B_L\, R_L\! \approx\!  3\! \times \! 10^{21}\, {\rm eV}$,
above the GZK cutoff. 

The analogy with AGN jets is one of the items originally inspiring
the CB model. Perhaps the Auger results may be used to close
this analogy into a loop. If AGNs are tracers of very dense environments,
it is conceivable that the jets of nearby SNe stop swiftly in mini-lobes,
much as the non-relastivistic ejecta of SNe are sometimes observed
to clash with molecular clouds \cite{W41, 443}. The CRs from these
mini-lobes would not be forward-collimated. This would make
core-collapse SNe in AGN environments potential sources of 
the UHECRs observed by Auger.

\subsection{TeV $\gamma$-rays from supernova remnants}
\label{gammaray}

Various observatories,
HESS, MAGIC, CANGAROO-III and VERITAS \cite{Ribo}
are currently charting the sky for `TeV' $\gamma$-ray sources,
following a multi-telescope Cherenkov technique pioneered by HEGRA
\cite{Pulhofer}. Their data may test, among other things, the earlier
claims from X-ray observations \cite{Koyama} that SNRs are `Tevatrons'.

The HESS system in Namibia has surveyed 
the Galactic plane at latitudes $\rm{|b|\le 6^o}$ and longitudes 
$\rm{-85\le |l|\le 60^o}$,
for $\gamma$-rays between 100 GeV and several tens of TeV.
It has found dozens of sources, many of them previously unknown,
and having no known counterpart at other wavelengths \cite{Hoppe}.
HESS also detected TeV $\gamma$ rays from three SNRs.
The collaboration concludes that ``The paradigm of CR acceleration
in Supernova remnants (SNRs) is consistent with our new findings"
but is cautious in adding ``but it seems clear that the new sources
are not drawn from a single population" \cite{Aharonian}.
If the cited ``paradigm" includes the consuetudinary contention that
the bulk of non-solar CRs below the knee(s) is accelerated by
SNRs (see, e.g.~Ref.~\cite{Drury}), we, and others \cite{Plaga}, 
beg to disagree.

In the conventional view, a
non-relativistic spherical shell ejected in a SN explosion 
drives a strong shock into the ambient medium, heating it up 
to a multi-keV temperature, and accelerating ions and electrons to 
CR energies. The ionized gas emits thermal X-rays via bremsstrahlung. 
The accelerated CR electrons emit non-thermal SR extending from the 
radio to the X-ray band. They also emit $\gamma$ rays by ICS of 
the locally-generated radiations. 
The accelerated CR nuclei emit $\gamma$ rays 
and neutrinos from the decay of mesons produced by hadronic collisions in 
the ejecta and in the ambient plasma.

Although SNRs are well established as non-thermal sources of radio, 
X- and $\gamma$-rays, as well as thermal sources of soft X-rays, the 
observations do not support the basic relations for the 
CR-accelerating shocks driven by the SN shells into the ionized ISM.
The shock velocity, $V_s$, and the ISM temperature behind the
shock, $T_s$, for instance, ought to be related by 
$k\, T_s=(3/16)\, \mu \,m_p\, V_s^2$, with $\mu\!\sim\! 0.6$ the mean
plasma mass in $m_p$ units. This expectation is badly disobeyed
by young SNRs, a result interpreted as evidence for an efficient
(though unobserved) acceleration by shocks of modified
structure and hydrodynamics \cite{Ellison}.

Radio SR has been observed from many    
Type II and Type Ib/c SNe, but no SN Ia has yet been detected as a
radio emitter, even when observed quite promptly 
or quite nearby \cite{Panagia}. This implies that efficient
CR acceleration and the generation of the required strong MFs
must involve collisions of
SN shells with targets much denser than the average ISM.  Such 
high-density targets, unlike the low-density ISM around SNe Ia,
can be provided by massive ejecta or
winds from the progenitor star of core-collapse SNe, 
or by the high-density molecular-cloud environment in which some 
Type II SNe take place. Collisions with such over-dense
targets can produce intense turbulent MFs in the expanding SN shell.
If the MF is in energy equipartition with ejecta of velocity 
$V_{\rm ej}\!\equiv\!\beta_{\rm ej}\,10^{-2}\,c$, impinging on a
medium of density $n\!\equiv\!n_1\,{\rm cm}^{-3}$, the MF
intensity is $B_{\rm eq}\!\sim\! 0.7\,\beta_{\rm ej}\,\sqrt{n_1}$ mG.
In principle, such strong magnetic fields can very efficiently accelerate 
ions and electrons to CR energies 
via the Fermi mechanism. The accelerated CR nuclei and electrons
can produce TeV $\gamma$ rays by hadronic interactions and
by ICS of ambient photons, respectively.

Two SNRs displaying a shell morphology at TeV energies have been found by 
HESS: RX J1713.7-3946 \cite{3946} and RX J0852-4622  \cite{4622}. 
The TeV shells seem to 
follow the SR keV emission pattern seen by X-ray satellites, suggesting 
ICS of ambient photons as the origin of the TeV emission. On the other 
hand, TeV emission has also been found in two SNRs, HESS J1834087/W41 
\cite{W41} and 
MAGIC J0616+225/IC 443 \cite{443}, in the interaction region between the SNRs and  
molecular clouds, suggesting a hadronic origin. Combining 
 this information with the absence of radio emission from SNe Ia, 
leads us to conclude that the molecular clouds do not only serve as 
gas targets converting hadronic CRs from adjacent SNRs to TeV 
$\gamma$ rays, but that the acceleration itself takes place in the collision 
region between the SN shell and the molecular cloud. Such $\gamma$-ray 
Tevatrons are very rare among the $\sim\!250$ SNRs observed in the Galaxy.
This implies that the total production rate of CR nuclei in SNRs falls 
very short of supplying the observed Galactic CR luminosity. 
This conclusion is consistent with two other facts: The distribution of Galactic 
CRs --as inferred from the distribution of diffuse $\gamma$ radiation from the 
Galactic disk-- has a scale hight and radius much larger than 
expected from the interactions of CRs from SNRs in the Galactic ISM. 
Non-solar CRs do not show the anisotropy of arrival 
directions near Earth expected from the location of the solar system, 
relative to the distribution of SNRs \cite{Smith}.

\subsection{TeV $\gamma$-rays from GRBs}

The typical or ``peak" energies of the prompt $\gamma$- rays of 
GRBs are of the order of a few hundred keV. This agrees with
the prediction of the CB model \cite {DD}, 
$(1+z)\,E_p\!\sim\!\gamma_0\,\delta_0\,T_i/2$, with $T_i\!\sim\!1$ eV
the pseudo-temperature of the glory's thin bremsstrahlung
spectrum. In this model there are two sources of higher-energy
photons, one during the prompt phase, the other during the
AG phase.

The CR electrons accelerated by a CB during the prompt
phase have an average Lorentz factor $\Gamma_e\!\sim\!\gamma_0^2$.
They Compton up-scatter the glory's photons to a distribution
with the same spectral shape as the bulk of the GRB's
prompt $\gamma$'s, but with a characteristic `second' peak energy
$(1+z)\,E'_p\!\sim\!\gamma_0^2\,\delta_0^2\,T_i/2$. This is typically a million
times larger than $(1+z)\,E_p$, that is, in the range of hundreds
of GeV \cite{DDS05}. The relative flux of this hard prompt component is
difficult to estimate. It may have been seen 
in the case of GRB 980425 \cite{DDS05}, observed very nearby
but at a relatively very large observer's angle (small $\delta_0$).
The satellite GLAST may be able to detect these prompt
high-energy photons.

In the AG phase,
CR nuclei accelerated by the CB, mainly protons, impinge
on ISM nuclei to produce $\pi^0$'s, and thus $\gamma$'s in their
$\pi^0\to\gamma\gamma$ decay. The calculation of an upper limit
to the $\gamma$ flux is straightforward. An estimate of the actual flux is
practically impossible, mainly because we cannot confidently ascertain the
broadening of the proton `beam' by magnetic fields, before the
beam encounters its target `beam dump'. We discuss in turn the flux upper limit,
its potential decrease by beam broadening, and an optimistic scenario
in which the CBs themselves are the beam dump.

At incident TeV energies, the total inelastic proton-nucleon cross section 
is $\sigma_{\rm in}\!\simeq\! 40$ mbarn. The column density of an effective beam 
dump --in which the fraction of protons that collide is large-- is a
reference $N_{\rm eff}\!\sim\! 1/\sigma_{\rm in}\!\simeq\! 2.5\times 10^{25}$
nucleons/cm$^2$, much larger than the values we shall deal with
(the astrophysical dumps relevant to our discussion are thin enough
for TeV photons not to be significantly absorbed, though, traveling for
longer cosmological
 distances, they may be attenuated by pair production in the cosmic infrared 
background).

The column densities extracted from the fitted X-ray absorption as a
function of energy, for GRBs observed by Swift \cite{SwiftAtt}, vary
within an order of magnitude of
$N_{\rm Swift}\!\sim\! 10^{21}$
nucleons/cm$^2$. The CR  protons made by a CB 
also traverse this dump, which is  ``thin". Indeed, the probability
of a TeV proton to interact while in the host galaxy, assuming  that all 
the extragalactic X-ray absorption took place in it, is 
$N_{\rm Swift}/N_{\rm eff}\! \ll \! 1$.
The proton beam  carries the energy of a
single average jet of CBs, 1/2 of the result of Eq.~(\ref{Ejets}). Its spectral index
 is that of the ``source'' flux, Eq.~(\ref{beautyfulindex}), since the effects of accumulation
in the magnetic field of the parent galaxy have not yet affected it.
Therefore, the proton number flux is:
\begin{equation}
{dn_p\over dE}\approx 1.25 \times 10^{53}\;{\rm GeV}^{-1}
\left( {{\rm GeV}\over E}\right)^{\beta_s}
\label{pnumberflux}
\end{equation}

To a very good approximation, the $\gamma$ flux generated by a power-law
proton beam, such as that of Eq.~(\ref{pnumberflux}), impinging on a thin dump,
has the same energy dependence as the proton beam, and 
a fraction $F_{p\to\gamma}\!\sim\! 0.04$ of its
normalization (per interacting proton) at a given energy \cite{nubeam}. 
For a GRB whose beam
dump's column density is $N_{_{\rm GRB}}$, the $\gamma$ number-flux is:
\begin{eqnarray}
{dn_\gamma\over dE}&\approx& N_{_{\rm GRB}} \, \sigma_{\rm in}\;
F_{p\to\gamma}\;{dn_p\over dE}
\nonumber\\
&\sim& {2.0\times 10^{48}\over{\rm GeV}}\,{N_{_{\rm GRB}}\over N_{\rm Swift}}
\left( {{\rm GeV}\over E}\right)^{\beta_s}
\label{gammanumberflux}
\end{eqnarray}

At a given point in time the number-flux of photons is enhanced in the
forward direction be the factor $\delta^2(\gamma,\,\theta)$ of 
Eqs.~(\ref{dist}, \ref{Doppler}). In search for an upper limit, we
shall illustrate the forward case, $\theta\!=\!0$, for which 
$\delta(t)\!=\!2\,\gamma(t)$, a good approximation in all cases
as soon as $\gamma(t)$ becomes small enough to satisfy 
$\theta\,\gamma(t)\!<\!1$.
In the same search for an upper limit, we integrate over all
of the trajectory of a CB, or all the CRs that it accelerates, till
$\gamma(t)\!\sim\!{\cal{O}}(1)$, with the help of Eq.~(\ref{gammadown}).
At $\theta\!=\!0$, this means that the instantaneous $\delta^2$ is to be traded 
for an integral:
\begin{equation}
\langle \delta^2 \rangle \simeq \gamma_0\,
\int_{\gamma_0}^{\sim 1} {(2\,\gamma)^2 \over \gamma^3}\,d\gamma
\simeq 4\,\gamma_0\,\ln{\gamma_0}
\label{trade}
\end{equation}

For a GRB at a luminosity distance $D_L(z)$, the $\gamma$ flux per unit area
and energy is:
\begin{eqnarray}
&&\!\!\!\!\!\!\!\!
{dF_\gamma\over dE}
\approx 
{(1+z)\,\langle\delta^2\rangle\over 4\,\pi\,D_L^2}\;
{dn_\gamma\over dE}
\biggl\vert_{E\rightarrow (1+z) E}
\nonumber\\
&&\!\!\!\!\!\!\!\!
\sim
{ 4.6 \times 10^{-4}  \over {(1+z)^{0.2}\,\rm GeV\,cm^2}}\;
{N_{_{\rm GRB}}\over N_{\rm Swift}}
\left( {{\rm GeV}\over E}\right)^{2.2}\left( {{\rm Gpc}\over D_L}\right)^2
\label{gammaflux}
\end{eqnarray}
where we have approximated $\beta_s\!\sim\!2.2$
and specified the result for $\gamma_0\!=\!10^3$. This flux is integrated
in time for the duration of the GRB's AG,
typically of order months.

The flux of Eq.~(\ref{gammaflux}) is not large. For example, for $z\!=\!1$ 
[$D_L(z)\!\approx\!7.1$ Gpc, for standard cosmological
parameters], and above a threshold of 250 GeV, it corresponds
to $\sim\!8.8\times 10^{-9}$ photons per cm$^2$, below the sensitivity
threshold of current Cherenkov telescopes, for a signal spread
over more than a few days. Moreover, there are
four other reasons why Eq.~(\ref{gammaflux}) is but an upper limit:

First, we have worked in the limits wherein all CRs
are generated before they hit the dump. 

Second, we have assumed that the primary proton flux 
is isotropic in the CB's rest system. For non-singular angular
distributions, this does not affect the derived spectral shape
of Eqs.~(\ref{beautyfulindex}, \ref{betapred}). But reasonable
distributions favouring forward proton-CB scattering in the CB's
system would result in a less forward-peaked proton beam in the
SN rest system, and in a beam of $\gamma$'s less forward-collimated
than reflected by the factor $\langle\delta^2\rangle$ in Eq.~(\ref{gammaflux}). 
This may reduce the flux by a factor of a few.

Third, the inter-stellar MFs encountered by TeV protons 
could widen the proton beam by many degrees,
before it interacts with the ISM to produce $\gamma$ rays.
 This would decrease the flux of Eq.~(\ref{gammaflux}) by orders of magnitude.
Yet, the energy density of the CR beam preceding a CB in its
trajectory is so very many orders of magnitude larger than the 
energy density of the typical ISM field, that the beam ought to wipe out
the MF within
a funnel of angular aperture $\sim\!1/\gamma$ \cite{DDMFWipeout}.
The CB-generated CRs exiting a CB at a given point in its
trajectory would travel straight in the MF-free domain produced by 
CRs having exited the CB before. That is only correct on average, for the CRs
emitted at a larger than average angle would escape into a domain with
conventional MFs. So would the CRs at the leading front
of the funnel, even if, before they escape, they are repeatedly caught up and 
reaccelerated by 
the CB \cite{Hillas}. This process is too complex to be understood in detail. But we
know observationally that CR electrons escape into the space surrounding
a beam of CBs, see, e.g.~Fig.~(\ref{Pictor}b).

Fourth, photon attenuation of TeV photons by 
pair production on the infrared background radiation
is very strong at redshifts in excess of $z\!\sim\! 0.1$ \cite{Stecker}.

A uniform-density CB with the initial baryon number $N_{\rm B}$
of Eq.~(\ref{typicalNB}), and
the radius $R_0$ of Eq.~(\ref{best}), has a maximum column density 
$N_{\rm CB}\!=\!3\,N_{\rm B}/(2\,\pi\,R_0^2)\!\simeq\! 4.8\times 10^{21}$
cm$^{-2}$. The probability of a proton crossing the CB along its diameter
to interact with a CB's proton
is  $N_{\rm CB}\,\sigma_{\rm in}\!\simeq\!1.9\times 10^{-4}$. 
To investigate a more favourable case than the one we have discussed, let
us assume that the MF within the CB is sufficiently
entangled for an entering ISM proton to travel some $10^{4}$
CB radii before it exits by diffusion. In that case, on average, every incoming
ISM proton interacts once with a CB proton as the CB is
decelerating to rest. The CB itself acts as an ``effective" dump.
The average collimation is again that of
Eq.~(\ref{trade}) and the factor $N_{\rm GRB}/N_{\rm Swift}$ 
is to be traded by $N_{\rm eff}/N_{\rm Swift}$ in
Eq.~(\ref{gammaflux}), so that the numerical value of the
flux would be enhanced by a factor 
$\simeq\!2.5\times 10^4$, bringing it
close to observable levels. Notice, moreover, that the first three caveats
discussed in the previous paragraphs do not apply to this case.

We conclude that Eq.~(\ref{gammaflux}) is an overestimate of the expected
$\gamma$ flux from CR interactions outside the CB,
 but we cannot ascertain by how much. Only in the case
of a very energetic, very close-by GRB, there is a slim chance 
to see these TeV $\gamma$ rays during the AG phase. On the other hand,
if the MFs within a CB are sufficiently entwined,  the $\gamma$ rays 
from within the CBs themselves may be observable
by GLAST and by Cherenkov telescopes. They would
originate from hadronic interactions of CR nuclei or from ICS of 
synchrotron radiation by electrons. It may turn out
to be productive to point a Cherenkov telescope to the location of an
intense GRB, for some time during the first days or weeks after its explosion,
to search for a possible TeV $\gamma$-ray afterglow.
No doubt GLAST, whose energy threshold
is much lower than that of the Cherenkov devices,
 will be pointed to GRB locations as fast as
possible, but it may be productive to extend its observations
well into the afterglow phase.

\section{High Energy Neutrinos}
\label{neutrinos}

A guaranteed source of UHECR neutrinos is the GZK effect. The protons
that interact with the MBR produce pions ($\pi$'s) and the consequent flux
of their decay products, including very energetic $\gamma$ rays and
neutrinos ($\nu$'s). Detecting these inevitable fluxes is hard 
\cite{GZKProducts}. Lower-energy
hadronic CRs also produce $\nu$'s and $\gamma$'s in their interactions
with matter targets, notably the atmosphere, the beam-dump responsible
for ``atmospheric" $\nu$'s. Galactic CRs interact with
the ISM to produce an observable $\gamma$-radiation, but the Galactic ISM,
even at locations where its density is enhanced, is
too distant and too thin a dump for the generated $\nu$'s to be observable.

The $\nu$ flux expected from a GRB is calculated along the same lines
as the $\gamma$ flux of Eq.~(\ref{gammaflux}). The relevant dumps have
sufficiently low densities for the parent charged pions and muons to decay.
Only the $\nu_\mu$ flux
is to be discussed, the $\bar\nu_\mu$ flux is similar, but its cross section on matter
is about 1/2 of that of $\nu_\mu$'s at the relevant energies, and the astrophysical
uncertainties are much larger than 50\%. The uncertainty also allows us
not to discuss $\nu$ oscillations, which reduce the $\nu_\mu$ flux by
a factor of $\sim\!3$. The detection of
electron or tau neutrinos is harder than that of their muon counterparts.

The approximate isospin invariance
of high-energy hadronic interactions implies the same production properties
for $\pi^+$, $\pi^-$ and $\pi^0$. 
In computing the $\nu_\mu$ flux in analogy with the $\gamma$ flux, the
production and decay of $\pi^0$'s is to be substituted by 
$\pi^+ \to \nu_\mu\,\mu^+$ and the chain $\pi^- \to \bar\nu_\mu\,\mu^-$,
$\mu^- \to \nu_\mu\,e^-\,\bar\nu_e$. The $\nu_\mu$ produced in $\pi^+$
decay is soft and that produced in $\mu^-$ decay is hard, resulting
in a longitudinal distribution of the two $\nu_\mu$'s fairly similar to that of
the two $\gamma$'s from $\pi^0$ production and decay.
All this results in a $\nu_\mu$ beam
almost identical to the $\gamma$ beam of Eq.~(\ref{gammaflux}).

\subsection{Neutrinos from GRBs}
\label{nuGRB}

The $\nu_\mu$ flux being approximately the same as that of the $\gamma$
flux of Eqs.~(\ref{gammanumberflux},\ref{gammaflux}), the calculation
of the expected number of neutrino interactions proceeds
along the same line as for hight-energy photons.

Large-area high-energy neutrino detectors, such as IceCube \cite{IceCube},
 search for
upward-going muons produced not only in the detector, but also in the
ice or the rock surrounding it, since the range of TeV-energy muons is larger
than the detector's depth. For IceCube, of surface $S\!\sim\! 1$
km$^2$, the $\nu_\mu$ conversion
probability into an observed muon,  $P_{\nu\to\mu}$, at TeV energies, 
is estimated \cite{IceCube} to be:
\begin{equation}
P_{\nu\to\mu} \sim 1.7\times 10^{-9} \left({E_\nu\over {\rm GeV}} \right)^{0.8}.
\label{nuconvert}
\end{equation}
We compute the number of events only in the optimistic limit
in which the CBs are ``effective dumps", discussed in the previous
section. The result is:
\begin{eqnarray}
&&\!\!\!\!
N_\mu = S\,\int_{E_{\rm min}}\!P_{\nu\to\mu}\;{dF_\nu\over dE}\,dE\sim
\nonumber\\
&&\!\!\!\!
{1.2\times 10^{-3} \over (1+z)^{0.2}}\;
{S\over {\rm km}^2}
\left[{{\rm TeV}\over E_{\rm min}}\right]^{0.4}
\left[ {{\rm Gpc}\over D_L}\right]^2\!\!\!,
\label{muobs2}
\end{eqnarray}
which is undetectably small, even for $D_L\!=\! 1$ Gpc ($z\!\sim\!0.19$), 
a distance below which very few GRBs have been detected.

\subsection{Energetic $\nu$'s and $\gamma$'s from Virgo}
\label{NGC}

The Virgo cluster, at a distance $D_{_{\rm V}}\!\sim\!17$ Mpc
($z\!\sim\!3.7\times 10^{-3}$),
is the most luminous nearby cosmological structure, and may be the
strongest extragalactic source of high-energy $\nu$'s and $\gamma$ rays.
We estimate their fluxes.

The CR flux escaping from a cluster's
constituent galaxies to permeate its IGM has the source spectral index,
$\beta_s\!\sim\! 2.2$, of Eq.~(\ref{beautyfulindex}).
The hypothesis that, like in our Galaxy, the MFs and CRs of clusters are
in energy equipartition, results in a good description of the properties
of clusters \cite{CDD, DDMF}. Equipartition allows
us to estimate the normalization of Virgo's CR number density, 
$dn_{_{\rm CR}}/(dE\,dV)$,
from the condition that its energy-weighed integral be equal to 
$\rho_B\!=\!B^2/(8\,\pi)\!\approx\! 1$ eV cm$^{-3}$. For $B\!\sim\! 5\,\mu$G,
the observational estimate at the cluster's core, the result is:
\begin{equation}
{dn_{_{\rm CR}}\over dE\,dV}\sim {1\times 10^{-10}\over {\rm cm^{3}\,GeV}}
\;\left({E\over {\rm GeV}}\right)^{-\beta_s}
\label{VirgoCR}
\end{equation}
For a CR density that traces the observed gas density \cite{CDD}, the average
CR number density is roughly $d\bar n_{_{\rm CR}}\!=\! dn_{_{\rm CR}}/2$,
the density at the cluster's core radius.

From an object of the size and MF 
strength of a cluster, CRs should not abundantly escape, but be
confined for times longer than their mean interaction time with the
cluster's gas.
The interaction rate, $c\,\sigma_{\rm in}\,d\bar n_{_{\rm CR}}/dV$,
is of the order of the Hubble expansion rate, $H_0$, so that the
effect of interactions on the CR flux is not negligible. But, to a 
good approximation, the `secondary' flux has the same 
energy dependence as that of the primary source flux, used in 
Eq.~(\ref{VirgoCR}). The reason is that, at its highest energies,
the secondary flux is dominated by the `leading' proton,
whose fractional energy $x_p$ is `forward-peaked' and
averages to $\sim\! 0.7$. The advantage of normalizing
the flux via equipartition with the MF (rather than via
an estimate of the primary flux luminosity) is that the 
primary and all subsequent secondary fluxes are automatically
included in the estimate.

The total mass of Virgo, including its dominant `dark' component,
is $M_{_{\rm V}}\!\sim\!1.2 \times 10^{15}\,M_\odot$,
of which 14\% is (mainly hydrogen) gas \cite{SBB}.
Thus, Virgo's gaseous baryon number is $B_{_{\rm V}} \!\sim\! 2\times 10^{71}$. 
 The neutrino flux is:
\begin{eqnarray}
{dF_\nu\over dE}&\sim&{F_{p\to \nu}\, \sigma_{\rm in}\,c\,B_{_{\rm V}}
\over 4\,\pi\,D_{_{\rm V}}^2}
\;{d\bar n_{_{\rm CR}}\over dE\,dV}\nonumber\\
&\sim&
{1.7\times 10^{-11}\over{\rm TeV\, s}}
\left( {{\rm TeV}\over E}\right)^{\beta_s}
\label{VirgoNus}
\end{eqnarray}

The number of $\nu_\mu\to\mu$ events pointing back to Virgo
(whose half-angle in the sky is $\sim\!4^{\rm o}$), with energy
$E_\mu\!>\!E_{\rm min}$ and
gathered in a time $\Delta t$, is estimated, as in Eq.~(\ref{muobs2}), to be:
\begin{eqnarray}
N_\mu &=& S\,\int_{E_{\rm min}}\!P_{\nu\to\mu}\;{dF_\nu\over dE}\,dE
\nonumber\\
&\sim&
{ 9 }\;
{S\over {\rm km}^2}
\left[{{\rm TeV}\over E_{\rm min}}\right]^{0.4}
\left[ {\Delta t \over 1\,\rm y}\right],
\label{muobsVirgo}
\end{eqnarray}
well below the atmospheric background at TeV energies.

The estimated flux of high-energy $\gamma$ rays from Virgo is approximately 
the same as the $\nu$ flux of Eq.~(\ref{VirgoNus}), and is above the expected 
detection threshold of GLAST. Perhaps it may even be detectable at higher
energies by the ground-based Cherenkov telescopes.

\section{Conclusions}

We have sketched a theory wherein
cosmic rays are ions of the interstellar medium, encountered by relativistic CBs and
magnetically kicked up to higher energies ---either  elastically, or `inelastically'
(i.e., after a succession of accelerating
encounters with the CBs' inner turbulent magnetic fields). The elastic component 
is entirely analogous to the mechanism which
---we contend--- generates the prompt $\gamma$-rays of a GRB:  `inverse'
Compton scattering, by the electrons comoving with a CB, of the `ambient light' 
they encounter around their parent exploding star. The inelastic CR component is also 
analogous to the high-energy tail of the spectra of
GRBs \cite{DD}, originating
from a small fraction of electrons that have been accelerated within a CB.
In this sense, our theory of CRs is but a straightforward generalization
 to cosmic rays of the very successful CB model of GRBs. All we have done
is to substitute the scattering of `ambient' light by the scattering of `ambient'
ions and electrons.

Our theory agrees with the classic proposal of Baade and Zwicky \cite{BZ} that
SN explosions are at the origin of CRs. But our mechanism is
different from that of the generally-accepted CR theory, in which it is
the non-relativistically-expanding SN shells ---as opposed to their
relativistic jets--- that accelerate relatively low-energy CRs. We have 
demonstrated how our simple {\it and single}
accelerators ---cannonballs--- are effective at all observed 
energies.

Our intention in this paper was not to reproduce the CR observations in
minute detail, or to refine at maximum those of the conventional inputs
that could be refined ---such as the details of the
photo-dissociation of UHECRs. Thus, we have chosen to minimize
the number of fit parameters to a grand total of one. The rest of the
required input has been gathered from information  independent of
the theory of the source of
CRs, or fixed from the simplest choices we could make at each point. 

Most of our results are `robust' in that ---within very large brackets---
they do not depend on the specific choices of parameters and priors:
\begin{itemize}
\item{} An all-particle power-law spectrum with four successive
features: two steepenings at the knee and the second knee, a
softening at the ankle, and an end-point at the GZK and proton-acceleration
cutoffs, which are roughly coincident. 
\item{} An UHECR flux above the ankle, which is predicted ---to 
within a factor of a few--- and otherwise parameter-free.
\item{} A composition dependence at 1 TeV with the observed trend,
so different from that of the ISM for the relative abundances of H and He
versus those of the heavier elements.
\item{} A very low-energy flux whose spectral shape is independent of any 
CB-model `prior' parameters.
\item{} Individual-element knees that scale like $A$ and occur at the 
predicted energies.
\item{} A non-trivial shape of the individual knees: an abrupt decrease in flux,
followed by a spectrum 
 steeper than that below the knee.
\item{} An ankle with the observed shape. The dominantly Galactic-Fe flux
below it and the dominantly extragalactic-proton flux above it are comparable
in magnitude at the estimated escape energy of Galactic protons. 
\item{} A composition dependence that is almost energy-independent below
the knee, becomes `heavier' from the knee to the second knee,
`lighter' again above it, and finally heavier at yet-unmeasured ultra-high
energies.
\item{} An `extended' distribution of CR sources along CB trajectories 
that emerge
from the central realms of the Galaxy, where most SN explosions take place, 
implying a CR flux at the Earth's position with a much 
smaller and less energy-dependent anisotropy than that of conventional 
SNR models of CRs.
\item{} 
Predictions for the values of a consistently related set of observables:
the CR luminosity, confinement time and volume of the Galaxy, the spectral
indices of CR electrons and of the diffuse GBR. 
\end{itemize}
Our results describe the observed properties of hadronic non-solar
CRs very well from the lowest energies  to
$\sim 10^{10}$ GeV. Above that energy and up to the highest observed
energies, $\sim 10^{11}$ GeV, our theory opts for the data gathered with
fluorescence detectors, corroborated by hybrid detectors such as
Auger. Overall, the energy range for which the theory is
successful covers ten decades
and the flux extends over  three times as many.

The CR theory we have discussed is currently incomplete in various respects.
The confinement of CRs, either in the Galaxy or in the
CBs themselves, is not well understood, but our assumptions on the subject 
---the simplest--- appear
to work very well. There is insufficient information on the Galactic CR wind
to model its effects with confidence; we have had to experiment with various
limiting possibilities. We have argued that CR diffusion need not be explicitly
considered, but we have not proved that to be the case, by considering it 
in detail. Moreover, our theory is based on a two-stage 
acceleration: that of CBs by core-collapse SNe, and that of CRs by CBs. For the
former, we have relied on observations, rather than on a deeper understanding.

In spite of the above limitations, our claims are supported by the simplicity of 
the theory, its extreme economy of free parameters, and by the good 
quality of its description of the fairly elaborate ensemble
of CR data. The theory is an item in a more general understanding of
high-energy astrophysical phenomena, including cooling
flows, large-scale magnetic fields, GRBs and XRFs. On the two later
topics, there has been great observational progress since the first
posting of this article, which very precisely corroborated the CB 
model of GRBs and XRFs.

The astronomy of non-thermal light sources, from radio frequencies 
to TeV energies, as well as high-energy neutrino astronomy,
are the studies of the interactions of CR nuclei and electrons
with ambient matter and magnetic fields. We have illustrated this point
by discussing the Gamma `Background' Radiation as a 
CR `secondary', and by briefly commenting on high-energy $\gamma$-ray 
and neutrino astronomy.

We contend that we have identified 
the acceleration mechanisms promoting the constituents of the 
interstellar medium of the Milky Way, and other galaxies,
 to become the bulk of non-solar cosmic rays of all 
 energies, and the `magnetic-racket' accelerators themselves:
the cannon balls emitted by a large fraction of ordinary core-collapse supernovae.
\\
\\
{\bf Aknowledgements}
We are indebted to Giuseppe Cocconi,
Andy Cohen, Shlomo Dado, Friedrich Dydak, 
Shelly Glashow, Karl-Heinz Kampert, Etienne Parizot and Rainer Plaga
for discussions,   and to the last of them for his comments on the
manuscript. This research was supported in part by the
Helen Asher Space Research Fund at the Technion Institute. AD thanks the 
TH division at CERN for its hospitality. ADR thanks the Technion Institute
at Haifa for the same reason.

\appendix

\section{More on the expansion of a CB}
\label{ExpansionApp}

Observed at X-ray wavelengths, the CBs emitted in various astronomical systems
appear ---within the limits of observational resolution--- not to expand sideways.
The example of Pictor A is shown in Fig.~\ref{Pictor}. Part of this effect may be a trivial
relativistic mirage. Consider an object expanding in its rest system at a fraction
$\beta_T$ of the speed of light and travelling with a large LF $\gamma(t)$.
A distant observer seeing the object move across the sky 
would see it trace a trajectory with an opening angle $\beta_T/\gamma$, as measured
from the trajectories origin: a very thin
`trumpet' if $\gamma(t)$ is originally large, and diminishes slowly with time.

We have first studied the expansion of CBs in \cite{AGoptical}.
We assumed  the CBs emitted by SNe to expand initially at a speed
comparable to that of sound in a relativistic plasma: 
$\beta_T\sim{\cal{O}}(1/\sqrt{3})$. We also assumed a large fraction of
the intercepted ISM particles to be elastically and rapidly scattered,
isotropically in the CB's rest system. The rate per unit
surface of the momentum carried by the exuding particles corresponds 
to a surface pressure $P_{\rm out}$. We assume that the dominant effect
of this pressure on the CB is to counteract its expansion. Then, in
the approximation of
a hydrogenic ISM and a Newtonian force law (to be justified a posteriori),
the CB's radius  as a function of CB's time $t$ satisfies:
\begin{eqnarray}
&&{3\over 4}\,{M_{_{\rm CB}}\over 4\pi\,R_{_{\rm CB}}^2}
\;{d^2R_{_{\rm CB}}\over dt^2}\approx - P_{\rm out}
\label{Newton}\\
 &&P_{\rm out}\approx {1\over 4}\,m_p\gamma^2\,n_p\,c^2,
 \label{Pout}
\end{eqnarray}
where the factor 3/4 is for an assumed homogeneous expansion.
For an assumed constant ISM density $n_p$ along the CB's trajectory,
the resulting $R_{_{\rm CB}}$ increases very fast (in minutes of GRB 
observer's time, for typical parameters) to a coasting value 
$R_{_{\rm CB}}(t)\propto [\gamma_0/\gamma(t)]^{2/3}$ \cite{AGoptical}, 
as illustrated in the (blue) dashed line of Fig. \ref{RofGamma}.

Here we also explore a different extreme, that the ISM particles are
phagocytized by the CB and exit it by diffusion in its entangled magnetic field,
rather than being immediately  and elastically scattered (the fraction that is
accelerated within the CB before they are re-emitted is small, as in \cite{AGoptical}
and in our current discussion of CRs).
We shall see anon that the functional form of $R_{_{\rm CB}}(\gamma)$ is that of 
the (red) continuous line of Fig. \ref{RofGamma}.

The characteristic diffusion time when the LF [radius] of the CB has reached a value 
$\gamma$ [$R_{_{\rm CB}}(\gamma)$] is given by Eq.~(\ref{tau1}) 
with $D=D(\gamma_{\rm in},\gamma)$ 
the diffusion coefficient of Eq.~(\ref{D1}). The rate at which the 
diffusing particles are exuded by the CB is $r=\beta_{\rm in}/\tau$. 

The rate of momentum loss per unit surface on a CB is proportional
to the average momentum $\sim m_p\,c\,\langle \gamma_{\rm in}\rangle$
of the particles exiting at time, LF and radius $t$, 
$\gamma(t)$ and $R_{_{\rm CB}}(\gamma)$, to wit: 
\begin{eqnarray}
P_{\rm out}&=&{m_p\,c\over 4\pi\,R_{_{\rm CB}}^2}
\int_{\gamma}^{\gamma_0}\,\beta_{\rm in}\,
\gamma_{\rm in} \,{1\over \tau}\;dn_{\rm in}(\gamma_{\rm in})\nonumber\\
&\propto& {M_0\,\gamma_0\over 4\pi\,R_{_{\rm CB}}^4\,\gamma}
\left({A\over Z}\right)^{\beta_{\rm conf}}\,F(\gamma),\nonumber\\
F(\gamma)&=&{1\over(\gamma-1)^{\beta_{\rm conf}}}\int_\gamma^{\gamma_0}
{d\gamma_{\rm in}\over (\beta_{\rm in}\,\gamma_{\rm  in})^{2-{\beta_{\rm conf}}}}\; ,
\label{Pout1}
\end{eqnarray}
where we have used $dn_{\rm in}$ as in Eq.~(\ref{gammadown}). Insert
this result into Eq.~(\ref{Newton}), with $M_{\rm CB}$ as in 
Eq.~(\ref{NRmass}), to obtain:
\begin{equation}
-R_{_{\rm CB}}^2\,\ddot R_{_{\rm CB}}\propto \gamma\,F(\gamma).
\label{differential}
\end{equation}
This equation, along with Eq.~(\ref{dtCBdgamma}), can be solved with various
initial conditions at $t=0$: $R_{_{\rm CB}}(0)\simeq 0$, 
$\beta_T\sim{\cal{O}}(1/\sqrt{3})$,
${\beta_{\rm conf}}\sim 0.5$, $\gamma_0\sim 10^3$. 
The results are very insensitive to reasonable variations
of these input values. An
example with the specified initial values is given in Fig.~\ref{RofGamma}
as the (red) continuous line. The coasting value of 
$R_{_{\rm CB}}(\gamma)$  is sensitive to the proportionality factors
in Eqs.~(\ref{NRmass},\ref{tau1},\ref{D1}), but its $\gamma$-dependence is not.
Only this last dependence plays a role in our study of CRs.

 The two solutions to 
Eqs.~(\ref{Newton},\ref{dtCBdgamma},\ref{differential})  shown in Fig.~\ref{RofGamma}
have very similar shapes. 
Neither shape is to be taken too `seriously', for the assumptions made
in deriving them are oversimplifications of a very complicated problem.
Moreover, the solutions we have presented are for a constant ISM
density $n_p$, an approximation that we know to be locally  incorrect, given
in particular the observed `bumps' in AG light curves 
\cite{DD,DDDSwift}. 

In our study of GRB AGs we have analysed several $\gamma$-dependences
of $R_{_{\rm CB}}(\gamma)$, including a constant radius. It is only the relatively
late AG that is sensitive to $R_{_{\rm CB}}(\gamma)$. For optical AGs, on which the
data were abundant for almost a decade, it was very difficult to decide on the
`best' $R_{_{\rm CB}}(\gamma)$, since the late AGs have contributions from the
GRB-associated SN and the host galaxy, and the corrections for absorption
are not negligible. None of this is the case for the recent X-ray AG data
of Swift. There, the `best' dependence \cite{DDDSwift} is that of 
\cite{AGoptical,AGradio}, given by Eq.~(\ref{best}).
Given its success in describing GRB data within the same CB model,
and for the sake of consistency, the above $R_{_{\rm CB}}$ 
dependence is the one we
adopt here, even though a best fit to CR data would result in a slightly
smaller power in Eq.~(\ref{best}). During the fast-rising part of 
$R_{_{\rm CB}}(\gamma)$
in Fig.~\ref{RofGamma}, Eq.~(\ref{best}) is very incorrect.
But the fractional CR production is small during this phase in which
the surface of the CB is also relatively small; we shall use Eq.~(\ref{best}) at
all values of $\gamma$. This makes the results of Sections \ref{elasticscatt}
and \ref{inelastscatt} simple and analytical.

A question arises in the `diffusive' case we have discussed, which did not
in the `fast elastic scattering' case of Eqs.~(\ref{Newton}) and ref.~\cite{AGoptical}:
if the intercepted ISM particles spend time diffusing within a CB, why do 
they not exert a pressure similar in magnitude and opposite in sign to that
of Eq.~(\ref{Pout})? Moreover, we contend that the CB's magnetic field
is in rough energy--density equipartition with the ISM particles it engulfs.
Also, why does the pressure of this field not contribute?
The short answer is that the CB may be `self-confining', that is, closer in a 
sense to a liquid or a solid than to a perfect gas or plasma. The longer answer
is the following:
a gas of magnetic dipoles, if polarized in a single direction, has a `positive'
contribution to its pressure from the repulsion between the dipoles. For the 
unpolarized case this effect vanishes on the average (it may even correspond
to an attraction: a collection of magnets allowed to coalesce at random
would form a bound state). A CB's MF is chaotic in its structure and
in the orientation of its coherent `cells', i.e.~`unpolarized', and
pressureless...~or even self-confining. There are low-energy
CRs confined to the MF lines of the Earth, spiralling North
to a higher-field position where they back up South, to reverse the process
periodically. These confined particles do not contribute a pressure on a
hypothetical surface enclosing their bound
trajectories. Once again, the high-energy
constituents of a CB may be similarly confined and, thus, `pressureless'.

\section{More priors not specific to the CB model}
\subsection{CR cross-talk between galaxies}
\label{cross-talk}

Meteorite records indicate that the CR flux on Earth has been steady for billions
of years \cite{Longair}, barring moderate fluctuations presumably 
due to the solar system crossing the spiral arms of the 
Galaxy \cite{Shaviv}. The confinement 
time of CRs to the Galaxy being much shorter than a billion years
at all energies, the production and escape of CRs is, to a good approximation, a 
steady-state phenomenon (we are neglecting here, for the sake of
a simpler discussion, CR interactions with anything but MFs). As a consequence,
the spectrum of CRs flowing out from the Galaxy has the shape of the source spectrum, $dF_s/dE$, as opposed to that of the locally observed spectrum
(the path lengths of CRs of different energy differ, the lower-energy ones 
cross our local neighbourhood more often, but eventually they 
escape the Galaxy at the same rate at which they are made).

The Galactic MFs are in rough energy equipartition with the CR
population, suggesting that the former are generated by the latter
\cite{Longair}. This ansatz can be successfully extended to the CRs and
MFs in galaxy clusters and in the intergalactic space
\cite{DDMF}. The CRs escaping a galaxy would thus be accompanied by a
magnetic-field `wind'. This wind should, to some extent, constitute a
Galactic `shield': it counteracts the income of CRs from other galaxies,
at energies below the ankle. The detailed energy--dependence of the effect of 
the shield is not crucial in our theory, as discussed in Section
\ref{versus1}.  Above the ankle, on the other hand, the flux
originating from other galaxies enters our Galaxy unhindered. Its shape is
that of the source spectrum, but for the tribulations of intergalactic travel,
which we discuss next.

When dealing with extragalactic CRs, it would be more adequate, contrary to
established custom, to refer to a `look-back' time rather than to a redshift or
distance to the source ---since the trajectories of CRs need not be straight---
some `extragalactic' CRs may even have originated in our own Galaxy and come
back to it after an extragalactic foray. In what follows, distances or redshifts
are to be understood as measures of look-back time.

 \subsection{Redshift effects on extragalactic CRs}
 \label{Redshift}
 
 The momentum --or, for relativistic energies, the energy-- of a CR emitted at 
 a redshift $z$ is degraded by a factor $1+z$ by the expansion of the Universe. 
 Assume, as is the case in our theory, that the energy dependence
 of the local source spectrum of CRs is the same at all times, and let
 $dF[{\rm EG}]/dE$ be the corresponding intergalactic flux. In our theory, as 
 well as in any other theory in which CRs are generated by SNe, the CR
 luminosity is proportional to the SN rate as a function of $z$. Since
 stars ending up as SNe have a very short life by cosmological 
 standards, the SN rate is proportional to the star formation rate 
 $R_{\rm SF}(z)$. The time--redshift relation is:
 \begin{eqnarray}
 {dt\over dz}&=&{1\over H_0}\,{1\over g(z)}\; ,\nonumber\\
 g(z)&\equiv& (1+z)\,\sqrt{\Omega_\Lambda+\Omega_M\,(1+z)^3}\; ,
 \label{Cosmotvsz}
 \end{eqnarray}
where, in  the current `standard' cosmology,
$H_0=100\, h$ km s$^{-1}$ Mpc$^{-1}$,
 $h\!\sim\!0.65$, $\Omega_\Lambda\!\sim\! 0.7$ and 
 $\Omega_M\simeq 1-\Omega_\Lambda$.
 
 The extragalactic flux currently impinging on the Galaxy has a spectral 
 distribution:
\begin{equation}
  {dF{\rm[EG]}\over dE}\propto
 \int_0^\infty {dF_s\over dE}\Big|_{E_z}\;
 {R_{\rm SF}(z)\over R_{\rm SF}(0)}\;{(1+z)\,dz\over g(z)},
 \label{z}
 \end{equation}
 where the $dF_s/dE$  is the source flux at $E_z=(1+z)\, E$,
`uncorrected' for the effect of Galactic confinement. 

  \subsection{The rate of supernova explosions}
  \label{rateSN}
 
In galaxies such as ours, the SN rate is approximately proportional
to the luminosity.  The measured SN rate in the local Universe~\cite{vdB}
is 2.8 y$^{-1}$, in a `fiducial sample' of 342 galaxies within the Virgo
circle, whose total B-band luminosity is $1.35\,h^{-2}\times
10^{12}\,L_\odot^B$, or $\sim 8.7\times 10^{-3}$ SN per year per
$10^{10}\, L^B_\odot$  
for $h=0.65$.  This ratio multiplied by 
the Galactic luminosity \cite{PvdB},
\begin{equation}
L_\star[{\rm MW}]=2.3\times10^{10}\;L_\odot=8.85\times10^{43}\;{\rm erg\, 
s^{-1}},
\label{MWlum}
\end{equation}
yields:
\begin{equation}
R_{\rm SN}\rm[MW]\!\approx\! 1/50 \;\rm y^{-1}.
\label{SNrate}
\end{equation}
The SN rate in the Milky Way, obtained  from the frequency and
spatial distribution of historical SNe and the measured galactic
extinction, is also approximately two per century \cite{vdB}.

We are also interested in the SN rate per unit volume in the Universe.
For $h=0.65$, the local-Universe luminosity density is estimated to be
\cite{Ellis}:
\begin{equation}
\rho_L\sim 1.2\times 10^8\,L_\odot\;{\rm Mpc}^{-3}.
\label{locallight}
\end{equation}
Multiplied by the measured  rate of SNe per luminosity 
in the local Universe,  
the average rate of SN explosions per unit volume in the current Universe 
is:
\begin{equation}
R_{\rm SN}\rm{[U]}\approx  10^{-4} \;{\rm Mpc}^{-3}\,y^{-1}.
\label{SNunivrate}
\end{equation}

\subsection{The star-formation rate}
\label{star}

\begin{figure}
\centering
 \epsfig{file=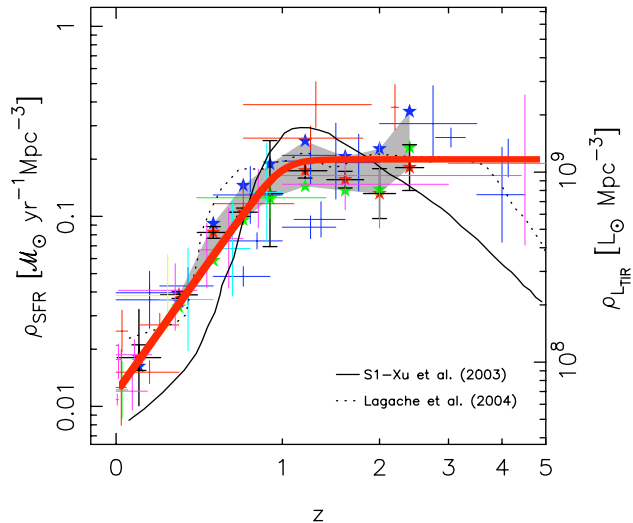,height=13.6cm,width=10cm}
\vspace*{-6cm}
 \caption{The observational data on the star formation rate 
as a function of redshift as compiled in \cite{SFRpaper}. The thick
(red) line is used in our calculations.}
\label{FigSFR}
\end{figure}

A compilation of the observational data
\cite{SFRpaper} on the function $R_{\rm SF}(z)$ is shown in Fig.~\ref{FigSFR}.
Its rough behaviour can be described as:
\begin{eqnarray}
R_{\rm SF}(z)\!&\simeq&\! R_{\rm SF}(0)\, (1+z)^4~~~~z\leq 1.2; 
\nonumber\\ 
&\simeq&\!  R_{\rm SF}(1.2)\,~~~~~~~~~~~1.4<z\leq 5. 
\label{SFR} 
\end{eqnarray}
In our calculations we have approximated $R_{\rm SF}(z)$ by the
function shown as a thick (red) line in Fig.~\ref{FigSFR}.
At $z>5$ the volume of the Universe ---or the function $1/g(z)$ in Eq.~(\ref{z})---
is relatively small and quenches the contribution of the corresponding
$R_{\rm SF}$, which is not known.

 \subsection{Spallation}
\label{Stable}

Spallation --the production of secondary stable 
and unstable CRs by interactions of primaries with the ISM--
 is a well studied and fairly well understood phenomenon 
\cite{Silberberg}.
We do not discuss the subject further, since our theory does not
significantly deviate from the standard lore on this subject.

\subsection{Pion photoproduction} 
\label{background}

\begin{figure}
\vspace{.5cm}
\centering
\epsfig{file=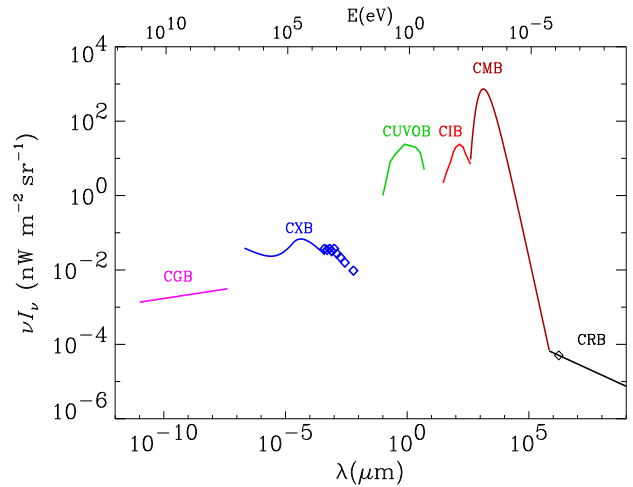,height=7cm,width=9cm}
\caption{Illustration of the observed spectral flux
of the CBR  in the radio (CRB), 
microwave (CMB), infrared (CIB), optical-ultraviolet (CUVOB), X-ray 
(CXB) and $\gamma$-ray (CGB) bands \cite{HD}.}
\label{IlCBR}
\end{figure}

The intergalactic space is permeated by very low-density ionized gas, 
MFs, photons 
and neutrinos, and perhaps by other relics from the Big Bang and
stellar evolution. The various `bands' of the flux of photons of
the CBR are
shown in  Fig~\ref{IlCBR} \cite{HD}. The intergalactic medium is  
extremely transparent to CRs, except at very 
high energies. Greisen and  Zatsepin \& Kuzmin (GZK) were
first to point out that the interactions of CR nuclei with the abundant 
but soft photons of the CMB  
would deplete the CR flux at energies above the pion-production threshold
\cite{GZK}. For 
nuclei of atomic number $A$ and energy: 
\begin{equation}
E_{_{\rm GZK}}(A)\sim A\times 10^{20}\;\rm eV,
\label{GZK}
\end{equation}
the energy loss length \cite{Stanev} on the CBR is about half the size
of the visible Universe, decreasing exponentially at higher energies.

\begin{figure}
\centering
\epsfig{file=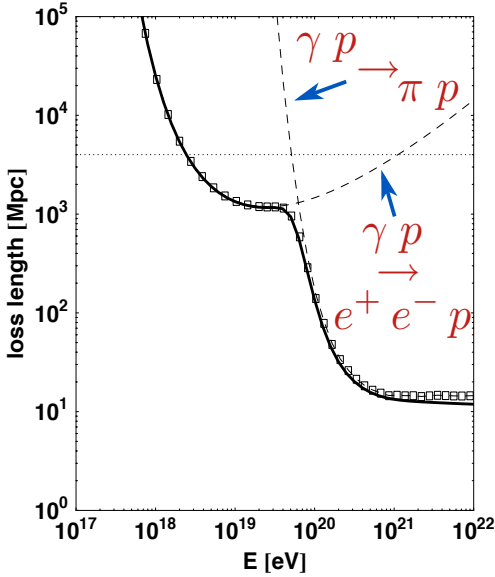,height=8.2cm,width=6.6cm}
\caption{Solid line: loss length for pion and $e^+\,e^-$
   photoproduction for CR protons on the CMB. The dashed
   lines are the separate contributions of the two processes. The
   dotted line shows the loss length for redshift losses \cite{Stanev}.}
\label{PhDisGZK}
\end{figure}

The GZK effect introduces a sharp cutoff on the flux of UHECRs
originating at large look-back times, as shown in Fig.~\ref{PhDisGZK}. To an
approximation sufficiently good for our purposes, we shall parametrize the
effect of the GZK cutoff by a probability for the overall time-integrated
flux of extragalactic nuclei to reach our Galaxy:
\begin{equation}
P_{\rm GZK}(E,A)=\exp \left[-\,{E\over E_{_{\rm GZK}}(A)}\right]\, .
\label{GZK2}
\end{equation}

\subsection{Pair production}
\label{pair}

The production of $e^+e^-$ pairs in the interactions of extragalactic CR nuclei 
with the CBR also results in a reduction of the momentum of the former.
The LF threshold for Bethe--Heitler pair production by CR nuclei on a CBR photon
of energy $\epsilon_\gamma$ is almost the same for all $A$:
\begin{equation}
\gamma_{\rm pair}
\geq {m_e\, c^2\over \epsilon_\gamma}\, \left [1+{m_e \over A\, 
m_p}\right].
\label{ppthrd}
\end {equation}
The most abundant background radiation is the CBR, for which
$\langle\epsilon_\gamma\rangle\simeq 0.63$ meV, and
$\gamma_{\rm pair}\approx  8.1\times 10^8$, 
corresponding to $E\approx 7.6\times 10^{17}$ eV 
for protons. The cross section abruptly increases to $\sim 3/4$ of
its constant high-energy value in a decade of energy and, not surprisingly,
detailed calculations \cite{Stanev} show that proton
attenuation due to pair production on the CMB  becomes important for 
energies around the ankle, at $3\times 10^{18}$ eV, as can be seen in
Fig.~\ref{PhDisGZK}.

De Marco and Stanev \cite{Stanev} have made a systematic study of the
effect of pair production and pion production  
on an extragalactic UHECR flux with various power source
spectra $\propto E^{-\beta}$, including our predicted $\beta\simeq 2.5$.
Their results can be very well reproduced by a multiplicative combination 
of the GZK cutoff of Eqs.~(\ref{GZK}), (\ref{GZK2}) and a factor describing
pair production:
\begin{eqnarray}
&&\!\!\!\!\!P_{\rm pair}(E)= 1-0.55\,
\exp\left\{-\, { \left[\log_{10}(E)-\log_{10}(E_{\rm pair})\right]^2
\over \alpha}\right\} \nonumber\\
&&E_{\rm pair}= 8\times 10^9, \;{\rm GeV}\;\;\;\;\;\; \alpha=1.4,
\label{PairProd}
\end{eqnarray}
which we adopt as our description of the effect of pair production
on the extragalactic proton flux.

The energy loss per pair-producing collision is  
${\cal{O}} (2\, \gamma m_e\, c^2)$ and the fractional energy loss is
${\cal{O}} (2\, m_e/A\, m_p)$, i.e. $A$ times smaller for a nucleus
than that for a proton. Since the pair-production cross 
section is proportional to $Z^2,$ the energy-loss rate for nuclei
is larger by a factor $Z^2/A$ than for protons. But, as we 
shall see in the next Section, photo-dissociation has a much larger
effect on the fate of extragalactic nuclei than pair production has.

\subsection{Photo-dissociation of nuclei}
\label{pd}

\begin{figure}
\centering
 \epsfig{file=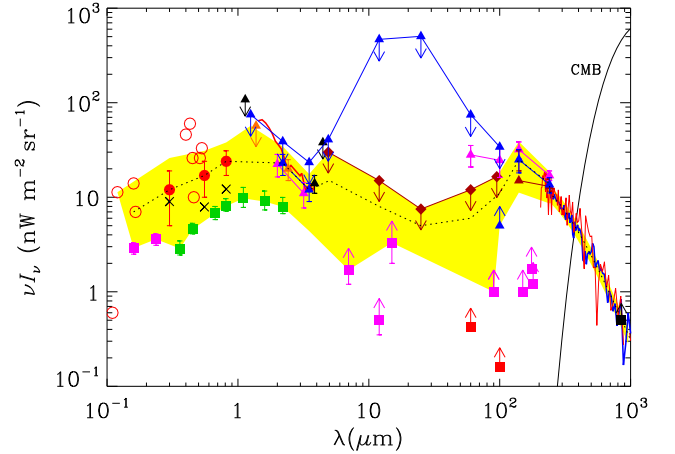,height=6.5cm,width=9cm}
 \caption{Summary of extragalactic background radiation measurements
in the ultraviolet (UV), visible, infrared and far infrared (FIR) 
\cite{HD}. Square symbols and X's are lower limits obtained by integrating
resolved sources. Diamonds and triangles are, respectively, $1\sigma$ and
$2\sigma$ upper limits obtained from fluctuation measurements. All other
symbols show absolute background measurements. The shaded region
represents the current uncertainty range of the CBR and the dotted line
the present best estimate. }
\label{UVOIR}
\end{figure}

The main mechanism of energy loss by UHECR nuclei is photo-dissociation
\cite{PSB} in collisions with the photons of the CBR. The frequency interval
most relevant to this process extends from the UV to the 
far infrared (FIR); the corresponding
observations \cite{HD} are summarized in Fig.~\ref{UVOIR}.
Let $dn_\gamma/d\epsilon$ be the number of CBR photons per unit energy
$\epsilon$. Its relation to $\nu\, I_\nu$, plotted in Figs.~\ref{IlCBR},\ref{UVOIR}, is:
\begin{equation}
\epsilon^2\,  {dn_\gamma\over d\epsilon}\,[{\rm  eV\, cm^{-3}}] 
= 2.62 \times 10^{-4}\,
\nu\, I_\nu\,[{\rm nW\, m^{-2}\, sr^{-1}}].
\label{CBRflux}
\end{equation}

To introduce some relevant quantities, it is convenient to discuss first
a steady-state Universe identical to the current one, and having existed
for a Hubble time (in the current cosmology, the Hubble time,
$\tau_{_{\rm H}}= 9.8\times 10^9\, h^{-1}=15$ Gy, coincides to a very good approximation
with the age of the Universe: $t_{_{\rm U}}\approx 0.96\, \tau_{_{\rm H}}$).
The photo-disintegration rate \cite{PSB} in the current-Universe's `rest' frame 
(in which the CMB is most isotropic) is:
\begin{equation}
R_0={c\over2}\,\int_{\epsilon_{\rm th}\over 2\gamma}^{\infty}
{d\epsilon \over \gamma^2\, \epsilon^2}\,
    {dn_\gamma\over d\epsilon} \int_{\epsilon_{\rm th}}^{2\gamma\epsilon} d\epsilon'\,
\epsilon'\sigma(\epsilon'),
\label{rateeq}
\end{equation}
where $\sigma$ is the total cross section, summed over the 
various nuclear break-up processes and $\epsilon_{\rm th}$ is the
reaction's energy threshold.
At low energies the cross section is dominated by the
giant dipole resonance which, in the nuclear rest frame, peaks at photon energies
ranging from 10 to 30 MeV and has a width of ${\cal{O}}(10)$ MeV. The 
cross section obeys the approximate Thomas--Reiche--Kuhn sum rule:
\begin{equation}
\int_{\epsilon_{\rm th}}^{\infty}\sigma(\epsilon)\, d\epsilon =
{2\, \pi^2\, e^2\, \hbar\over m_p\, c}\, {N\, Z\over A}\approx
 15\, A\, {\rm mb\,\, MeV}, 
\label{TRK}
\end{equation}
where $N=A-Z$ is the neutron number, 
$\epsilon$ is the photon energy in the
rest system of the nucleus and, in the last equality, $N\,Z/A$ 
has been approximated by $A/4$. Upon substitution of the  sum rule
in Eq.~(\ref{rateeq}) we obtain: 
\begin{equation}
R_0\approx {4.5\times 10^{-3}\, A\, {\rm cm^3\, eV^2\, s^{-1}}  \over 
\gamma^2}\, 
    \int_{\epsilon_\gamma}^\infty {d\epsilon\over \epsilon^2}\, 
     {dn_\gamma\over d\epsilon},
\label{rateeq1}
\end{equation}
where we have roughly approximated $\sigma$ by a peak at 15 MeV, so that 
$\epsilon_\gamma\approx 1.5\times 10^7\;{\rm eV}/\gamma$. 
We have explicitly checked that this is a good approximation,
and so is the neglect of the fact that $\epsilon^2\, dn_\gamma/d\epsilon$
drops abruptly at frequencies beyond the UV.

The spectral energy density is
roughly constant in the UV to FIR interval and has 
a value  $\epsilon^2\, dn_\gamma/d\epsilon \sim
6\times 10^{-3}\, {\rm eV\, cm^{-3}}$,  as can be seen in Fig.~\ref{UVOIR}.
Consequently, Eq.~(\ref{rateeq}) yields a nuclear photo-dissociation rate:
\begin{equation} 
R_0\approx 2.7\times 10^{-27}\, \gamma\,
A\, {\rm s^{-1}}, 
\label{Rfinal}
\end{equation}
The current mean photo-dissociation time coincides with $\tau_{_{\rm H}}$
when the energy of the CRs, independently of their $A$-value, is:
\begin{equation}
E_{\rm PhD}\simeq 7\times 10^{17}\;\,\rm eV,
\label{PhDisCut}
\end{equation}
or with $\sim \tau_{_{\rm H}}/5$ when $E=E[{\rm ankle}].$ In the fake steady-state
Universe, the flux of CRs generated at look-back time $t$ is depleted as:
\begin{equation}
-{dn\over n} = R_0\,dt\approx {E\over E_{\rm PhD}}\,{dt\over \tau_H}\; ,
\label{deplete1}
\end{equation}
and the corresponding attenuation is 
$a(E,t)=\exp[-(E/E_{\rm PhD})\,(t/\tau_{_{\rm H}})]$.

To deal with our actual Universe we must paraphrase the above calculation
for an expanding Universe in which the CR production rate and the 
amount of (accumulating, non-primordial) background radiation vary with time.
The spectral index of UHECRs above the ankle, without any type of 
attenuation, is predicted to be $\beta\simeq 2.5$.
 Attenuated only by redshift and expansion in the standard 
Universe, such a flux maintains its spectral index, but is reduced in magnitude
by an overall factor:
\begin{equation}
I_{\rm SU}\simeq H_0\,\int {R_{\rm SF}(z)\over R_{\rm SF}(0)}\, {dt\over 
dz}\,
{dz\over (1+z)^{\beta-1}}\approx 3.15.
\label{COSInt}
\end{equation}
The time evolution of the spectral energy density in the UV to FIR range is:
\begin{equation}
\epsilon^2\,{dn_\gamma(z)\over d\epsilon}=(1+z)^3\, 
\int_z^\infty \epsilon'\, \mathcal{L}_\epsilon(\epsilon',z'){dt\over 
dz'}\, {dz'\over 1+z'}\, ,
\label{CBRevol}
\end{equation}
where $\mathcal{L}_\epsilon$ is the spectral luminosity density 
in a comoving unit volume at $\epsilon'=(1+z')\, \epsilon$.
In the approximation of no spectral evolution of the luminosity sources and
an evolution of their numbers described by the star formation rate, 
Eq.~(\ref{CBRevol}) can be rewritten as:
\begin{eqnarray}
\epsilon^2\,{dn_\gamma(z)\over d\epsilon}&\approx& (1+z)^3\,
\epsilon^2\,{dn_\gamma(0)\over d\epsilon}\;{F(z)\over F(0)},
\nonumber\\
F(z)&\equiv&\int_z^\infty {R_{\rm SF}(z')\over R_{\rm SF}(0)}\, 
 {dt\over dz'}\, {dz'\over (1+z')}\, .
\label{CBRevole}
\end{eqnarray} 

In analogy with Eq.~(\ref{deplete1}), the attenuation of the flux of CRs observed at 
energy $E$ and generated at redshift $z$ satisfies: 
\begin{eqnarray}
-{dn\over n} &=& {E\over E_{\rm PhD}}\, H(z),\nonumber\\
H(z)&\equiv&(1+z)^4\,{F(z)\over F(0)}\,
{dz\over g(z)}\; ,
\label{atteneq}
\end{eqnarray}
where $F(z)$ is as in Eq.~(\ref{CBRevole}) for the past background radiation,
 and $g(z)$ is the function involved in the time--redshift relation,
 Eq.~(\ref{z}). The corresponding CR attenuation factor is:
\begin{equation}
a(z,E)\simeq \exp\left[-\,{ E\over E_{\rm PhD}}
\,\int_0^z H(z')\,dz'\right]\, .
\label{atteneq1}
\end{equation}
Attenuated by this extra photo-dissociation factor, the reduction factor of 
Eq.~(\ref{COSInt}) becomes:
\begin{equation}
I_{\rm PhD}[E] =H_0\,\int a(z,E)\, 
      {R_{\rm SF}(z)\over R_{\rm SF}(0)}\, {dt\over dz}\, 
      {dz\over (1+z)^{\beta-1}}.  
\label{Atten}
\end{equation}
The attenuation of the flux of UHECR nuclei due exclusively
to photo-dissociation is given by the $A$-independent
ratio ${A}_{\rm PhD}(E)=I_{\rm PhD}[E] /I_{\rm SU}$. The actual 
result of the calculation
of Eqs.~(\ref{COSInt}) to (\ref{Atten}) is well described by:
\begin{equation}
{A}_{\rm PhD}(E)\approx  {1\over\sqrt{ 1+(I_{\rm SU}\,E/ E_{\rm PhD})^2}}\, .
\label{PhDisCut1}
\end{equation}
This result is affected by the uncertainty in the current and 
past UV, visible and infrared CBR, but is otherwise a sufficiently
good approximation for our purposes.

Photo-dissociation is a multiple-step process in which the debris 
eventually end up as protons and neutrons which $\beta$-decay to 
protons. Their individual energies are 
$\simeq 1/A$, the energy of the parent nucleus whose CR abundance 
has been enhanced by a factor $Z^{\beta_{\rm conf}}$ by CB acceleration.  
Consequently, a complete photo-dissociation 
of the UHECR nuclei would increase the proton flux by 
a factor $\Sigma_{Z} A[Z]\, X_{\rm SB}[Z]\, Z^{\beta_{\rm conf}}\simeq 1.6$.

At CR energies of order $E_{\rm PhD}$, photo-dissociation is not
complete. The average reduction of the parent-nucleus atomic mass
in a single photo-dissociation process at the relevant laboratory
energies is observed to be $\Delta A\simeq 1.2$
for He, $\Delta A\simeq 3.6$ for the CNO group and, $\Delta A\simeq 3.7$
for elements ranging from Na to Fe.
Consequently, He is efficiently photo-dissociated in a couple of steps
at energies above $E_{\rm PhD}$, and we have simply treated the fraction
of the He flux that is photo-dissociated as an addition to the proton flux:
\begin{equation}
\Delta F^{ p}_{[{\rm He}\to p]}(E)\,dE^{ p}=
F^{\rm He}_{\rm PhD}(4\,E)\,dE^{\rm He},
\label{Hetop}
\end{equation}
where the $dE$ factors are reminders of the fact that it is baryon number 
which is conserved. Similarly, in $n(A)$ photo-dissociations, the heavier 
elements have their flux reduced by a factor
\begin{equation}
r\sim
\left[{A-n(A)\,\Delta A\over A}\right]^{\beta - 1}.
\label{r}
\end{equation}
Only traces of relatively heavy fragments remain in the UHECR flux,
since their ab-initio relative abundances are small. We have 
simply described the photo-dissociation of the corresponding primary fluxes 
by the substitution $E_{\rm PhD}\rightarrow n(A)\,E_{\rm PhD}$ in
Eq.~(\ref{PhDisCut1}), with $n(A)=2$ for $A<8$,  increasing linearly
thereafter up to $n(A)=15$ at $A=56$.
These values of $n(A)$ are estimates of the number of photo-dissociations
required for the value of $r$ in Eq.~(\ref{r}) to represent a significant reduction
(scaling up $E_{\rm PhD}$ by a factor $n(A)$ is tantamount to reducing
the nuclear mean free path by the same amount).  

\section{Jets in Astrophysics}
\label{Jets}

A look at the sky, or a more economical one at the {\it web}, results in the
realization that jets are emitted by many astrophysical systems: forming
stars, binary stars, planetary nebulae, pulsars, radio galaxies, quasars,
and microquasars. High-resolution radio, optical and X-ray observations 
indicate that (apparently `superluminal') relativistic jets 
are fired by quasars, microquasars and SN explosions. These jets consist 
of a sequence of plasmoids (CBs) of
ordinary matter whose initial expansion in their rest frame
---presumably at a speed close to that of sound in a relativistic plasma--- 
stops shortly after launch.
One impressive instance~\cite{WYS} is that of the quasar
Pictor A, shown in Fig.~\ref{Pictor}. {\it Somehow}, the active galactic
nucleus of this object is discontinuously spitting {\it something} that
does not appear to expand sideways before it stops and blows up, having by
then travelled for a distance of several times the visible radius of a
galaxy such as ours. Many such systems have been observed. They are very
relativistic: the LFs  of their ejecta are
typically of ${\cal{O}}(10)$. The mechanism responsible for these mighty
ejections ---suspected to be due to episodes of violent accretion into a
very massive black hole--- is not understood.

\begin{figure}
\vspace{-5cm}
\centering
 \epsfig{file=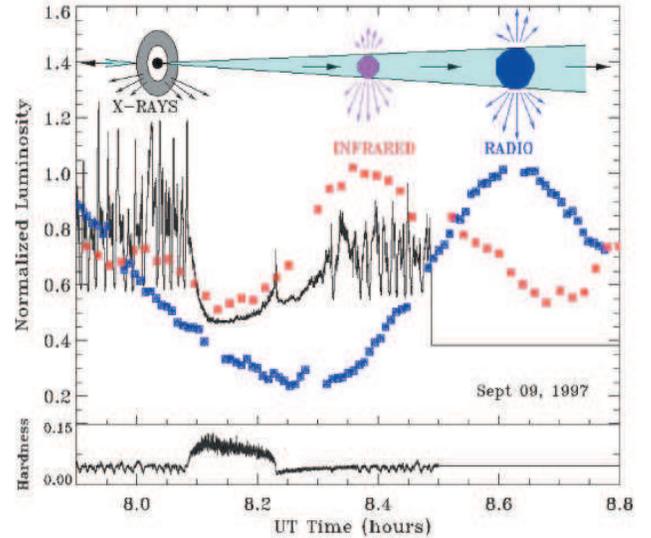,height=14cm,width=9cm}
\vspace*{-2cm}
 \caption{Illustration of one side of a CB-shooting event in GRS 
 1925+105~\cite{Felix}. The X-ray emission ---attributed to an unstable
accretion disk--- temporarily decreases when the CBs are ejected \cite{DMR}. 
How part of the accreting material ends up ejected along the
system's axis is not understood.}
\label{FigFelix}
\end{figure}

Microquasars are binary systems
consisting of a stellar-mass black hole or a neutron star 
accreting mass from a normal-star companion and displaying
in miniature some of the main properties of quasars.  The matter lost from
the companion temporarily stations in a fast-spinning accretion disk, heated 
to millions of degrees. Aperiodically, a fraction of the disk falls towards
the compact object, and a fraction of it is axially emitted as a pair of 
relativistic CBs. Some dozen microquasars have been found in the Milky Way. 
The first, GRS 1915+105, 40,000 light-years away in
Aquila, was discovered in 1994 by the GRANAT X-ray satellite. It
consists of a main-sequence star orbiting around the heaviest
stellar black hole found to date, with a mass of $14\, M_\odot$. Already
in its year of discovery, it was observed to shoot out, a few times a
year, aperiodically, pairs of CBs with one-third the mass of the Moon 
and a $v\sim 0.92\, c$~\cite{Felix}. Some properties of one
of its CB-firing events are shown in Fig.~\ref{FigFelix}.
                                                                                
The continuous collision of CBs with the ISM produces in the latter
turbulent magnetic fields, which gather and scatter the ionized ISM
particles on their path.  The collisions result
 in bremstrahlung, line emission and
synchrotron radiation. Atomic lines from many elements have been 
observed in the optical~\cite{Eiken} and X-ray \cite{Kotani} emission 
from the CBs of microquasar SS 433, indicating that the jetted ejecta are 
in this case ---and reasonably in all cases--- made of ordinary matter, and 
not of some fancier substance such as $e^+e^-$ pairs.
                                                                                
\subsection{The motion of CBs}
\label{CBmotion}

In analogy with the `hot-spots' of quasars such as Pictor A, two
infrared and radio sources appear symmetrically located with
respect to GRS 1915+105, aligned with the position angle of the
relativistic ejecta~\cite{RM}. They were presumably created by the
plasmoids from GRS 1915+105, which finally stop and blow up nearly
$60$ pc away from their ejection point. Even the mildly
relativistic CBs from microquasars appear to travel a very long distance 
until their gradual deceleration in their interactions with the ISM finally
stops them. In the case of  XTE~J1748--288 an originally ballistic jet                                                                     
was observed to stop over the course of a few
weeks, presumably following a collision with denser environmental material
\cite{kot00}. 
 
\begin{figure}
\vspace{10pt}
\epsfig{figure=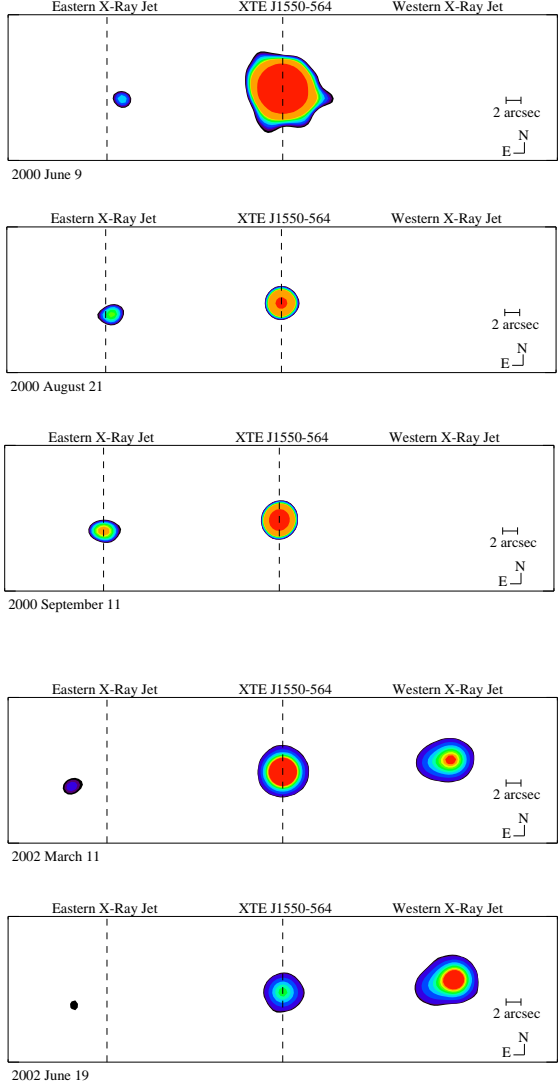,width=8.2cm}
\caption{Five {\it Chandra} 0.3--8~keV images showing the microquasar
XTE~J1550--564 and the
evolution of the eastern and western X-ray emitting CBs  between June 
2000 and June
2002. The observations are ordered chronologically from top to bottom, and
each image is labelled with the observation date. The dashed lines mark the
positions of XTE J1550-564 and the eastern X-ray jet on 11 September 2000
 \cite{Corbel}.}
\label{XTE}
\end{figure}

\begin{figure}
\epsfig{figure=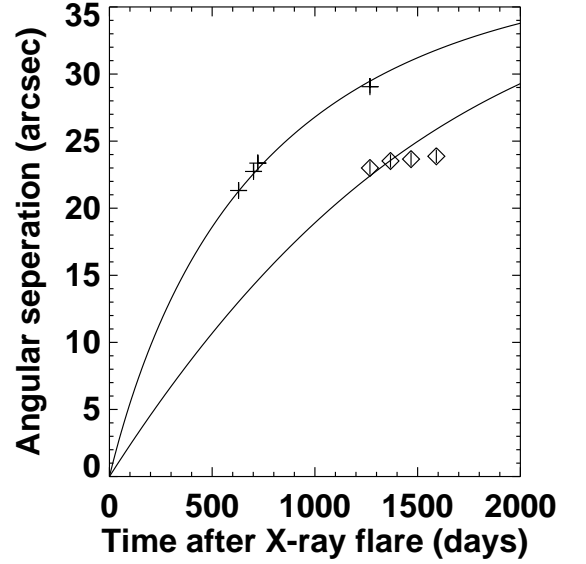,width=8.4cm}
\caption{The decelerating motion of the CBs of the $\mu$-quasar
XTE~J1550--564 \cite{Corbel}.}
\label{Motion}
\end{figure}

 In one case, that of  XTE~J1550--564, the gradual deceleration of CBs    
 has been observed.  A major radio-flare took place in September 
1998. The resulting eastern CB was observed with {\it Chandra}
between June 2000 and June 2002 \cite{Corbel}, as shown in
Figs.~\ref{XTE} and \ref{Motion}. 
The emission from the western CB was first detected on 11
September 2000 after it flared up. Probably this CB
moved through a very low density region before it
encountered a denser region and flared up.

\subsection{CBs or conical jets?}
\label {conical}
                                                                                                                                       
The conical or trumpet-like radio and optical images of some astrophysical
jets are often interpreted as being produced by conical ejecta. 
Conical-like images are often produced by precession of the ejection axis 
around the rotation axis, as in the case of SS433 \cite{Margon}.

The conical flows that are produced along the motion of the decelerating 
CBs can also generate conical-looking trails.
Radiation and charged  particles (swept in ionized ISM particles),
which are emitted isotropically in the CBs' rest frame,
are beamed forward by its motion. 

Let primed quantities denote their values in the CB's rest frame and
unprimed quantities their corresponding values in the SN frame. The
angle $\theta'$ ---relative to the CBs' direction of motion--- 
of the particles emitted in the CBs' rest frame, and 
the corresponding angle $\theta$ in the SN's rest frame, 
are related through:
\begin{equation}
  \cos\theta' = {\cos\theta-\beta \over 1-\beta\, \cos\theta}\, .
\label{thetaprime}
\end{equation}
Applied to an isotropic distribution of emitted particles
in the CBs' rest frame, Eq.~(\ref{thetaprime}) results in a distribution
in the SN's frame:
\begin{equation}
{dn\over d\Omega}={dn\over d\Omega'}\, {d\cos\theta'\over d\cos\theta}
\approx {n\over 4\, \pi}\, \delta^2\approx {n\over 4\, \pi}\, \left[
{2\gamma\over 1+\gamma^2\, \theta^2}\right]^2\, ,
\label{dist}
\end{equation}
where the Doppler factor $\delta$ of the CB's radiation, as viewed from an
angle $\theta$, was approximated by:
\begin{equation}
\delta = {1\over \gamma\, (1-\beta\, \cos\theta)}\approx
{2\gamma\over 1+\gamma^2\, \theta^2}\; ,
\label{Doppler}
\end{equation}
which is well satisfied for CBs with 
$\gamma^2 \gg 1\, ,$  and  $\theta^2 \ll 1$.
The ISM particles, which are  isotropized elastically within the moving CB,
 are emitted with energy
$\langle\epsilon\rangle =\gamma\, \delta\, m\, c^2.$  Hence, the energy 
outflow from a  CB is:
\begin{equation}
{dE\over d\Omega}={n\, m\, c^2\, \gamma\,  \delta^3 \over 4\, \pi}\,
\approx {n \, \gamma\, m\, c^2\over 4\, 
\pi}\, \left[
{2\, \gamma\over 1+\gamma^2\, \theta^2}\right]^3\, .
\label{EPdist}
\end{equation}

The distribution of the energy outflow
from a decelerating CB is collimated into the narrow `beaming cone' 
of Eq.~(\ref{EPdist}), along
the direction of motion of the CB. 
Subsequently the emitted charged particles
are isotropized by the ambient MF
and slowly diffuse away from the CBs' trajectories.  Most of the
electrons' energy is radiated via synchrotron emission in the MFs and inverse
Compton scattering of CBR photons. This secondary radiation originates from 
outside the CB's original beaming cone, diminishing with distance to 
the CBs' trajectories:
it may look like a much wider cone or domain. The narrow geometry of the relativistic
ejecta reveals itself only in observations at much higher frequencies,
 the observed emission requiring much stronger MFs than
those present in the ISM and IGM: the MFs within the CBs.

In our theory of GRBs and CRs, our Galaxy and its halo are at
any point in time permeated by thousands of traveling CBs.
Why have they not been observed? The answer is simple,
and provided by Eq.~(\ref{EPdist}). Their radiation at all
wavelengths is tiny, except extremely close to their direction
of motion: $\delta^3$ decreases dramatically at angles
larger that $1/\gamma$, a few milliradians.

\section{More on the CB Model}
\label{TheCB}

The `cannon' of the CB model is analogous to the ones
responsible for the ejecta of quasars and microquasars.
{\it Long-duration} GRBs, for instance, are produced in
{\it ordinary core-collapse} SNe by jets of CBs, made of {\it
ordinary-matter plasma}, and travelling with high LFs,
$\gamma\sim{\cal{O}}(10^3)$. An accretion disk or torus is produced around
the newly born compact object, either by stellar material originally
close to the surface of the imploding core and left behind by the
explosion-generating outgoing shock, or by more distant stellar matter
falling back after its passage~\cite{ADR,GRB1,DD}. 
A CB is emitted, as observed in microquasars~\cite{Felix,DMR}, when part 
of the accretion disk falls abruptly onto the compact object 
\cite{GRB1,DD}.

\begin{figure}[]
\centering
\hspace{.5cm}
\hbox{
\epsfig{file=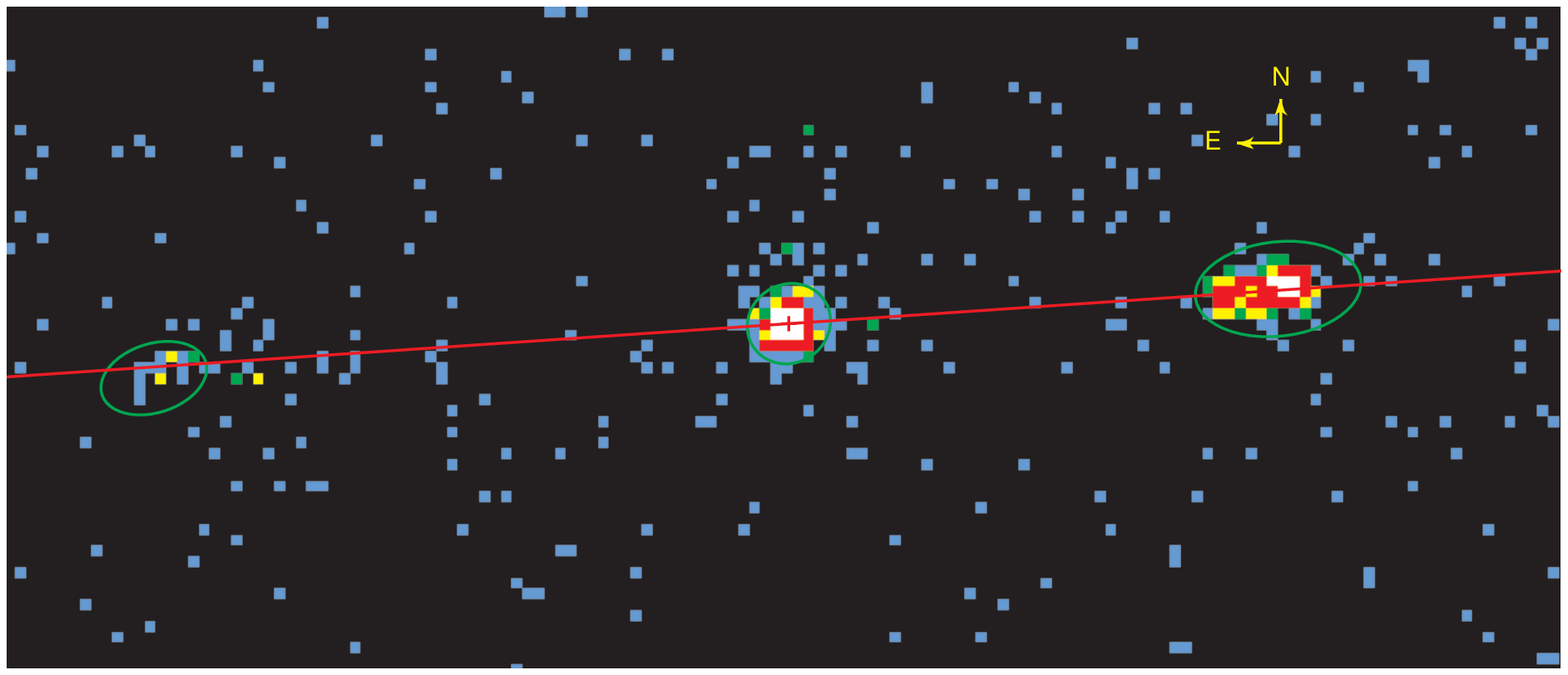,width=5.8cm,angle=90}
\epsfig{file=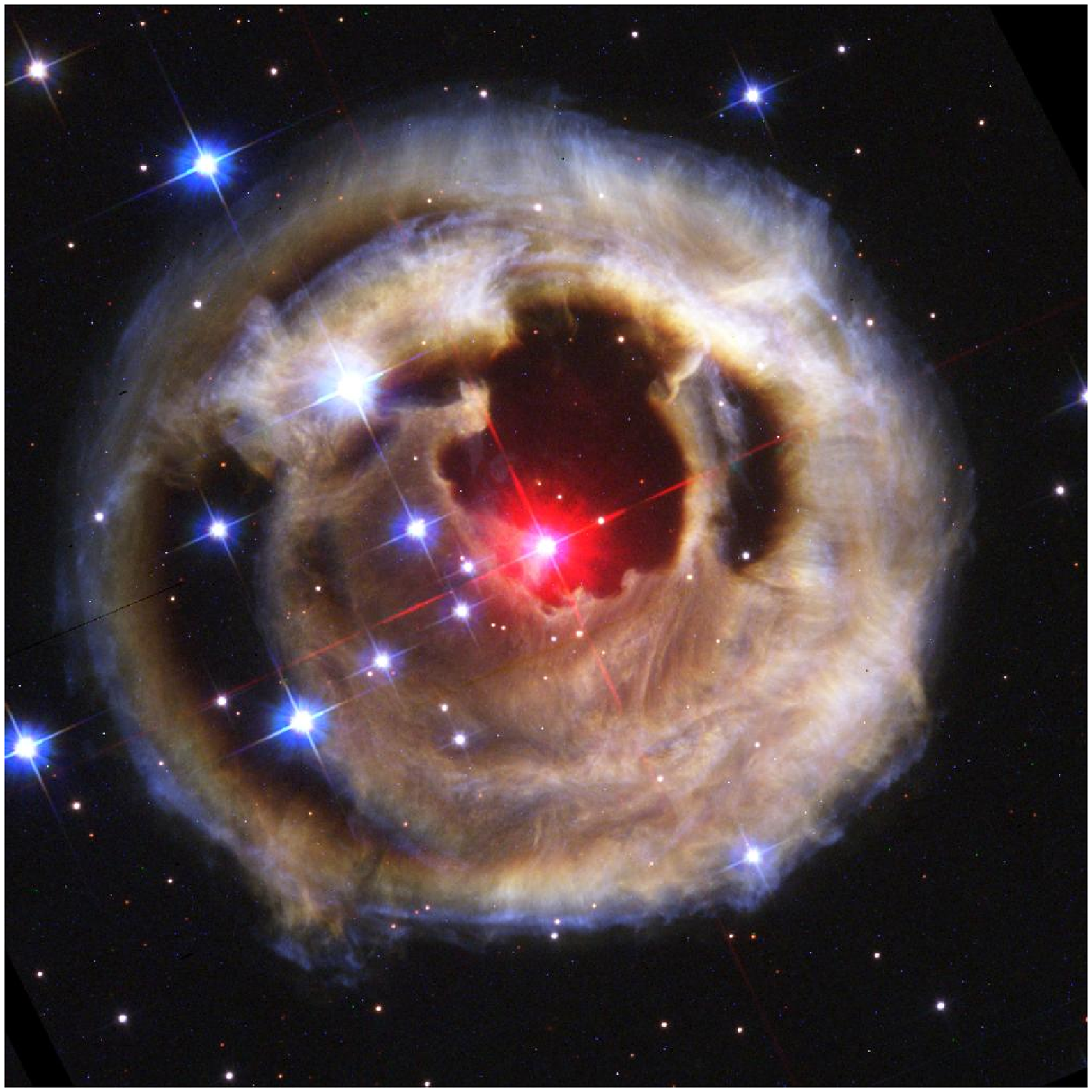, width=5.8cm}
}
\vspace*{8pt}
 \caption{Left: Two relativistic CBs emitted in opposite directions
 by the microquasar XTE J1550-564, seen in X-rays \cite{Corbel}. Right:
 HST picture from 28 October 2002 of the {\it glory}, or light echo,
 of the outburst of the red supergiant V838 Monocerotis
 in early January 2002 \cite{Bond}. The light echo was formed by 
 scattering off dust shells from previous ejections.}
 \label{CBGlory}
\end{figure}

Massive stars shed much of their matter in their late
life, in the form of stellar `winds'. Even before they die as SNe,
they undergo occasional explosions and rebrightenings, that
illuminate their semi-transparent `wind-fed' circumstellar material,
creating a light echo, or `glory'.
The example of the red supergiant V838 Monocerotis~\cite{Bond} 
 is shown in the right panel of Fig.~\ref{CBGlory}. As a SN explodes, it
also illuminates its surroundings, producing a scattered,
non-radially-directed {\it ambient light}
that permeates the semi-transparent circumburst material, previously 
ionized by the early extreme UV flash accompanying the explosion, or 
by the enhanced UV emission that precedes it.

The $\gamma$-rays of a GRB are produced by inverse Compton scattering 
of the ambient light permeating the vicinity of the exploding 
star by the electrons enclosed in the CB. 
To produce, in the CB model, a GRB pulse by ICS of ambient light,
it suffices to `superimpose the two halves' of 
Fig.~\ref{CBGlory} \cite{Corbel,Bond}, and to work
the result out in detail for the specific SN 
environment. The CBs electrons, comoving with it at a
LF $\gamma \sim {\cal{O}}(10^3)$, Compton up-scatter the ambient photons
of energy $E_i$ to energies of ${\cal{O}}(\gamma^2\,E_i)$,
while beaming them forward at angles of ${\cal{O}}(1/\gamma)$.  The
collimated GRB is seen by a distant observer only when the jet points
fairly precisely in her direction.

The time structure of GRBs ranges from a single pulse of $\gamma$-rays to
a complicated succession or superposition of many pulses.  A single pulse
is generated as a CB coasts through
the ambient light.  The timing sequence of the successive individual
pulses (or CBs) reflects the chaotic accretion process; its properties
are not predictable, but those of the single pulses are.

\begin{figure}[]
\epsfig{file=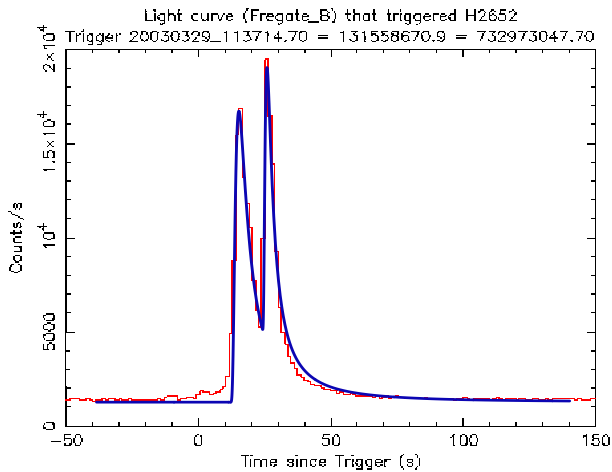, width=8cm}
\caption{The $\gamma$-ray number count $dN/dt$ of GRB 030329,
as measured by HETE II \cite{Vander}, showing two dominant 
pulses, or CB contributions. The continuous line is the CB-model's
description of the line shapes~\cite{DDD329a}.}
 \label{fig329NC}
\end{figure}

Each pulse lasts from a fraction of a second to tens of seconds. 
The two-pulse $\gamma$-ray number count as a function of time
for GRB 030329 is given, as an example, in Fig.~\ref{fig329NC},
which also shows the CB-model description of its
two pulses \cite{DDD329a}. The CB-model predicts the properties of 
XRFs and long-duration GRBs remarkably well \cite{DD}, as
summarized in Appendix  \ref{GRBXRF} and briefly updated in Section
\ref{Swift}.

Long-duration GRBs have `afterglows',  long term emissions that
are often observable at frequencies ranging from radio to X-rays, for
months after their $\gamma$-rays are seen. The rapid expansion of the CBs
stops shortly after ejection by their interaction with the ISM 
\cite{AGoptical}. As a CB pierces the ISM with a roughly constant radius, its
emission is
dominated by synchrotron radiation from swept-in ISM electrons, which
spiral in the CBs' enclosed MF. This picture yields an excellent
description of the entire AG phase and allows us to infer the 
parameters of CBs \cite{AGoptical,AGradio}, as discussed in Appendix
\ref{GRBAGs}. These parameters and their distributions, employed as
`priors', can be used to predict the properties of the prompt
$\gamma$-ray phase of GRBs \cite{DD}, much as we use them here to predict 
the properties of CRs.

\subsection{Do supernovae emit cannonballs?} 
\label{emit}
\begin{figure}
\centering
\epsfig{file=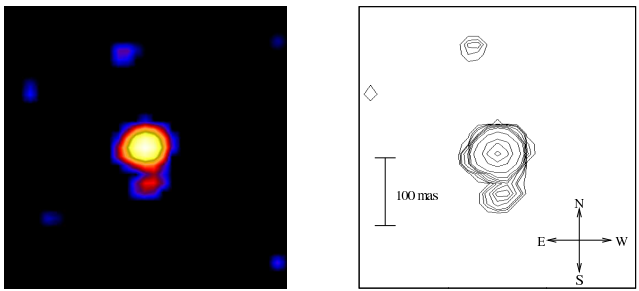, width=8.4cm}
\vspace*{8pt}
\caption{The two CBs emitted by SN1987A in opposite axial
directions \cite{NP}. The northern and southern
bright spots are compatible with being jets of CBs emitted at
the time of the SN explosion and travelling at a velocity equal,
within errors, to that of light. One of the {\it apparent} velocities is superluminal.
The corresponding GRBs were not pointing in our direction, which
may have been a blessing.}
\label{figCostas}
\end{figure}

Up to quite recently, there was only one
case in which the data were good enough to tell: SN1987A, the core-collapse
SN in the LMC, whose neutrino emission was detected \cite{TWOBANGS}. 
Speckle interferometry
measurements made 30 and 38 days after the explosion~\cite{NP} 
did show two relativistic CBs, 
emitted in opposite directions, as shown in Fig.~\ref{figCostas}. 
The apparent motion of the approaching CB is `superluminal':
it appears to have moved faster than light and further than the receding 
CB in the same time, even if their real speeds are comparable, 
$v\!\simeq\! c$.

Another resolved image of what appears like relativistic CBs
emitted in a SN explosion was obtained recently
with the Spitzer Space Telescope \cite{OK}.  Two FIR images 
of Cassiopeia A, the
youngest observed SNR in our Galaxy (about 325 years old) show
discrete compact structures at a distance of more than 20 arcmin from the
SNR, moving in opposite directions at roughly 
the speed of light. The trail of the CBs inside and outside the SNR is 
still visible in an X-ray image of Cassiopeia A obtained by Chandra \cite{UH}.

\begin{figure}  
\centering 
\vbox{\epsfig{file=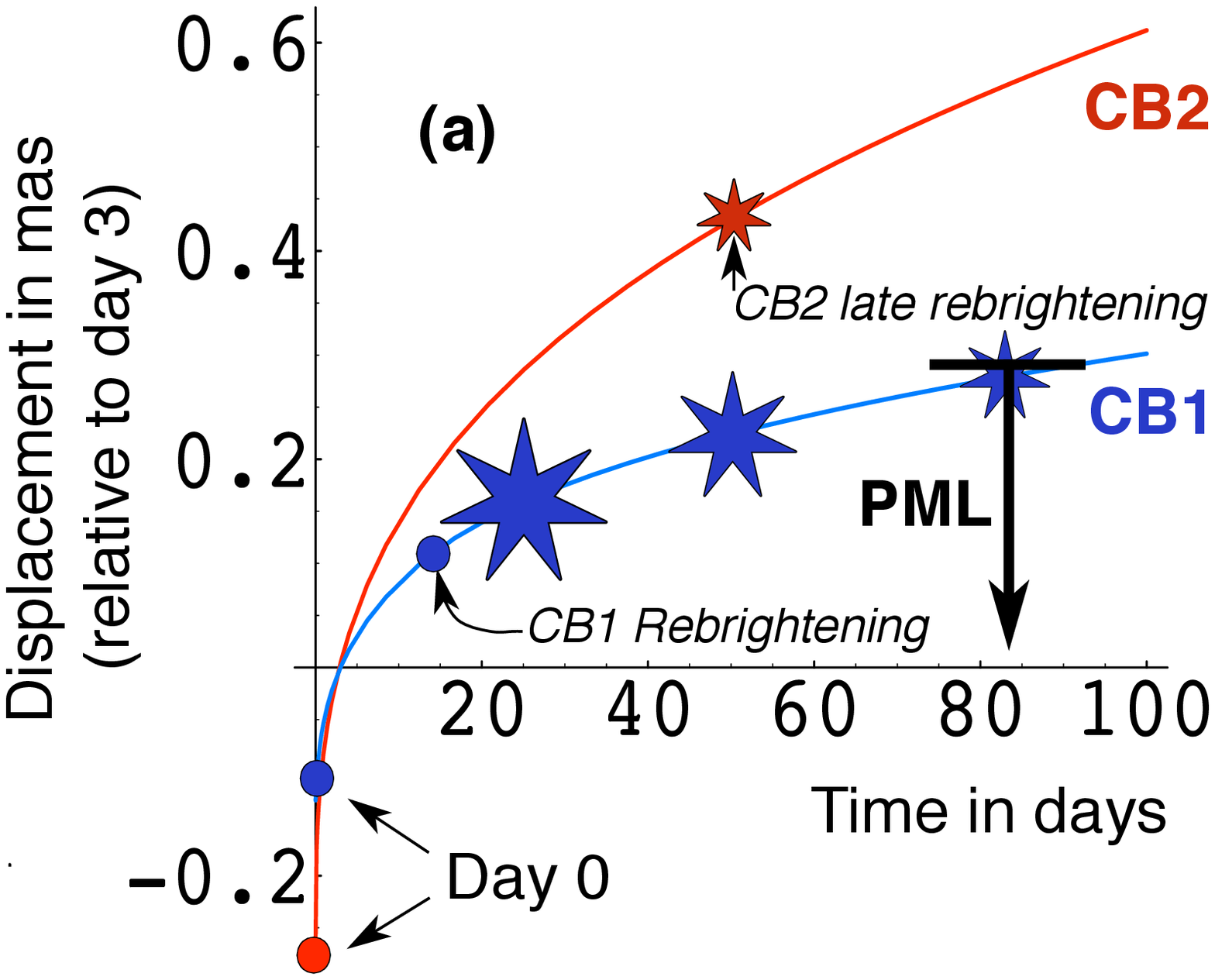, width=7cm}}
\vbox{\epsfig{file=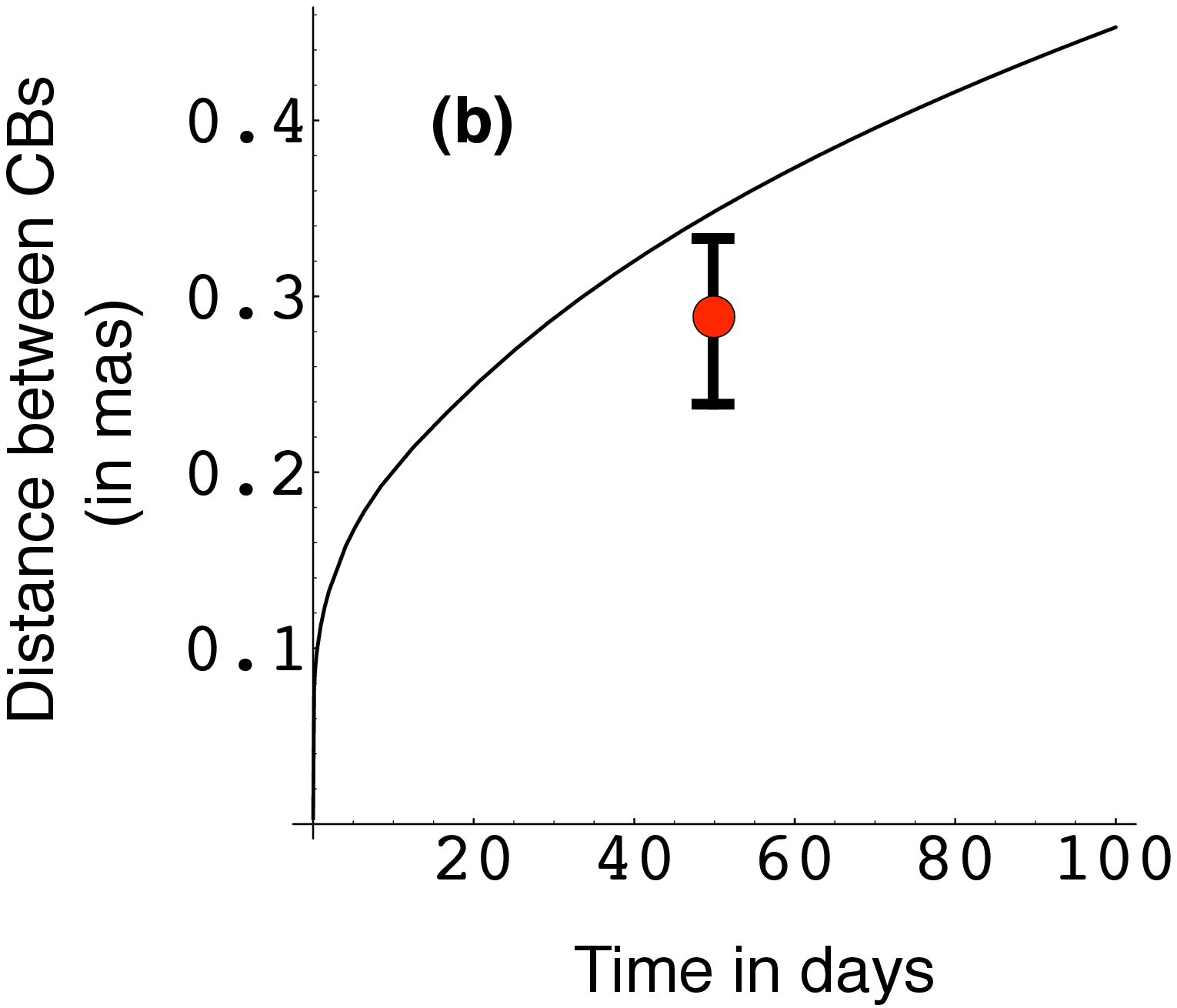, width=7cm}}
\caption{{(a)} The predicted angular displacement 
in the sky (in mas)  of the two CBs of GRB 030329,
as a function of observer's time from the 
first day of radio observations, day $\sim 3$. The positions at day 0, the
start-up time of the successive predicted rebrightenings of the slower CB1, 
 the observed time of the intense late
rebrightening of the faster CB2, as well as the fluences at 15.3 GHz on day 51
(70\% and 30\% of the total) are illustrated.
The proper motion limit (PML) of Taylor et al.~\cite{Taylor} is also
shown. {(b)} The expected angular distance between the two
CBs as a function of time, and its measurement at day 51~\cite{Taylor}.}
\label{tororo}
\end{figure}

Cosmic GRBs are usually too far away to provide resolved (radio) images of
CBs or their trails, except for relatively nearby GRBs such as GRB 980425, at
$z=0.0085$ \cite{Galama} and GRB030329, at $z=0.16$ \cite{Greiner}.  In the 
case of GRB 980425, the possibility was overlooked.  The situation concerning
GRB 030329 is debated.  The supernova explosion SN2003dh that produced 
GRB 030329 was first detected spectroscopically on day 10 after the GRB
\cite{SH}, as predicted by the CB model \cite{DDD329a}. The properties of its
complex AG are also understood \cite{DDD329b}. However, while we 
claim that the two CBs of this two-pulse GRB have been seen, in high-resolution
radio observations, moving apart at an apparently superluminal velocity, at
the predicted angular separation \cite{DDD329c}, as shown in
Fig.~\ref{tororo}, the authors of the corresponding radio observations claim
that their observed superluminal velocity does not agree with that
predicted \cite{Taylor}.

\subsection{Are GRBs made by SNe?} 
\label{GRBSN}

For long-duration GRBs, the answer is
affirmative~\cite{AGoptical,DD,AD2004}.  The first spectroscopic evidence 
for a GRB--SN
association came from the discovery of SN1998bw~\cite{Galama}, at redshift
$z\!=\! 0.0085$, within the directional error cone~\cite{Soffitta} towards GRB
980425. The time of the SN explosion was within $-\,2$ to $+\,0.7$ days of
the GRB~\cite{Iwamoto}.  The observations did not fit at all into the
framework of the `standard' {\it fireball} model.  This GRB's fluence
was `normal', but the total `equivalent isotropic' $\gamma$-ray energy
was $\sim\! 10^5$ times smaller than that of `classical' GRBs, with
$z={\cal{O}}(1)$, transported to $z\!=\! 0.0085$.

In the CB model the GRB emission is {\it very} narrowly forward-peaked,
with a characteristic opening angle $\sim\! 1/\gamma\!\sim\! 1$ mrad along the 
opposite jets of CBs. Distant GRBs are only detectable if the observer is within an
angle $\theta\!\sim\! 1/\gamma$ relative to the emission axis. 
GRB 980425 was seen unusually {\it far} off-axis, its close location
resulting in a `normal' fluence. Its associated SN was seen unusually
{\it close} to its axis of rotational symmetry. Both the GRB and the
SN were otherwise fairly `normal'~\cite{AGoptical,AGradio,DD}. 

The optical luminosity of a 1998bw-like SN 
peaks at $\sim 15\,(1+z)$ days. The SN light 
competes at that time and frequency with the AG of its 
GRB, and it is not always easily detectable.
 In the CB model, it makes sense {\it to test} whether long-duration GRBs 
are associated with a `standard torch' SN,  akin to SN1998bw, 
`transported' to their respective redshifts. The test works optimally:
{\it for all cases in which such a SN could be seen, it was seen (with varying 
degrees of significance)} and {\it for all cases in which the SN could not be seen,
it was not seen}; the redshift
establishing in practice the transition to SN undetectability was 
$z\! \sim\! 1.1$~\cite{AGoptical}. 

\begin{figure}[]
\centering
\vbox{
\epsfig{file=                  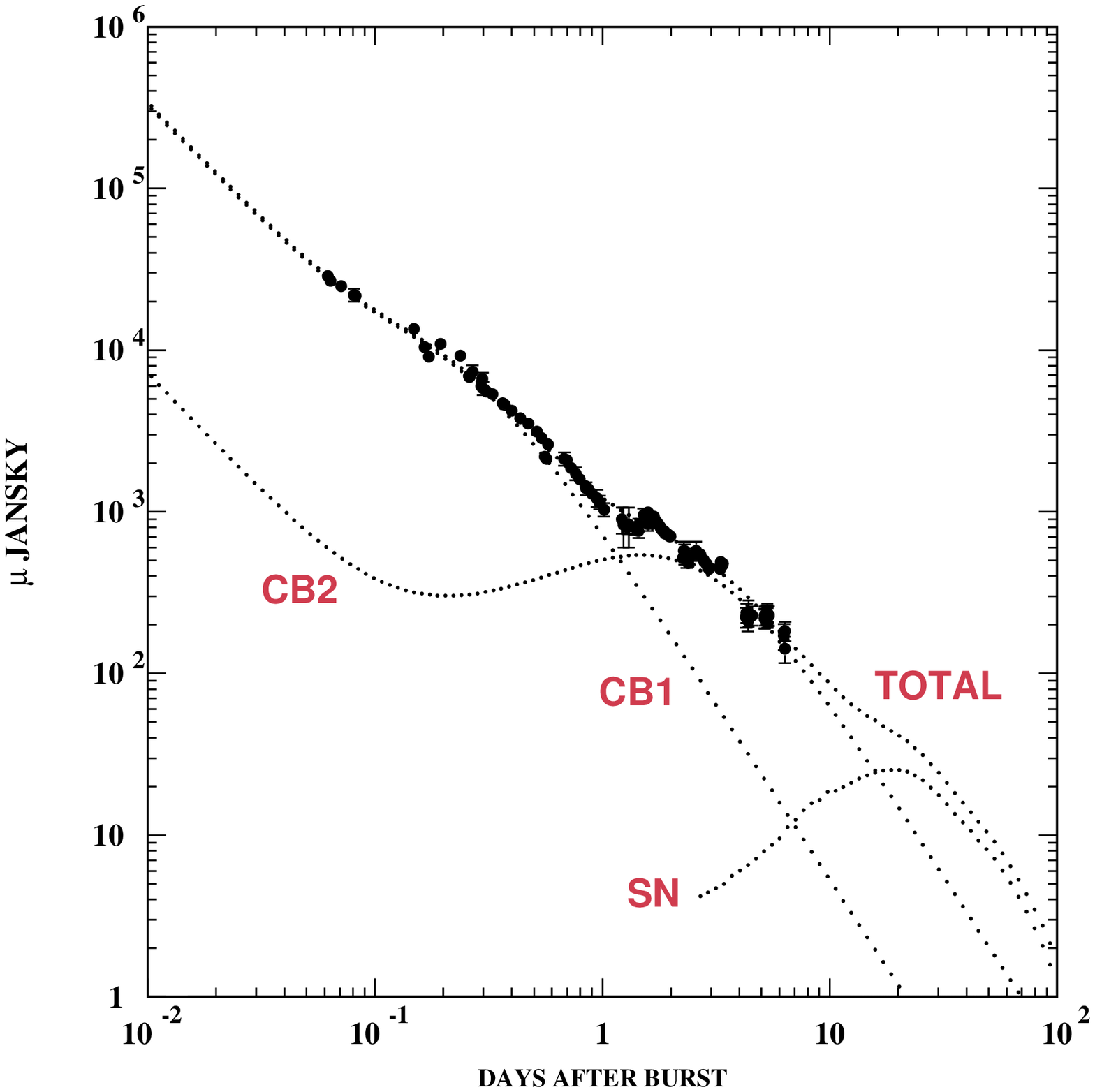,height=1\linewidth}
\epsfig{file=                  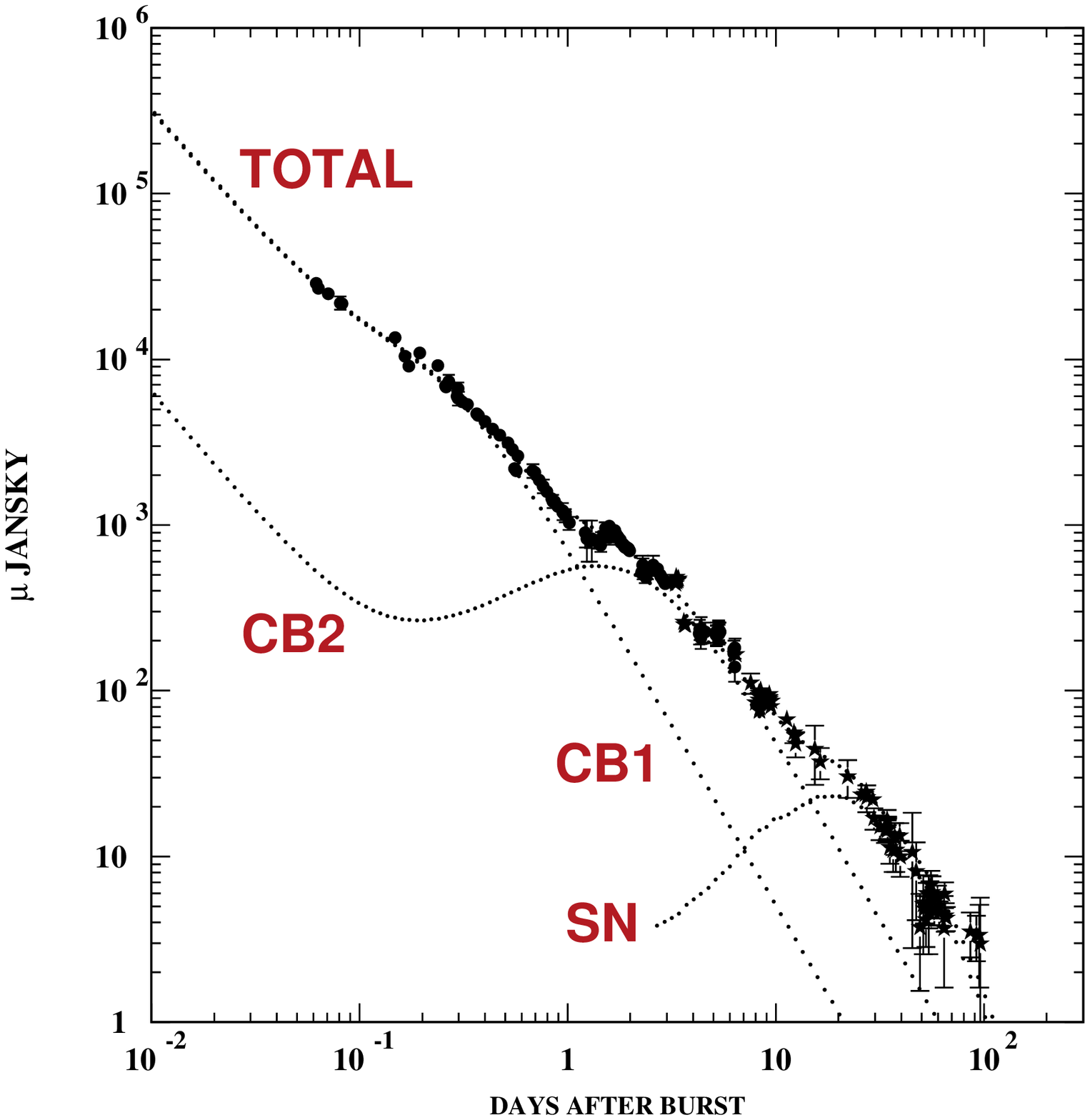,height=1\linewidth}
}
\vspace*{8pt}
\caption{Left: The R-band AG of GRB 030329, 
used along with other optical data to
predict, in the CB model, the presence of a SN akin to SN1998bw.
 Right: The subsequent data (the $\star$ symbols) are added.}
 \label{329red}
\end{figure}

Naturally, truly `standard torches'' do not exist, but SN1998bw
made such a good job of it that it was possible to 
{\it predict}~\cite{DDDSN,DDD329a}
the  SN contribution to the AG in all recent cases of
early detection of the AGs of nearby GRBs 
(000911, 010921, 010405, 012111, 021211 and 030329). 
Besides the  980425--1998bw pair,  
the most convincing pre-Swift associations were provided by the  spectroscopic
discoveries of a SN in the AGs of GRBs 030329~\cite{SH} 
and 021211~\cite{DV}. For GRB 030329, shown in Fig.~\ref{329red},
even the exact date when the SN 
would be bright enough to be discovered was foretold~\cite{DDD329a}. 

By now, the association between
GRBs and core-collapse SNe (perhaps only of Types Ib,c) is fully
established. The long history of this conclusion is reported
in some detail in \cite{DDDNoSN}.

\subsection{What fraction of SNe produce GRBs?}
\label{association}

From a CB-model analysis of GRBs and their 
AGs~\cite{AGoptical,AGradio,DDDSN,DDD329a} we conclude
that GRBs more distant than GRB 980425 are observable, with
past and current instruments, only for $\theta\leq$ 2--3 mrad. 
With two jets of very precisely collinear CBs per GRB,  only a fraction
\begin{equation}
f\sim 2\, {\pi\, \theta^2 / (4\, \pi)}\sim
(2\;\rm{to}\; 4.5)\times 10^{-6}
\label{fraction}
\end{equation}
of SN-generated GRBs are observable. 

The local rate of long-duration GRBs is estimated to be~\cite{Schmidt} 
$(2.5\pm 1.0) \times 10^{-10}\; {\rm Mpc^{-1}\, y^{-1}}$
for the current cosmology. The local rate of core-collapse SNe~\cite{Capellaro}
 is $(7.5\pm 3.8)\times 10^{-5}\; {\rm Mpc^{-1}\, y^{-1}}$. The ratio
of these rates, $(3.3\pm 2.1)\times 10^{-6}$, is consistent with the
fraction $f$ of observable GRBs. Thus, within the pervasive cosmological 
factor of a few, the long-GRB--SN association would be 1 to 1. Yet, the
CBs of GRBs may not be emitted {\it precisely} in the same
direction; a distribution of emission angles of a few mrad width, for instance, 
would not affect the results of the CB model, but would significantly increase $f$ in 
Eq.~(\ref{fraction}), diminishing the fraction of GRB-generating SNe.
Current data are insufficient to determine whether long-duration
GRBs are associated with all core-collapse SNe (some 70\% of all SNe,
including Type II) or only with Type Ib/c SNe (some 15\% of all core-collapse
SNe). Most well observed GRB-associated SNe are compatible with the
latter of these possibilities. 

Limits on the rate of SNe which produce GRBs, derived from wide-field deep
surveys for radio-transient sources \cite{Levinson} 
or in a targeted survey of optically selected SNe Ib/c \cite{Soderberg}, have 
relied on the accepted but probably incorrect assumption  
\cite{Rhoads,Sari,Frail} that the relativistic jets that produce GRBs are conical or
trumpet-like in shape, stop shortly near their
ejection site, and become an isotropic radio source, unlike in
our discussion in Appendix
\ref{Jets}. As noted by Gal-Yam et al.~\cite{Levinson}, the limits are invalid,
if the jets from SN explosions are similar to those fired by
microquasars, as advocated by the CB model.

\subsection{CB-model evidence for a SN--GRB association}
\label{GRBXRF}

Perhaps the best evidence that long-duration GRBs from SN explosions are
produced by narrowly collimated relativistic jets, which are ejected in the
explosion, comes from the remarkable success of the CB model in predicting
 the fluence, spectral and temporal properties of GRBs and of their
AGs, despite their apparent complexity and large diversity.

For the GRB phase the CB-model consequences of a SN--GRB association
---based exclusively on Compton scattering as the $\gamma$-ray-generating
mechanism and on the hypothesis that the `windy' material is 
less dense than average in the  `polar' directions--- are essentially the 
list of properties of GRBs~\cite{DD}.
To wit:
\begin{itemize}
\item
The characteristic peak energy 
of the $\gamma$-rays: $E={\cal{O}}(250)$ keV, as
observed by BATSE \cite{Preece}
and BeppoSAX \cite{Amati}.

\item
The distribution of the `peak'  energies of the GRB
spectra \cite{Preece}. 

\item
The duration of the single pulses of GRBs: a median $\Delta t\sim 1/2$ s 
FWHM \cite{McBreen}.

\item
The typical (spherical equivalent) number of photons per pulse, 
$N_\gamma\sim 10^{59}$
on average, which, combined with the characteristic $\gamma$ energy, 
yields the average total (spherical equivalent) fluence of a 
pulse: $\sim 10^{53}$ erg \cite{DD}.

\item
The general {\it FRED} pulse-shape: a very `fast rise' followed by a fast 
decay $N(t)\propto 1/t^2$,
inaccurately called `exponential decay' \cite{Nemiroff,McBreen}. 

\item
The $\gamma$-ray energy distribution, $dN/dE\sim E^{-\alpha}$,
with, on average, $\alpha\sim 1$ exponentially evolving into $\alpha\sim 2.1$,
and generally well fitted \cite{Band} by the so-called  `Band function''.

\item
The time--energy correlation of the pulses: the pulse duration 
decreases like $\sim E^{-0.4}$ and peaks earlier the higher the energy 
interval \cite{Fenimore}.

\item
Various correlations between pairs of the following observables:
photon fluence, energy fluence, peak
intensity and luminosity, photon energy at peak intensity or luminosity,
and pulse duration \cite{DD,DDS05}. 

\item
The possibly large polarization of the $\gamma$-rays \cite{Coburn}.
\end{itemize}

In the CB model, X-ray flashes are simply GRBs viewed at larger angles,
which makes their fluence and the energy of their quanta
smaller, and their time structure less rugged~\cite{DDDXRF}.
The recent progress in the understanding of GRBs and XRFs is
discussed in Section \ref{Swift}.

\subsection{The ``prompt" phase of a GRB or XRF}
\label{promptphase}

The distinction between a ``prompt" and an ``after"-glow phase
---in terms of a given transition time--- is arbitrary. Inverse
Compton scattering dominates the early $\gamma$- and X-ray
production, while synchrotron radiation (SR) from swept-in 
ISM electrons spiraling in the CB's enclosed magnetic field dominates the
 X-ray signal at late times (the ``late" $\gamma$-ray flux is too weak
to be observable). For X-rays ``prompt" and ``ICS-dominated" are
equivalent, and so are ``afterglow" and ``SR-dominated". But at
optical frequencies there are cases in which SR dominates during the 
``prompt" X-ray phase, and cases of ICS dominance in the
 X-ray ``afterglow" phase.
We summarize ICS in the this subsection, SR in the next.

The $\gamma$-rays of a single pulse of a GRB are 
produced as a CB coasts through the glory. The electrons enclosed in the 
CB boost the energy of the glory's photons, via ICS, 
to $\gamma$-ray energies. The initial fast expansion of the 
CBs and the radially-increasing transparency of the windy environment 
result in the exponential rise of a GRB pulse.  As a CB proceeds, the distribution of 
the glory's light becomes more radially directed, its density decreases. 
Consequently, the energy of the observed photons is continuously shifted 
to lower energies as their number plummets. 
During a GRB pulse the spectrum softens and the 
peak energy decays with time, tending to a power law. This is also the behaviour of 
the X-ray ``flares" of a GRB, which are either the low-energy tails of 
$\gamma$-ray pulses, or fainter and softer signals with the same origin.

The above effects can be explicitly analized \cite{DD}, and 
summarized to a good approximation
in a {\it master formula}
 for the temporal shape and spectral evolution of the energy fluence
of an ICS-generated $\gamma$-ray pulse (or X-ray flare). For a
single pulse starting at time $t=0$:
\begin{eqnarray}
&&\!\!\!\!\!\!\!\!\!\!\!\!\! F_E\propto E\, {d^2N_\gamma\over dt\,dE} \propto
\nonumber \\
&&\!\!\!\!\!\!\!\!\!\!\!\!\! \Theta[t]\;
e^{-[\Delta t/(t)]^m}\,
 \left\{1-e^{-[\Delta t/(t)]^n}\right\}\, E\,{dN_\gamma(E,t) \over dE}\; .
\label{GRBlc}
\end{eqnarray}
The time scale is set by $\Delta t$, with $\gamma\, \delta\, c\, \Delta t/(1+z)$  
the radius of transparency of the glory, within which its photons are
approximately isotropic. In $\Delta t$ time units, a pulse rises 
as ${\rm exp}[-1/t^m]$, $m\!\sim\!1$ to 2, and decreases as $1/t^n$, $n\!\sim\!2$. Finally,
$E\, dN_\gamma/dE$ is the spectral function of the glory's 
photons, up-scattered by  the CB's electrons, and discussed anon.

The glory has a thin thermal-bremsstrahlung spectrum:
$E_i\, {dn_\gamma / dE_i} \!\sim\!
(E_i/T_i)^{1-\alpha_g}\,e^{-E_i/T_i}$,
with a typical (pseudo)-temperature
$T_i \!\sim\!1$ eV, and index $\alpha_g\!\sim\!1$.
During the $\gamma$-ray phase of a GRB,
the Lorentz factor $\gamma$ of a CB stays put at its initial value,
for the deceleration induced by
the collisions with the ISM has not yet had a significant effect.
Let $\theta_i$ be the angle of incidence of the initial
photon onto the CB, in the parent star's rest system.
The energy of an observed photon, 
Compton scattered in the glory by an electron comoving with a
CB at redshift $z$, is given by
$E\!=\!\gamma\, \delta\, E_i\, (1\!+\!\cos\theta_i)/(1+z)$.
The predicted GRB prompt spectrum is \cite{DD}:
\begin{equation}
E\, {dN\over dE} \sim \left({E\over T}\right)^{1-\alpha_g}\,
 e^{-{E\over T}}+ b\,(1-e^{-{E\over T}})\, 
 \left({E \over T}\right)^{-{\beta_{\rm ac}\over 2}}.
\label{GRBspec}
\end{equation}
The first term, with $\alpha_g\!\sim\! 1$, is the result of ICS
by the bulk of the CB's electrons, which are comoving with it.
The second term in is induced by a very small fraction of
`knocked on' and Fermi-accelerated electrons, whose initial spectrum
(before Compton and synchrotron cooling) is 
$dN/dE_e\!\propto\! E_e^{-\beta_{\rm ac}}$,
with an index assumed to be the same for electrons and
nuclei, $\beta_{\rm ac}\!\approx\! 2.2$, see Eq.~(\ref{Acc}).
Finally, $T$ is the effective (pseudo)-temperature
of the GRB's photons,
$T\!\equiv\!{4\, \gamma\, \delta\,T_i\,
\langle 1\!+\!\cos\theta_i\rangle / [3\, (1+z)]}$.
For a semi-transparent glory $\langle\cos\theta_i\rangle$ would be
somewhat smaller than zero.

For $b={\cal{O}}(1)$,
the energy spectrum predicted by the CB model, Eq.~(\ref{GRBspec}),
bears a striking resemblance
to the Band function \cite{Band} traditionally used to model the
energy spectra of GRBs \cite{Wigger}. For many Swift GRBs the spectral observations
do not extend to energies much bigger than $T$, or the value of $b$
in Eq.~(\ref{GRBspec}) is relatively small, so that the first term
of the equation provides a very good approximation.
At later times, the CB is sampling the glory at distances
for which its light is becoming increasingly radial, 
$\langle1\!+\!\cos\theta_i\rangle\!\to\!1/r^2\!\propto\!1/t^2$.
For a pulse starting at $t\!=\!0$, the
value of $E_p(t)$ consequently decreases as: 
\begin{equation}
E_p(t)\approx E_p(0)\,
\left[1-{t\over \sqrt{\Delta t^2+t^2}}\right]\, .
\label{Epi}
\end{equation}
The light-curve of a single CB is well approximated by: 
\begin{eqnarray}
&&\!\!\!\!\!\!\!\!\!\!\!\!\! F_E \approx \nonumber\\
&&\!\!\!\!\!\!\!\!\!\!\!\!\!  \Theta[t]\;
e^{-\left[{\Delta t\over t}\right]^2}
 \left\{1-e^{-\left[{\Delta t\over t}\right]^2}\right\}
 \left[{E\over E_p(t)}\right]^{1-\alpha_g} 
e^{-\left[{E\over E_p(t)}\right]}\,
\label{GRBXlc}
\end{eqnarray}
until ICS is overtaken by synchrotron radiation. The generalization
to a multi-pulse GRB is straightforward.

\subsection{The Synchrotron Radiation ``afterglow"}
\label{GRBAGs}

In the CB model, the AGs of GRBs and X-ray flashes 
(XRFs) consist of three contributions,
from the CBs themselves, the concomitant SN, and the host galaxy:
\begin{equation}
F_{\rm AG}=F_{_{\rm CBs}}+F_{_{\rm SN}}+F_{_{\rm HG}}\, .
\label{sum}
\end{equation}
The latter contribution is usually determined
by late-time observations, when the CB and SN contributions become
negligible, or from measurements with sufficient angular resolution
to tell apart $F_{_{\rm CBs}}+F_{_{\rm SN}}$ from $F_{_{\rm HG}}$.

The first convincing observation of a GRB--SN association was that
of GRB 980425 with SN1998bw, at a record-low redshift $z_{bw}=0.0085$ 
\cite{Galama}. In the CB model, we often used this SN as the `template',
or candidate `standard candle' associated with GRBs.
Let the unattenuated energy flux density of SN1998bw   
be $F_{bw}[\nu,t]$. For a similar SN placed at a redshift $z$ 
\cite{GRB1,AGoptical,AD1999}:   
\begin{equation}
F_{\rm SN}[\nu,t] = {1+z \over 1+z_{bw}}\;
{D_L^2(z_{bw})\over D_L^2(z)}\, A(\nu,z)\, F_{bw}[\nu',t']\, ,
\label{bw}
\end{equation}
where $A(\nu,z)$ is the attenuation along the line of sight,
$\nu'=\nu\, (1+z)/ (1+z_{bw})$,  and $t'=t\, (1+z_{bw})/(1+z)$.     
The simple ansatz that {\it all} long-duration GRBs would be
associated with SN1998bw-like SNe \cite{AD1999,DP,GRB1,AGoptical} 
proved to be unexpectedly precise and successful;
see Appendix~\ref{GRBSN}.

The time 
dependence of $\gamma(t)$ ---and, consequently, of the Doppler factor $\delta(t)$ 
of Eq.~\ref{Doppler}--- is obtained from 
Eq.~(\ref{stop}) and the relation between the {\it observer's} time, $t$,
and the travel-distance 
in the SN rest frame \cite{DD}, $dx =c \,\gamma(t)\, \delta(t)\, dt/(1+z)$ 
[notice how $dx$ and $c\,dt$ may differ by a factor ranging up to ${\cal{O}}(10^6)$].
Typically, within minutes of observer's time,  a CB reaches its roughly
constant `coasting' radius, $R_0\!=\!{\cal{O}}(10^{14}\,\rm cm)$, which increases
slowly until the CB finally stops and blows up, as in Eq.~(\ref{best}).
Up to the end of the coasting phase, and in a constant density ISM, $\gamma(t)$
obeys:
\begin{eqnarray}
&&\!\!\!\!\!\!\!\!\!\!\!\!\!\!\!({\gamma_0/ \gamma})^{4}+
2\,\theta^2\,\gamma_0^2\,(\gamma_0/\gamma)^{2}
= 1+2\,\theta^2\,\gamma_0^2+t/t_0\,,
\nonumber\\
&&\!\!\!\!\!\!\!\!\!\!\!\!\!\!\!{t_0\over 1+z} = {N_{_{\rm B}}\over
8\,c\, n_p\,\pi\, R_0^2\,\gamma_0^3}=\nonumber\\
&&\!\!\!\!\!\!\!\!\!\!\!\!\!\!\!(1300\,{\rm s}) 
\left[{10^3\over \gamma_0}\right]^{3}\,
\left[{10^{-2}\, {\rm cm}^{-3}}\over n_p \right]
\left[{10^{14}\,{\rm cm}\over R_0}\right]^{2}
\left[{N_{_{\rm B}}\over 10^{50}}\right] 
\label{deceleration}
\end{eqnarray}

We have assumed that a CB's magnetic field is
in approximate energy equipartition with the energy of the
intercepted ISM, $B\approx \sqrt{\pi\, n\, m_p\, c^2}\, \gamma$.
In this field, the intercepted
electrons emit synchrotron radiation. The SR, isotropic in the CB's
rest frame, has a characteristic frequency, $\nu_b(t)$,
the typical frequency radiated by the
electrons that enter a CB at time $t$ with a relative Lorentz
factor $\gamma(t)$. In the observer's frame:
\begin{equation}
\nu_b(t)\simeq  {\nu_0 \over 1+z}\,
{[\gamma(t)]^3\, \delta(t)\over 10^{12}}\,
\left[{n_p\over 10^{-2}\;\rm cm^3}\right]^{1/2}
{\rm Hz}.
\label{nub}
\end{equation}
where $\nu_0\!\sim\! 1.8\times 10^{16}\, \rm Hz \simeq 112\, eV$. 
The spectral energy density of the SR
from a single CB at a luminosity distance $D_L$  is given by \cite{AGoptical,AGradio}:
\begin{eqnarray} 
&&\!\!\!\!\!\!\!\!\!\!\!\!F_{_{\rm CB}} \simeq {\eta\,  \pi\, R_0^2\,n_e\, m_e\, c^3\,
\gamma(t)^2\, \delta(t)^4\, A(\nu,t)
\over 4\,\pi\, D_L^2}\,S(\nu,t)
\nonumber\\
&& \!\!\!\!\!\!\!\!\!\!\!\!S(\nu,t)\approx {1\over \nu_b(t)}\;
{\beta_{\rm ac}-2\over \beta_{\rm ac}-1}
\left[{\nu\over\nu_b(t)}\right]^{-{1\over2}}
\left[1 + {\nu\over\nu_b(t)}\right]^{1-\beta_{\rm ac}\over2}
\label{Fnu}
\end{eqnarray}
where $\eta\!\approx\!1$ is the fraction of the impinging ISM 
electron
energy that is synchrotron re-radiated by the CB, and $A(\nu, t)$ is
the  attenuation of photons of observed frequency $\nu$ along the
line of sight through the CB, the host galaxy, the IGM        
and the Milky Way \cite{pfootnote}.

At all times, X-rays are above the frequency $\nu_b$ 
in Eq.~(\ref{nub}).
It then follows from Eq.~(\ref{Fnu}) that
 the {\it unabsorbed} X-ray spectral energy density has the form:
\begin{eqnarray}
F_{_{\rm CB}} &\propto& R_0^2\, n_e^{(\beta_{\rm ac}+2)/4}\,
\gamma^{(3\beta_{\rm ac}-2)/2}\, \delta^{(\beta_{\rm ac}+6)/2}\,  
\nu^{-\beta_{\rm ac}/2}\nonumber\\
&=&
R_0^2\, n_e^{\Gamma/2}\,
\gamma^{3\,\Gamma-4}\, \delta^{\Gamma+2}\, \nu^{-\Gamma+1}\, ,
\label{Fnux}
\end{eqnarray}
where we used the customary notation $dN_{\gamma}/dE\!\approx\!E^{-\Gamma}$.
Notice that the time evolution of the entire AG ---via its $\gamma(t)$ and 
$\delta(t)$ dependences--- is linked to its spectral behaviour. When testing
this relation, which has a simpler asymptotic form discussed in the next
paragraph, we fit the entire AG evolution in time \cite{DDDDecline}.

The functions $\delta(t)/\delta_0$ and $\gamma(t)/\gamma_0$
evolve slowly, up until a time $t_b\!=\!(1+2\, \theta^2\,\gamma_0^2)\,t_0$,
with $t_0$ as in Eq.~(\ref{deceleration}). 
The quantity $t_b$ characterizes the {\it deceleration bend-time}
of the CB model; Eq.~(\ref{deceleration}) for $\gamma(t,t_0,\theta,\gamma_0)$
describes the gradual character of this bend or `break'.
At later times  Eq.~(\ref{deceleration})
implies that $\gamma\to\gamma_0\,(t/t_0)^{-1/4}$, and Eq.~(\ref{Doppler})
that $\delta\to 2\,\gamma$. Thus, at $t\! \gg \!t_b$,
Eq.~(\ref{Fnux}) yields:  
\begin{eqnarray}
F_{_{\rm CB}}(t)&\propto&
 t^{-1/2-\beta_{\rm ac}/2}\,\nu^{-\beta_{\rm ac}/2}= 
t^{-\Gamma+1/2}\, \nu^{-\Gamma+1},
\nonumber\\
\beta_{\rm ac}&=&2\,(\Gamma -1),
\label{Asymptotic}
\end{eqnarray}
with a predicted power decay in time half a unit steeper than in
frequency, as long as the ISM has an approximately constant
density. Density inhomogeneities complicate the shape of AGs.
In an ISM with an approximate $1/r^2$ density profile, such
as the galactic halos that CBs may reach late in their motion,
the X-ray and optical time- and frequency- 
indices would differ by one unit, as opposed to one-half unit.

The prompt ultraviolet-to-infrared AG depends critically on the complex 
density profile along the CB's trajectory and on the extinction along the line 
of sight. 
These complex environments can produce very bright and fast
declining early-time AGs \cite{Akerlof,DDDdens}, or strongly extinct the
early AG \cite{Rykoff}, or even decelerate completely the CB, producing a
`dark GRB' \cite{Malessani}.

In the radio, the AG spectrum
is also affected by self-absorption in the CBs themselves,
characterizable by one parameter per CB: a 
`free--free' absorption frequency $\nu_a$ \cite{AGradio}.
At late time the radio emission from CR electrons ejected into the ISM 
along the CB trajectory can become important and even dominate the radio AG
\cite{DDD329c}.

We had previously posited and concluded \cite{DDD2003Lines,DDDSwift} 
that three mechanisms successively dominate the 
radiation of a GRB: ICS in the prompt phase, 
thermal bremsstrahlung and line emission in the fast-declining X-ray 
phase, SR thereafter. The line emission phase was 
supported by the claimed observations of X-ray lines in early GRB 
afterglows \cite{Xlines} and their very natural CB-model interpretation 
\cite{DDD2003Lines}. But these observations were of very limited statistical 
significance, and a phase during which line-emission significantly 
contributes may not, after all, be inevitably required.

 The above description of GRB synchrotron radiation
AGs is very simple and successful, 
and provides support to the CB model description 
of SN jets \cite{AGoptical,AGradio,DD,DDD329b}.
We extract from the corresponding
fits to the data the typical values of the CB parameters needed
as inputs in our analysis of CRs and of the prompt $\gamma$-rays
of GRBs. This is entirely analogous to what we did to predict
the properties of the prompt ICS-dominated phase of a GRB 
from the parameters extracted from their AGs, and some other
independent input ``priors" \cite{DD}.

\section{Short Hard GRBs}
\label{SHB}

The origin of Short Hard Bursts (SHBs) is not 
established \cite{Nakar}, in contrast to that of the 
longer-duration softer-spectrum GRBs. 
The SHB spectra and pulse-shapes
are akin --except for hardness and duration-- to those of long GRBs.
The X-ray light curves of some well-sampled SHBs \cite{Burrows}
are `canonical'.
The similarities suggest common mechanisms generating the GRB and SHB 
radiations. This is expected in the CB model, wherein both burst types are 
produced by  jets of CBs \cite{Dar1995}. The `engine' 
is different; it is a core-collapse SN for GRBs and XRFs, in SHBs it may be
the result of mass accretion onto a compact object in a close binary system.
In a SHB, the ambient light may be scattered by prior ejecta of the
progenitor system (as in GRBs) or emitted by an accretion 
disc, or by the companion star. 

In the CB model, the expressions
that describe the promt and AG 
emissions of long GRBs are directly applicable to SHBs, provided the 
parameters of the CBs, of the glory, and of the circumburst environment, are 
replaced by those adequate for SHBs \cite{DDDSGRB}.

The fluence of SHBs localized by Swift is 
2 to 3 orders of magnitude smaller than for ordinary GRBs \cite{Berger}.
Their rate, measured (with different efficiency)
by BATSE and the interplanetary network, is
$20\%$ of the rate of long-duration GRBs.
Assuming the same conversion efficiency of jet energy into $\gamma$-rays
in SHBs and long GRBs, the CR luminosity produced by SHBs
is  also much smaller than that produced by core-collapse SNe.

\section{Power supply by other cosmic accelerators}
\label{Accelerators}

{\it Pulsars} are born with typical periods of $P \gtrsim 30$ ms 
\cite{Lorimer}.  With
a moment of inertia $I\sim 10^{45}\, {\rm gm\, cm^2}$ corresponding to a
typical mass $\sim 1.4\, M_\odot$ and a radius $\sim 10$ km, their typical
rotational energy is $E_{\rm rot} \simeq I\, \Omega^2/2=2 \pi^2 I/ P^2\simeq
2\times 10^{49}$ erg. Most of this energy, which is two orders of
magnitude smaller than the typical kinetic energy release in a core-collapse SN
explosion, is radiated as magnetic dipole radiation and only a small
fraction of it can be used to accelerate CR nuclei.

{\it Soft Gamma-Ray Repeaters} (SGRs) are slowly rotating ($P\sim$ 8--12
s), newly born pulsars which produce repeated `soft' $\gamma$-ray bursts.
Their rotational energy is too small to power either their persistent
emission or their soft and hard $\gamma$-ray activity. 
Less than once in 30 years, they erupt in a hyperflare, such as that of the
Galactic SGR 1806-20 on 27 December 2004, whose entire electromagnetic
energy release was concentrated in a short spike of hard
$\gamma$-rays that could have been interpreted as a normal
SHB, had it
taken place in an external galaxy within a distance of $\lesssim 30$ Mpc
\cite{Hurley}. But, if the hyperflare 
was relativistically beamed and was viewed slightly off axis, near-axis
hyperflares from SGRs in external galaxies can be seen from much
larger distances and may also be a source of SHBs \cite{DDDSGRB}. 
As for SHBs in general, the contribution of SGRs to CRs would be negligible.

{\it Neutron-star mergers} (NSMs)  are likely to
produce relativistic jets which can be an origin of both SHBs
\cite{GDN} and UHECRs \cite{DLS}.  The estimated Galactic
rate of NSMs \cite{Phinney} is $\sim 1.8\times 10^{-4}\, {\rm y^{-1}}$,
which is smaller than the Galactic SN rate by $\sim 2$ orders of magnitude;
see Appendix~\ref{rateSN}. Since, in addition, the kinetic energy of the jets
that produce SHBs is smaller than for
 GRBs, the contribution of
NSMs to the CR luminosity is negligible.

{\it Microquasars} fire mildly relativistic CBs.  Power-law radio emission
from their CBs provides evidence that they accelerate high-energy CR 
electrons \cite{Felix}. Thus, microquasars were suggested as sources of 
relativistic CRs \cite{HS}. But the total kinetic power of the jets 
of a dozen or
so microquasars in our Galaxy \cite{Felix,Kaaret} is smaller than that of
SN jets by 2 to 3 orders of magnitude, consistent with estimates
\cite{HS} of their relative contribution to the Galactic CR luminosity.


\begin{thebibliography}{999}
\bibitem{cosmopriors}%
The distinction between parameters and
priors has become established in the analysis of the CMB.
\bibitem{pfootnote}
The normalization in Eq.~(\ref{nub}) is only correct for $\beta_{\rm ac}\!>\!2$,
for otherwise the norm diverges. The cutoffs for the $\nu$ distribution are
time-dependent,
dictated by the acceleration and SR times of electrons  and their `Larmor'
limit. The discussion of these processes being complex,
we satisfy ourselves here with the statement that for
$\beta_{\rm ac}\!\leq \!2$ the AG's normalization is not predicted.
\end{thebibliography}
\end{document}